\documentclass{emulateapj}

\usepackage{psfig}

\newcommand{\uu}	{$U$}%
\newcommand{\bb}	{$B$}%
\newcommand{\vv}	{$V$}%
\newcommand{\rr}	{$R$}%
\newcommand{\ha}	{H$\alpha$}%
\newcommand{\hb}	{H$\beta$}%
\newcommand{\etal}	{et~al.}%
\newcommand{\figref}[1]	{Fig.~\ref{#1}}%
\newcommand{\figrefm}[2]{Figs.~\ref{#1}--\ref{#2}}%
\newcommand{\figreftwo}[2]{Figs.~\ref{#1} and \ref{#2}}%
\newcommand{\eqref}[1]	{equation~(\ref{#1})}%
\newcommand{\mum}	{$\mu$m}%
\newcommand{\betav}	{$\beta_V$}%
\newcommand{\bzero}	{$\beta_{V,{\rm 0}}$}%
\newcommand{\av}	{$A_V$}%
\newcommand{\HII}	{\mbox{H\,\textsc{ii}}}%
\newcommand{\HH}	{\mbox{H$_2$}}%
\newcommand{\Paa}	{\mbox{Pa$\alpha$}}%
\newcommand{\ch}{\colhead}%
\begin{document}

\title{MAPPING THE SPATIAL DISTRIBUTION OF DUST EXTINCTION IN NGC\,959
  USING BROADBAND VISIBLE AND MID-IR FILTERS}

\shorttitle{The Two-Dimensional Dust Extinction in NGC\,959}

\author{K. Tamura\altaffilmark{1}, R. A. Jansen\altaffilmark{2,1} 
and R. A. Windhorst\altaffilmark{2,1}}

\altaffiltext{1}{Department of Physics, Arizona State University, 
  Tempe, AZ 85287-1504, USA; \texttt{ktamura@asu.edu}}

\altaffiltext{2}{School of Earth and Space Exploration, Arizona State 
  University, Tempe, AZ 85287-1404, USA; \texttt{rolf.jansen@asu.edu, 
  rogier.windhorst@asu.edu}}

\accepted{\footnotesize for publication in AJ 2009 September 24}

\email{ktamura@asu.edu}
\shortauthors{Tamura et al.}

\begin{abstract}

We present a method to estimate and map the two-dimensional distribution of 
dust extinction in the late-type spiral galaxy NGC\,959 from the theoretical 
and observed flux ratio of optical \vv\ and mid-IR (MIR) 3.6\,\mum\ images. 
Our method is applicable to both young and old stellar populations for a 
range of metallicities, and is not restricted to lines-of-sight toward 
star-formation (SF) regions.  We explore this method using a pixel-based 
analysis on images of NGC\,959 obtained in the \vv-band at the Vatican 
Advanced Technology Telescope (VATT) and at 3.6\,\mum\ ($L$-band) with 
\emph{Spitzer}/IRAC.  We present the original and extinction corrected 
\emph{GALEX} far-UV (FUV) and near-UV (NUV) images, as well as optical 
\emph{UBVR} images of NGC\,959.  While the dust lanes are not clearly 
evident at \emph{GALEX} resolution, our dust map clearly traces the dust 
that can be seen silhouetted against the galaxy's disk in the 
high-resolution \emph{HST} images of NGC\,959.  The advantages of our method 
are: (1) it only depends on two relatively common broadband images in the 
optical \vv-band and in the MIR at 3.6\,\mum\ (but adding a near-UV band 
improves its fidelity); and (2) it is able to map the two-dimensional 
spatial distribution of dust within a galaxy.  This powerful tool could be 
used to measure the detailed distribution of dust extinction within higher 
redshift galaxies to be observed with, e.g., the \emph{HST}/WFC3 
(optical--near-IR) and \emph{JWST} (mid-IR), and to distinguish properties 
of dust within galaxy bulges, spiral arms, and inter-arm regions.  
\end{abstract}

\keywords{dust, extinction --- galaxies: individual (NGC\,959) --- 
galaxies: spiral --- galaxies: structure}

\section{INTRODUCTION}

Dust extinction is a longstanding problem when studying stellar populations 
within our Galaxy and in extragalactic objects \citep[e.g.,][]{trumpler30, 
mathis77, viallefond82, caplan85, roussel05, driver08}.  Variations in the 
amount and the spatial distribution of interstellar dust have different 
effects on the light from background stellar populations  \citep[e.g.,][]{
elmegreen80, walterbos88, waller92, witt92, calzetti94, deo06}.  Measuring 
and correcting for dust extinction in individual galaxies is important to 
understand the true nature of their stellar populations, especially when 
one aims to study small-scale structures such as \HII\ regions of spiral 
galaxies, which tend to be particularly dusty \citep[e.g.,][]{petersen97,
scoville01, calzetti05}. 

Many different methods are used to estimate dust extinction within target 
galaxies.  Some commonly used methods involve: (1) the ratios of Hydrogen 
recombination-line fluxes for \HII\ regions, e.g., \ha/\hb\ 
\citep[e.g.,][]{caplan85, caplan86, maiz04, roussel05, relano06} or of 
\ha/\Paa\ \citep[e.g.,][]{petersen97, scoville01, calzetti05, calzetti07}; 
(2) the ratio of the FIR and UV fluxes \citep[e.g.,][]{buat96, calzetti00, 
calzetti05, calzetti07, charlot00, witt00, bell02a, boselli03, panuzzo03, 
boissier04}; (3) the UV spectral slope \citep[e.g.,][]{calzetti94, 
meurer99, bell02b}; and (4) the CO column density from (sub-)millimeter 
observations of, e.g., $^{12}$CO($J$\,=\,1--0) line and $J$\,=\,1--0 and 
$J$\,=\,2--1 transitions of $^{13}$CO and C$^{18}$O \citep{encrenaz75, 
dickman78, frerking82, harjunpaa96, hayakawa99, hayakawa01, harjunpaa04}.  

Even though all these methods measure dust extinction in some way, each 
has its limitations.  The Balmer decrement, \ha/\hb, and the \ha/\Paa\ 
flux ratios are generally limited to lines-of-sight toward \HII\ regions.  
These \HII\ regions can be distributed all over a galaxy, but cover only 
a small fraction of an entire galaxy disk \citep[e.g.,][]{scoville01}.  
Accurate measurements of the underlying Balmer absorptions from 
moderate-resolution spectra are also required \citep{petersen97, 
boissier04}.  The 
FIR/UV flux ratio can be used throughout a galaxy, although the spatial 
resolution in the FIR is generally poor: about $\sim$\,20\arcsec\ for the 
Midcourse Space Experiment (MSX) at 4.2--26\,\mum\ \citep[][]{price02}, 
$\sim$\,40\arcsec\ for \emph{Spitzer}/MIPS at 160\,\mum\ \citep{rieke04}, 
and even larger ($\sim$\,100\arcsec) for IRAS at 100\,\mum\ \citep{xu96, 
boissier04}.  This becomes a significant limitation in studying small-scale 
structural features, and limits access to only the very nearest galaxies. 
Unlike the FIR images, the UV images can have much higher resolution 
(FWHM\,$\lesssim$\,0\farcs1) with the \emph{HST} WFPC2 and ACS \citep{
trauger94, sirianni05}, or FWHM\,$\lesssim$\,4--6\arcsec\ for \emph{GALEX} 
\citep{morrissey05}.  A major complication of methods involving UV filters 
(FIR/UV flux ratio and UV spectral slope) is that the UV emission is 
significantly affected not only by dust, but also by the age and metallicity 
of stellar populations.  Finally, sub-millimeter and millimeter observations 
are used to measure the column density of CO molecules directly, and hence 
--- under certain assumptions --- that of molecular hydrogen, \HH.  The CO 
(and hydrogen) column density is converted to extinction using the 
correlation found between CO and dust-extinction in nearby galaxies 
\citep[e.g.,][]{dickman78, bachiller86, komugi05}.  The typical resolution 
at millimeter wavelengths is poor \citep[FWHM\,$\simeq
$\,20\arcsec--2\farcm6;][]{encrenaz75, dickman78, frerking82, harjunpaa96, 
hayakawa99, hayakawa01, harjunpaa04}, although ALMA \citep{brown04} will 
soon greatly improve on this.  Studies such as that of \citet{boissier07} 
also show that the relationship between the dust column density and the 
amount of extinction is complicated, and that further study is necessary 
to derive the relationship between the two values.

Without using these methods, \citet{regan00} investigated dust extinction 
using the more commonly used ground-based optical \emph{BVRI} and near-IR 
(NIR) \emph{JHK} filters to obtain various color maps.  The advantage of 
using optical--NIR filters is that the extinction measurement is independent 
of the dust temperature \citep{regan00}, and that the images can have much 
higher spatial resolution at these wavelengths \citep[e.g.,][]{regan99}.  
Combining theoretical models and color maps from optical--NIR images, the 
spatial dust distribution can then be derived.  Fig.~7 of \citet{regan00} 
shows that this method can reconstruct extinction-free galaxy images.  
However, as \citet{regan00} points out, there are several issues with 
this particular method.  The first is that all of the bandpasses involved 
are affected by the dust to some degree: no single optical--NIR filter 
directly maps either stellar or dust morphology by itself.  The second 
issue is that, while the intrinsic colors of the underlying stellar 
populations are well known for the older stellar populations, the colors 
of younger stellar populations in star-formation (SF) regions vary, 
depending on their environment and specific properties.  Since \citet{
regan00} treated all stellar populations as old stellar populations, the 
extinction measurements for stellar populations in SF-regions are less 
accurate and should be treated with care, as their paper points out.   

In this paper, we will map the two-dimensional spatial distribution of dust 
extinction in NGC\,959, using images from commonly used optical--MIR filters.
Many images observed with optical--MIR filters from both ground and space 
are now readily available for large samples of galaxies though publicly 
available archives.  We will also treat the younger stellar populations 
separately from older stellar populations to measure dust extinction 
in SF-regions.  We initially combine data spanning the \emph{GALEX} FUV 
and NUV, the ground-based optical \emph{UBVR} filters, the 2MASS NIR 
$JHK_s$ filters, and the \emph{Spitzer}/IRAC 3.6, 4.5, 5.8, and 8.0\,\mum\ 
(MIR) filters.  The IRAC 3.6 and 4.5\,\mum\ filters are not nearly as 
affected by dust extinction as the NIR filters, and therefore are commonly 
used to trace the distribution of stellar populations \citep[e.g.,][]{
willner04}.  We use the library of spectral energy distributions (SEDs) by 
\citet[][hereafter AF03]{anders03}, to study possible combinations of 
filters and colors in the analysis of dust extinction.  Using a pixel-based 
analysis \citep[e.g.,][]{bothun86, abraham99, eskridge03, lanyon07, 
welikala08} on 
NGC\,959, we will demonstrate that our method can reveal a two-dimensional 
distribution of dust extinction. 

This paper is organized as follows.  In \S2, we present our method to 
estimate dust extinction using the SED models.  In \S3 and \S4, we test 
and apply this method to all available images for the nearby late-type 
spiral galaxy NGC\,959.  We give a discussion of our results in \S5, and 
present our conclusions in \S6.

\section{MODELS AND CONCEPT}

Before we analyze any observed images, we first evaluate theoretical SED 
models for single stellar populations (SSPs) at different ages and 
metallicities.  In this section, we describe the SED library that we used 
in our study, and then select the optimal filters for the subsequent SED 
analysis.  Once the filters are selected, we describe how to estimate 
the dust extinction, using the theoretical SEDs and observed images, 
through flux ratios of an optimal set of filters.

\subsection{Simple Stellar Population Models}

Among published SSP SED libraries \citep[e.g., AF03;][]{bruzual03, 
maraston05}, we elected to use the SED library by AF03.  While other SED 
libraries do not contain information about emission lines, this library 
includes both spectral and gaseous emission for young stellar populations.  
AF03 has created multiple sets of SED libraries\footnote[3]{
  Since the publication of \citet{anders03},       
  there have been several updates and additions    
  to their models, resulting in slightly different 
  combinations of IMFs and metallicities from the  
  original description.  We used the version dated 
  28 October, 2007.\\
  \texttt{http://www.galev.org/}
}                            
of SSPs, using the Padova \citep{bertelli94} and Geneva \citep{lejeune01} 
isochrone models with \citet{scalo86}, \citet{salpeter55}, and 
\citet{kroupa02} Initial Mass Functions (IMFs).  These SED libraries 
contain models for metallicities of Z\,=\,0.0004, 0.004, 0.008, 0.02 
(Solar), and 0.04.  Both isochrone models include the thermally pulsing 
asymptotic giant branch (TP-AGB) phase and model atmosphere spectra from 
\citet{lejeune97}.  The main difference between the two libraries with 
different isochrone models is the time resolution, $\Delta t$.  While the 
library with the Padova isochrones have an age coverage from 4~Myr to 
14~Gyr with $\Delta t$\,=\,4 Myr up to an age of 2.35~Gyr, and $\Delta 
t$\,=\,20~Myr for older ages, the library with the Geneva isochrones have 
variable time steps.  Among the available models, we will focus on the ones 
computed with Padova isochrones and the Scalo IMF --- adopting an upper 
mass limit of $\sim$\,50\,$M_{\odot}$ for super-solar metallicity, and 
$\sim$\,70\,$M_{\odot}$ for all other metallicities.  We refer the reader 
to AF03 and references therein for a detailed description of this library.  
The SEDs of SSPs change rapidly at all wavelengths, but especially so in 
the UV for ages younger than $\sim$\,100~Myr (see \figref{model}).  Once a 
stellar population reaches the age of $\sim$\,500~Myr, the rate of change 
in SED diminishes with increasing age.  As a practical subset for our 
analysis, we consider 10 different ages ($t$\,=\,4, 8, 12, 52, 100, 
500~Myr, 1, 5, 10, and 13.5~Gyr) for each metallicity, resulting in a 
model grid of 50 different SEDs.  In the main panel of \figref{model}, we 
plot the SEDs for six ages and two metallicities that span the full range 
of our adopted model grid.

\subsection{Optimal Filter Selection}

\begin{figure}
\centerline{\psfig{file=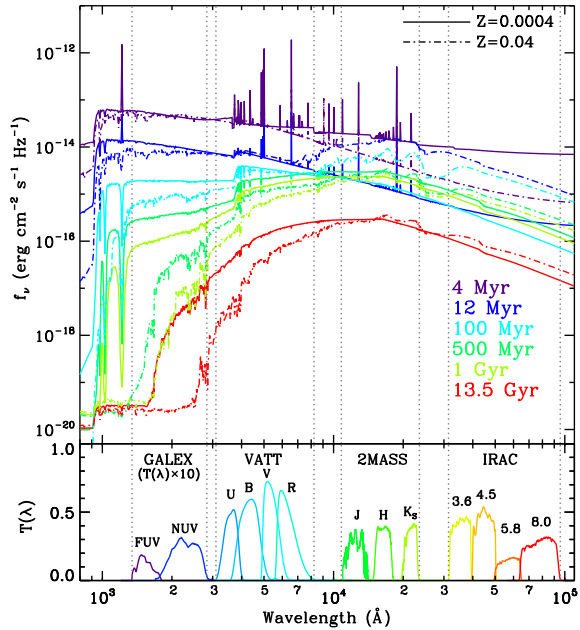,width=0.45\textwidth}}
\caption{
\emph{Top panel}: Theoretical SEDs from the library of \citet{anders03}.
Among the different sets of isochrones and IMFs, SEDs for
Padova isochrones assume a Scalo IMF with metallicity of 0.0004 (solid
curves) and 0.04 (dash-dot curves) at 6 different ages as indicated.
The age-metallicity degeneracy is clearly visible for wavelengths
shortward of 4000\,\AA.   On the other hand, SEDs at
$\lambda$\,$\gtrsim$\,6000\,\AA\ are less affected by metallicity,
except for extremely young ($t$\,$\lesssim$\,100~Myr) stellar
populations.
\emph{Bottom panel}: Total throughput curves, T($\lambda$), for different
telescope-filter combinations.  Throughput for the \emph{GALEX}
filters are scaled up by a factor of 10 for better visibility.  Vertical
dotted lines indicate the wavelength coverage for each telescope-filter
combination.
}\label{model}
\end{figure}

The bottom panel of \figref{model} shows the total throughput curves, 
$T(\lambda)$, for various FUV--MIR filters considered in this study, as 
well as in the subsequent multi-wavelength studies on NGC\,959 and on a 
sample of 45 galaxies (Tamura \etal\ 2009b,c, in preparation).  Total 
throughput curves for the \emph{GALEX} filters --- published in 
\citet{morrissey05} and on the \emph{GALEX} web page\footnote[4]{
  Goddard Space Flight Center, the 
  \emph{GALEX} Post Launch Response Curve Data: \\
  \texttt{http://galexgi.gsfc.nasa.gov/docs/galex/Documents/ PostLaunchResponseCurveData.html}
} 
--- are multiplied by a factor of 10 for better visibility.  The throughput 
curves for the VATT filters are the total responses of the filters, CCD, 
and telescope\footnote[5]{
  Instrumentation for VATT: \\
  \texttt{http://vaticanobservatory.org/vattinst.html}\\
  The response curve used in this paper is for the VATT 2k CCD.
}.  
The throughput curves for the 2MASS $JHK_s$ filters are the total responses 
of the filters, CCD, telescope, and atmosphere\footnote[6]{
  2MASS All-Sky Data Release Explanatory Supplement
  Facilities and Operations: \\
  \texttt{http://www.ipac.caltech.edu/2mass/releases/allsky/doc/ sec3\_1b1.html}
}. 
Finally, for the \emph{Spitzer}/IRAC filters, the total throughput curves 
are calculated using all components of the \emph{Spitzer} 
instrument\footnote[7]{
  Spitzer Science Center: IRAC: Spectral Response: \\
  \texttt{http://ssc.spitzer.caltech.edu/irac/spectral\_response.html}
}.
The vertical dotted lines in \figref{model} indicate the wavelength ranges 
covered by different telescopes and instruments.

The largest change in SEDs with the increase in age is the reduction of 
flux for wavelengths shortward of 4000\,\AA.  Especially the UV flux --- 
as measured by, e.g., the \emph{GALEX} FUV and NUV filters --- decreases 
by up to $\sim$\,6 dex relative to the maximum flux level, as an SSP ages 
from 4~Myr to 13.5~Gyr.  SEDs are also significantly affected by 
metallicity.  For SSPs older than 100~Myr, the metallicity mainly affects 
the wavelengths $\lesssim$\,4000\,\AA, where the flux decreases by up to 
$\sim$\,2 dex from Z\,=\,0.0004 to Z\,=\,0.04.  Since our primary goal is 
to measure dust extinction and to map its spatial distribution, filters 
covering wavelengths shorter than 4000\,\AA\ should be avoided to minimize 
effects from the age-dust-metallicity degeneracy.  

According to AF03, the model SEDs become uncertain longward of 
$\sim$\,5\,\mum.  Also, the observed stellar continuum at MIR wavelengths 
redward of $\sim$\,5\,\mum\ is contaminated by emission from polycyclic 
aromatic hydrocarbons \citep[PAHs;][]{leger84} and silicates \citep{
willner04}.  We thus exclude \emph{Spitzer}/IRAC 4.5, 5.8, and  8.0\,\mum\ 
filters.  On the other hand, the effects from dust extinction and 
emission by PAHs and silicates reach a minimum near 3.5\,\mum\ 
\citep[$L$-band; e.g.,][]{fazio04, willner04}.  The IRAC 3.6\,\mum\ filter 
therefore provides the most reliable stellar population tracer \citep[see, 
e.g.,][]{kennicutt03, helou04}.  

From the filters with 0.4\,$\lesssim$\,$\lambda$\,$\lesssim$\,3.5\,\mum\ 
--- optical \emph{VR} and NIR $JHK_s$ in our study --- we need to select (at 
least) one more filter to trace dust extinction.  Since dust extinction is 
much smaller in the NIR than in the optical \vv- and \rr-bands 
\citep[e.g.,][]{vanhouten61, cardelli89, calzetti94, gordon03}, NIR filters 
are not optimal for this purpose.  Another reason to avoid ground-based 
NIR observations is absorption due to atmospheric water vapor.  The amount 
and uncertainty due to this absorption depend on atmospheric conditions, 
as well as the location of the observations \citep[e.g.,][]{nitschelm88, 
cohen03}.  Dust extinction is stronger in the \vv-band \citep[e.g.,][]{
calzetti94, gordon03}, while the metallicity effects are slightly weaker 
(see Fig.~1).  We therefore choose the \vv\ and 3.6\,\mum\ filters to 
globally trace the dust extinction.  In the following section, we will 
explore the theoretical \vv-to-3.6\,\mum\ flux ratio in detail, and explain 
how we can use this ratio to estimate dust extinction. For other studies, 
in general, it should be noted that filters like SDSS $g$, \emph{HST} 
F555W, F550M, and F606W would be adequate substitutes for \vv-band after 
calibration of the theoretical flux ratios.

\subsection{Theoretical \vv-to-3.6\,\mum\ Flux Ratio Map}

\begin{figure}
\centerline{\psfig{file=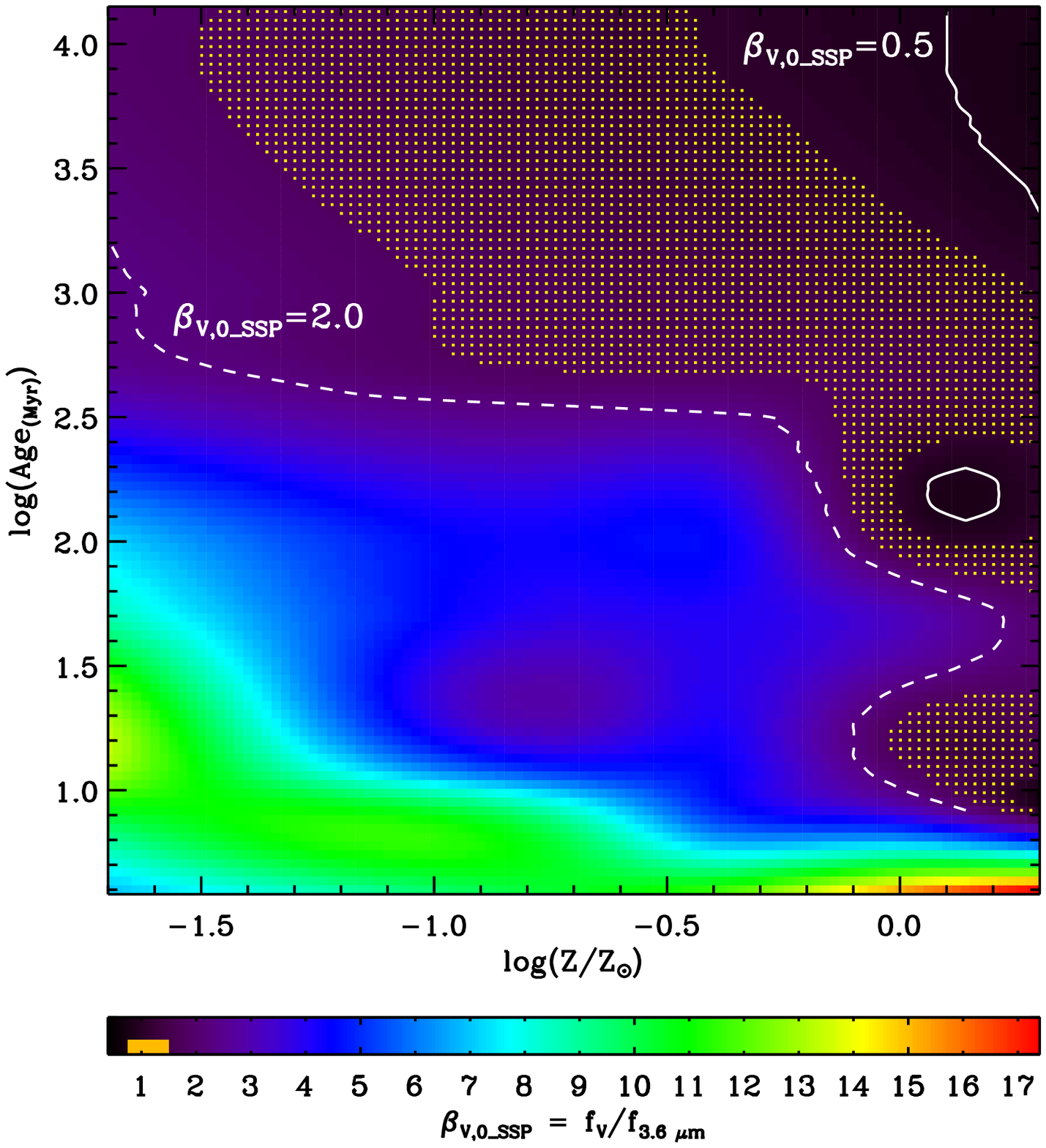,width=0.45\textwidth}}
\caption{
Map of the theoretical \vv-to-3.6\,\mum\ flux ratio, \betav, calculated
from an SSP library of \citet{anders03} for varying age and metallicity.
Contours are plotted at \betav\,=\,0.5 (solid curves) and 2.0 (dashed
curve).  The orange stippled region (and the orange band in the color bar)
indicates the regions in the metallicity-age space where the flux ratio
is 0.75\,$\leq$\,\betav\,$\leq$\,1.50.  While \betav\
varies rapidly for young --- $t$\,$\lesssim$\,500 Myr or
$\log(t_{\rm Myr})$\,$
\lesssim$\,2.7 --- stellar populations, it is relatively stable and
insensitive to age and metallicity for older stellar populations.
}\label{betamap}
\end{figure}

Given the SEDs from the AF03 library and the throughput curves of the 
optical \vv\ and IRAC 3.6\,\mum\ filters (see \figref{model}), we explore 
the range in the theoretical flux ratio, \betav\,=\,$f_V$/$f_{\rm 
3.6\,\mu m}$, as a function of age and metallicity.  The resulting flux 
ratio map in age-metallicity space is shown in \figref{betamap}.  From the 
input of 50 SEDs described above, both age and metallicity are resampled 
logarithmically using 100 steps with Spline interpolation to map the 
variation in \betav\ values over the full parameter ranges.  The minimum 
ratio, $\beta_{V,{\rm min}}$\,=\,0.39, occurs for the most metal-rich 
(Z\,=\,0.04\,=\,2Z$_{\odot}$) and oldest ($t$\,=\,13.5~Gyr) SSP, while the 
maximum ratio, $\beta_{V,{\rm max}}$\,=\,17.4, is found for the most 
metal-rich and the youngest ($t$\,=\,4 Myr) SSP.  Contours for 
\betav\,=\,0.5 (white solid curves) and 2.0 (white dashed curve) are drawn 
in \figref{betamap}.  A band of orange dots --- also included as an orange 
band in the color bar --- indicates the region where the \vv-to-3.6\,\mum\ 
flux ratio spans the range 0.75\,$\leq$\,\betav\,$\leq$\,1.50\,.  

For stellar populations older than 500~Myr, or log($t_{\rm Myr}$)\,$
\gtrsim$\,2.70, the theoretical flux ratio changes little as a function of 
age and metallicity.  \figref{betamap} shows that \betav\ ranges mostly 
between $\sim$\,0.5 and $\sim$\,2.0\,.  On the other hand, for stellar 
populations younger than 500~Myr, \betav\ changes more dramatically, 
depending on their age and metallicity.  Especially for SSPs with 
super-solar metallicity, \betav\ changes from $\sim$\,17 (red) for the 
youngest SSPs to $\lesssim$\,0.5 (dark purple/black) in a matter of only a 
few 100~Myr.  For the young SSPs with sub-solar metallicities, the value 
of \betav\ is more stable, but still changes much more rapidly than for 
SSPs older than 500~Myr.  Among these young sub-solar metallicity SSPs, 
\figref{betamap} shows two distinctive ranges of \betav\ values: (1) a 
band color-coded in green (9\,$\lesssim$\,\betav\,$\lesssim$\,13) for SSPs 
younger than 10--100~Myr; and (2) a band coded in blue (4\,$\lesssim
$\,\betav\,$\lesssim$\,7) for SSPs up to $\sim$\,500~Myr.  

The distribution of \betav\ values in \figref{betamap} confirms the 
assumption of, e.g., \citet{regan00}: that old stellar populations have 
relatively constant color, while young stellar populations change their 
colors more rapidly depending on their properties and environments.  An 
important corollary of the former is that a mixture of SSPs with ages 
larger than 500~Myr will have a \betav\ value that is very similar to that 
of a single SSP at those ages.  Our analysis of the theoretical 
\vv-to-3.6\,\mum\ flux ratios furthermore shows that --- at least for 
young stellar populations with sub-solar metallicities --- the theoretical 
flux ratio can be approximated as being constant for stellar populations 
in either of  the two age ranges ($t$\,$\lesssim$\,10--100~Myr or 
10--100\,$\lesssim$\,$t$\,$\lesssim$\,500~Myr) described above.  In the 
following section, we will derive a method for using these theoretical 
and observed \vv-to-3.6\,\mum\ flux ratios to estimate the amount and 
spatial distribution of dust extinction in the \vv-band.

\subsection{Estimating the \vv-band Dust Extinction, \av}

In the previous section, we showed that the theoretical \vv-to-3.6\,\mum\ 
flux ratio, \betav, is well-behaved, particularly for sub-solar 
metallicities.  To estimate the dust extinction for an observed stellar 
population, we first have to determine its approximate age and metallicity.  
The appearance of horizontal bands of different \betav\ values in 
\figref{betamap} indicates that an age determination is more important 
than the determination of metallicity.  Based on the \betav\ values and 
corresponding age range for different metallicities as described above, 
the accuracy required in determining age is $\sim$\,0.5 dex for extremely 
metal poor stellar populations, and $\sim$\,1 dex for sub-solar to solar 
metallicity and older ($t$\,$\gtrsim$\,500~Myr) stellar populations.  For 
stellar populations with super-solar metallicities, the accuracy required 
is also $\sim$\,0.5--1 dex for most ages, unless they are extremely young 
($t$\,$\lesssim$\,10~Myr).  The use of color-magnitude diagrams (CMDs) and 
color-color diagrams, therefore, should suffice to estimate the age of a 
stellar population and select a theoretical \betav\ value from 
\figref{betamap}.  
By comparing the theoretical and observed flux ratios, we can infer the
amount of flux missing in the V-band observation due to dust.  The missing
flux is a very robust property of the dust distribution, independent of
the geometry and fine structure of obscuration.  The amound of extinction
by dust along the line-of-sight that this stellar pouplation suffers can
then be estimated from this missing V-band flux.  

The dust extinction, \av, is defined as
\begin{equation}\label{av_def}
  A_V = (m_V - m_{V,{\rm 0}})\,,
\end{equation}
where $m_V$ and $m_{V,{\rm 0}}$ denote the observed and intrinsic 
(extinction-free) \vv-band magnitudes, respectively.  Even though the 
extinction-free magnitude is not known for an observed stellar population, 
we can use the selected \betav\ value and observed 3.6\,\mum\ flux to 
estimate the predicted $m_{V,0}$.  Since we assume that the observed 
3.6\,\mum\ flux is unaffected by dust extinction \citep[e.g.,][]{fazio04, 
willner04}, we can estimate the extinction-free \vv-band flux by 
multiplying the selected theoretical \betav\ value with the observed 
3.6\,\mum\ flux: 
\begin{equation}\label{f_0}
  f_{V,{\rm 0}} = \beta_V \times f_{\rm 3.6\,\mu m}\,.
\end{equation}
Therefore, \eqref{av_def} can be rewritten as 
\begin{equation}\label{av_calc}
  A_V = m_V - [-2.5 \log(\beta_V \times f_{\rm 3.6\,\mu m})-V_{\rm zp}]\,,
\end{equation}
where $V_{\rm zp}$ is the zero-point magnitude for the \vv-band.  Compared 
to popular methods such as the UV spectral slopes \citep[e.g.,][]{bell02b, 
kong04} and the FIR-to-UV flux ratio \citep[e.g.,][]{dale01, buat05}, 
this method is much simpler.  

Even though our prescription is simple, there remain many possible sources 
of error.  In addition to the usual uncertainties due to observational 
measurements, there are several sources of uncertainty in the selected 
theoretical \betav\ value.  Only when the age and metallicity are known 
for a resolved stellar population can we determine \betav\ for this 
stellar population with minimal error.  For unresolved stellar populations, 
determining the age and metallicity becomes much harder as a result of 
light-blending from intermixed and superposed stellar populations.  As we 
have shown above, the metallicity dependency is not as strong as the age 
dependence in determining the likely \betav\ value.  We therefore are 
still able to estimate the \betav\ value, if an approximate age of the 
stellar populations can be determined.  Even though stellar population 
composites span a range in age, their luminosity tends to be dominated by 
their younger components at most wavelengths (see \figref{model}).  From 
the distribution of \betav\ values in \figref{betamap}, we find a typical 
uncertainty in the determination of \betav\ for a stellar population of a 
factor of $\sim$\,1.4 (e.g., assuming a theoretical \betav\,=\,1.1, an old 
stellar population may actually show a range of 0.75\,$\lesssim
$\,\betav\,$\lesssim$\,1.5).  This corresponds to an error in dust 
extinction, $\sigma_{A_V}$, of up to $\sim$\,$\pm$0.37~mag, or 
mag\,arcsec$^{-2}$ in the case of surface brightness.  If this stellar 
population is either extremely metal-poor or metal-rich, the uncertainty 
is a factor of $\sim$\,2, or $\sigma_{A_V}$\,$\simeq$\,$\pm$0.75~mag.  
On the other hand, if an extremely young (old) stellar population is 
treated as an old (extremely young) stellar population, the error in the 
estimated theoretical \betav\ value will become a factor of $\sim$\,5 or 
larger.  This corresponds to a significant error in estimating the dust 
extinction, with $\sigma_{A_V}$\,$\gtrsim$\,$\pm$1.75~mag.  Since the 
extremely young and old stellar populations usually have distinct 
characteristics in the CMDs and color-color diagrams, it is highly 
unlikely that one would mistake a young stellar population for an old one.  
In \S3.3 below, we will use \uu-band observations to robustly separate 
pixels dominated by the light of younger and older stellar populations.

\section{DATA ANALYSIS}

In this section, we apply the above method to the observational data.  
Before applying it to a large sample of galaxies, we assess its 
reliability in this paper for one galaxy: NGC\,959.  In future papers 
(Tamura \etal\ 2009b,c, in preparation), we will reconstruct and analyze 
the extinction-corrected images of NGC\,959 and a sample of 45 galaxies,
spanning a wide range of elliptical--spiral galaxies, with \emph{GALEX} 
FUV and NUV, optical \emph{UBVR}, and \emph{Spitzer}/IRAC images using the 
present method.  Unless indicated otherwise, the color composite images 
and the pixel coordinate maps shown in the following sections are oriented 
such that North is up and East is to the left.  The magnitudes and colors 
used in this analysis are on the AB-magnitude system \citep{oke74, oke83}.

\subsection{Data Sets}

The galaxy we will use for our proof of concept, NGC\,959, is a nearby
late-type spiral galaxy classified as Sdm in the Third Reference Catalog
of Bright Galaxies \citep[RC3;][]{devaucouleurs91}, at a distance of 
9.9\,$\pm$\,0.7 Mpc\footnote[8]{
  This value is based on its recession 
  velocity including the influence of the Virgo 
  cluster, the Great Attractor, and the Shapley 
  supercluster, and taken from the NASA/IPAC Extragalactic 
  Database (NED) as of September 24, 2008 
} 
\citep{mould00}, with an inclination of 50$\degr$ \citep{esipov91}.  It 
has been observed from UV (\emph{GALEX} FUV) through MIR 
(\emph{Spitzer}/IRAC 8.0\,\mum) wavelengths.  Optical \emph{UBVR} images, 
observed with the Vatican Advanced Technology Telescope (VATT), were taken 
from \citet{taylor05}.  The pipeline-processed \emph{Spitzer}/IRAC images 
were retrieved from the \emph{Spitzer} Archive\footnote[9]{
  SSC: Data Archives/Analysis:\\
  \texttt{http://ssc.spitzer.caltech.edu/archanaly/}
} 
via Leopard.  The \emph{GALEX} FUV and NUV images were obtained from the 
Multi-Mission Archive at Space Telescope Institute\footnote[10]{
  Galaxy Evolution Explorer, GR4/GR5 Data Release:\\
  \texttt{http://galex.stsci.edu/GR4}
} 
(MAST).

In this study, we use a pixel-based analysis, first introduced by 
\citet{bothun86} in the form of pixel Color-Magnitude Diagrams (pCMDs). 
This has recently become a popular technique to study stellar populations 
in nearby galaxies \citep[e.g.,][]{abraham99, eskridge03, lanyon07, 
welikala08}.  This technique is performed in the same way as regular 
aperture photometry, but simply with an individual pixel as a source of 
flux.  The advantage of using a pixel-based analysis is that this 
technique allows a two-dimensional analysis throughout a galaxy, or any 
part of a galaxy, without any object overlap or gaps as would be created 
in regular aperture photometry \citep[see figures by, e.g.,][]{scoville01, 
calzetti05, calzetti07}.  

To perform this pixel-based analysis on images from different telescopes 
and instruments, we first have to resample the pixel-scales and convolve 
all the images to a matching resolution.  IDL\footnote[11]{
  IDL is distributed by ITT Visual Information Solutions
  (Research System Inc.), Boulder, Colorado:\\ 
  \texttt{http://rsinc.com/idl/}                 
}                                              
and IRAF\footnote[12]{
  IRAF is distributed by National Optical 
  Astronomy Observatory (NOAO), which is 
  operated by the Association of Universities 
  for Research in Astronomy, Inc., under 
  cooperative agreement with the National 
  Science Foundation (NSF):\\
  \texttt{http://iraf.net}  
} 
routines were used to match the pixel-scales and point spread functions 
(PSFs) of all images to the 1\farcs5~pixel$^{-1}$ and $\sim$\,5\farcs3~FWHM 
of the \emph{GALEX} NUV image, because these have the coarsest pixel-scale 
and PSF among the filters considered for further panchromatic (FUV--MIR) 
studies (see \figref{model} and Tamura \etal\ 2009b,c, in preparation).  At 
the distance of NGC\,959 (D\,=\,9.9\,$\pm$\,0.7\,Mpc), each 1\farcs5~pixel 
spans 72\,pc.  As a result, the light from different stellar populations is 
mixed together within a single pixel, and the observed flux ratios cannot 
be compared directly to the SSP-derived \betav\ values in \figref{betamap}.  
While the pixels with most of the light coming from older stellar 
populations are not affected as much, the pixels with light from younger 
stellar populations will be affected significantly.  Even though the light 
is smoothed over a much larger area than the area covered by a single 
pixel, the peak of the light distribution stays at the same pixel 
coordinate as before the smoothing.  In the following, we therefore 
statistically analyze the observed images to estimate the intrinsic, 
dust-free flux ratios for both younger and older stellar populations, 
instead of a direct comparison to SSP models in the previous section.  We 
will only use pixels with signal-to-noise ratio of S/N\,$\geq
$\,S/N$_{\rm min}$\,=\,3.0 in all FUV--MIR filters used.

\subsection{Assumptions about the Dust Distribution}

The dust is not uniformly distributed across an entire galaxy.  Instead, it 
is distributed in complicated patterns of wisps, lanes, and bands of thin 
and filamentary structures, as well as in small clumps \citep[e.g.,][]{
waller92, deo06}.  Using stellar radiative transfer models, \citet{
elmegreen80} and \citet{witt92} showed that depending on different spatial 
distributions of the dust, e.g., a cloud or a slab, the effect of dust 
extinction --- including both absorption and scattering --- varies along 
different lines-of-sight.  \citet{walterbos88} 
estimated the variations in extinction values calculated from the 
``symmetry argument'' \citep[e.g.,][]{lindblad41, elvius56}. \citet{
calzetti94} compared models of five different dust distributions to the 
observational data: (1) a uniform dust screen; (2) a clumpy dust screen; 
(3) a uniform scattering slab; (4) a clumpy scattering slab; and (5) an 
internal dust model (see their Fig.~8).  The importance of these models 
is that the resulting extinction is different for each model although the 
amount of dust is same.  Even though \citet{calzetti94} could not find 
satisfactory agreement between the observed data and these models using 
Large Magellanic Cloud (LMC) or Milky Way (MW) dust-extinction curves, 
models 2 and 4 with clumpy dust distributions show in general a better 
fit to the data than the other models.  We will assume that the dust 
affecting the observed light is distributed in wisps and clumps, and is 
mostly in front of the observed stellar populations --- with a much larger 
effect of absorption than light scattering \citep{byun92} --- following 
the observations of, e.g., \citet{waller92} and \citet{deo06}.  The true 
distribution of the dust would be more complex, but representable by a 
combination of different types of simple geometrical distributions, with 
the dust distributed not only in front, but also intermixed with stellar 
populations.  Dust located mostly behind the stars would not be detected 
at the wavelengths considered here.  Decoding the distribution of dust in 
different geometries is beyond the scope of the present paper, and hence 
deferred to a subsequent analysis (Tamura \etal\ 2009c, in preparation),
where we will have a larger sample of galaxy types and inclinations.

Another important issue is the filling factor of dust within a pixel.  
While each pixel subtends 1\farcs5, the dust features can span either a 
larger or smaller area.  Given a single average value of the dust 
extinction in each pixel, there are two extreme possibilities for the dust 
distribution within that pixel: (1) an extended distribution of a 
relatively thin layer of dust; and (2) small high-density clumps of dust 
covering only a fraction of the area in that pixel.  Since the light 
observed in a single pixel is a mixture of light from different stellar 
populations, the effect that we observe is a weighted average of the light 
from those stellar populations.  For the former type of dust distribution, 
the dust is affecting the light of all stellar populations contributing to 
a single pixel equally.  For the latter case, the dust is only affecting a 
small fraction of the total light in that pixel.  While the light from 
stellar populations behind a dust clump is reduced, the light from 
unextincted stellar populations is observed at its full strength, reducing 
the average effect of dust extinction within a pixel.  If the intrinsic 
\vv-band flux is estimated from the observed NIR-flux --- which is still 
affected by dust extinction to some degree --- this partial coverage of any 
dust extinction might cause a large uncertainty.  Our method, on the other 
hand, estimates the intrinsic \vv-band flux from the observed MIR 
(3.6\,\mum) flux, 
which is usually considered as extinction free \citep[e.g.,][]{fazio04, 
willner04}.  The observed 3.6\,\mum\ flux is therefore the \emph{total} 
light from all stellar populations along the line-of-sight, ranging from 
the front to the far side of a galaxy.  Since the intrinsic \vv-band 
flux is estimated based on this 3.6\,\mum\ flux, our method estimates the 
\emph{total} amount of missing \vv-band flux.  Hence, even though the 
``exact'' effect of dust extinction depends on the true dust geometry, the 
\emph{total} missing \vv-band flux and the corresponding dust extinction 
should not be affected by the dust geometry, unless the extinction in 
individual knots or filaments is $\gg$\,1.0~mag, thereby also affecting 
the observed MIR flux, and preferentially affecting only young and highly 
concentrated (on scales $\ll$\,72\,pc) stellar populations.  In retrospect, 
this does not appear to be the case (see e.g., \figref{av_hr}), although 
one has to be careful that this does not become a circular argument.

\subsection{Separating Younger and Older Stellar Populations}

\begin{figure}
\centerline{\psfig{file=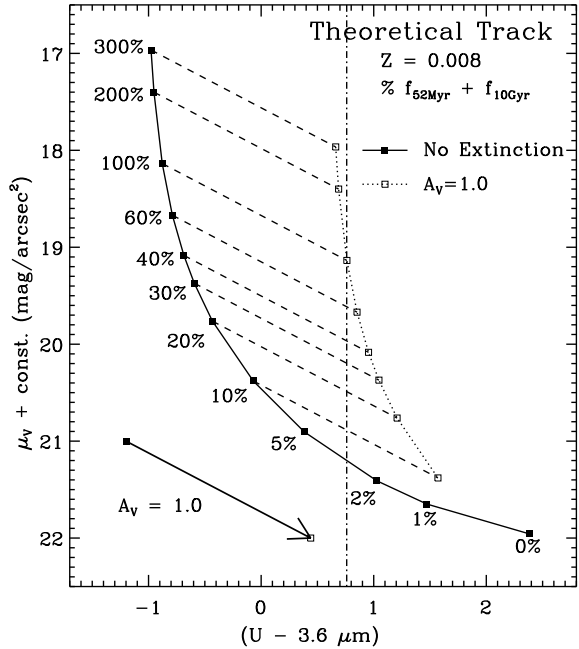,width=0.45\textwidth}}
\caption{
The theoretical $\mu_V$ vs.\ (\uu$-$3.6\,\mum) color-magnitude diagram.
The solid curve is the theoretical track produced by adding different
fractions of a young (52~Myr) SSP SED to an old (10~Gyr) SSP SED:
the ``100\%'' indicates a mass-ratio between young and old SSPs of 1,
and ``200\%'' means the mass-ratio is two-to-one.
The effect of an average visual extinction of 1.0~mag\,arcsec$^{-2}$ is
shown as a reddening vector in the lower left, and by dashed lines for
selected data-points.  The dotted curve traces a theoretical track for
the color and surface brightness if the entire pixel (with
a filling factor of 100\%) is affected by this reddening.  The vertical
dot-dashed line at (\uu$-$3.6\,\mum)\,=\,0.76~mag represents
the color where the mass-weighted SEDs of young and old SSPs are
contributing equally, and are both affected by a dust extinction of
\av\,=\,1.0~mag\,arcsec$^{-2}$.
}\label{pcmd_model}
\end{figure}

Even though the effect of light-blending is significant in NGC\,959, some 
pixels are still dominated by the light from younger stellar populations.  
Therefore, the first step in our data analysis is to separate these pixels 
from pixels whose light mostly comes from older stellar populations.  
\figref{pcmd_model} shows theoretical tracks of $\mu_V$ vs.\ 
(\uu$-$3.6\,\mum) using 52~Myr and 10 Gyr SSP SED models with a 
metallicity of Z\,=\,0.008.  These metallicity and ages are selected 
because: (1) Z\,=\,0.008 is the central metallicity among five 
metallicities available for the SED library by AF03; (2) $t$\,=\,10~Gyr 
represents an old stellar population; and (3) $t$\,=\,52~Myr represents a 
young, but not an extremely young ($t$\,$\lesssim$\,10~Myr) stellar 
population (see \figref{betamap}).  Since the \uu-band light is sensitive 
to younger stellar populations and 3.6\,\mum\ light traces older stellar 
populations, the (\uu$-$3.6\,\mum) color indicates the luminosity-weighted 
average age of stellar populations within a given pixel.  The solid curve 
shows a track of the surface brightness and color as the fraction of 
light from the young stellar population increases.  The fraction 
indicated in the figure is the mass fraction of stellar populations, i.e., 
``100\%'' indicates that the mass ratio between young and old stellar 
populations is one-to-one.  A reddening vector corresponding to a visual 
extinction of \av\,=\,1.0~mag\,arcsec$^{-2}$ is drawn in the lower left of 
\figref{pcmd_model}, and is applied to selected data-points.  Even though 
this is an extreme case, \figref{pcmd_model} shows that, once the young 
stellar population dominates, the color of mixed stellar populations tend 
to an asymptotic value of (\uu$-$3.6\,\mum)\,$\simeq$\,$-$1.0~mag.  In a 
real situation, younger stellar populations are known to associate with a 
larger amount of dust \citep[e.g.][]{vanhouten61, knapen91, regan04, 
barmby06}, and the observed color will therefore most likely shift away 
from the no-extinction track to redder colors.  

\begin{figure*}
\centerline{
\psfig{file=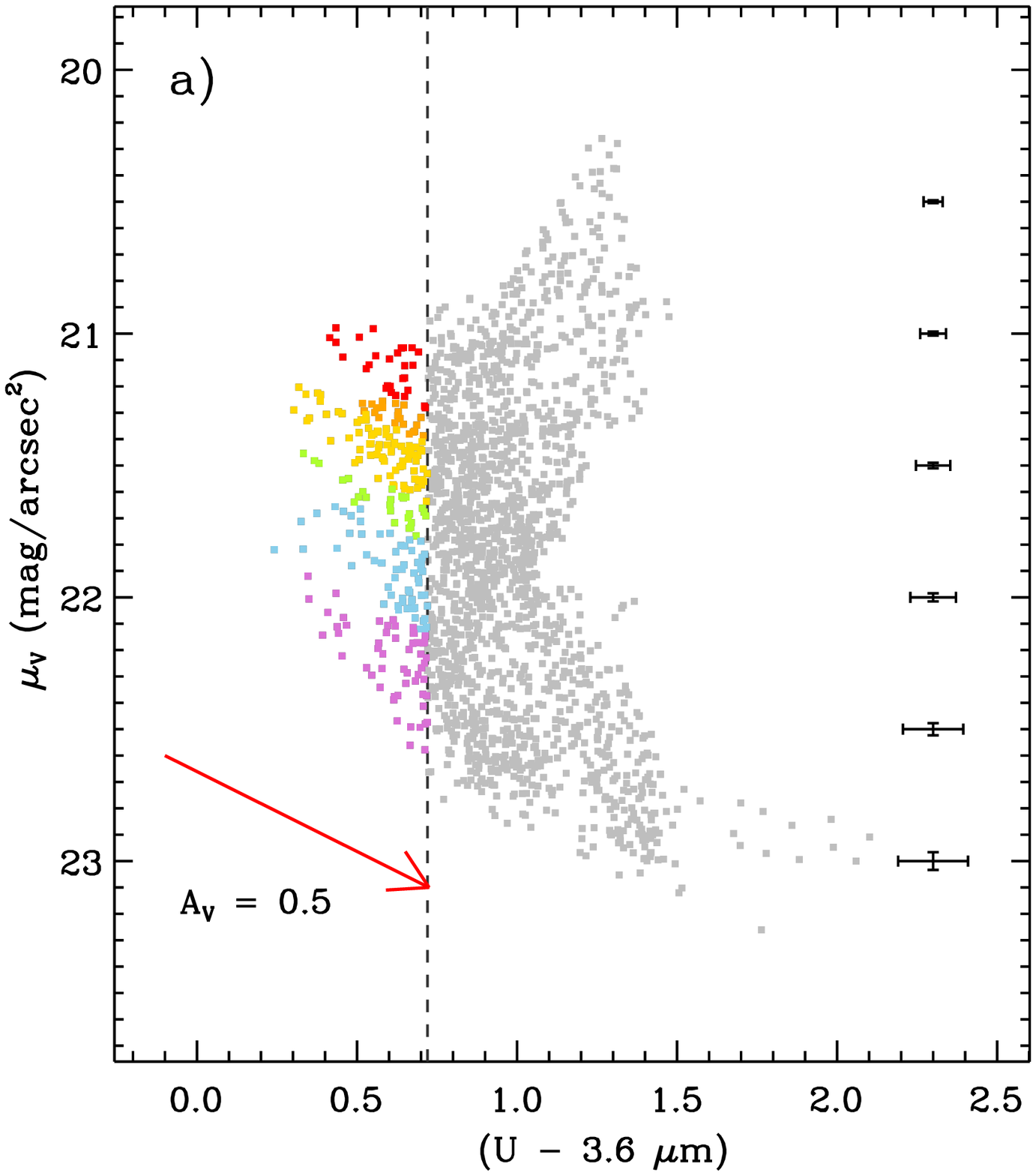,width=0.45\textwidth} 
\psfig{file=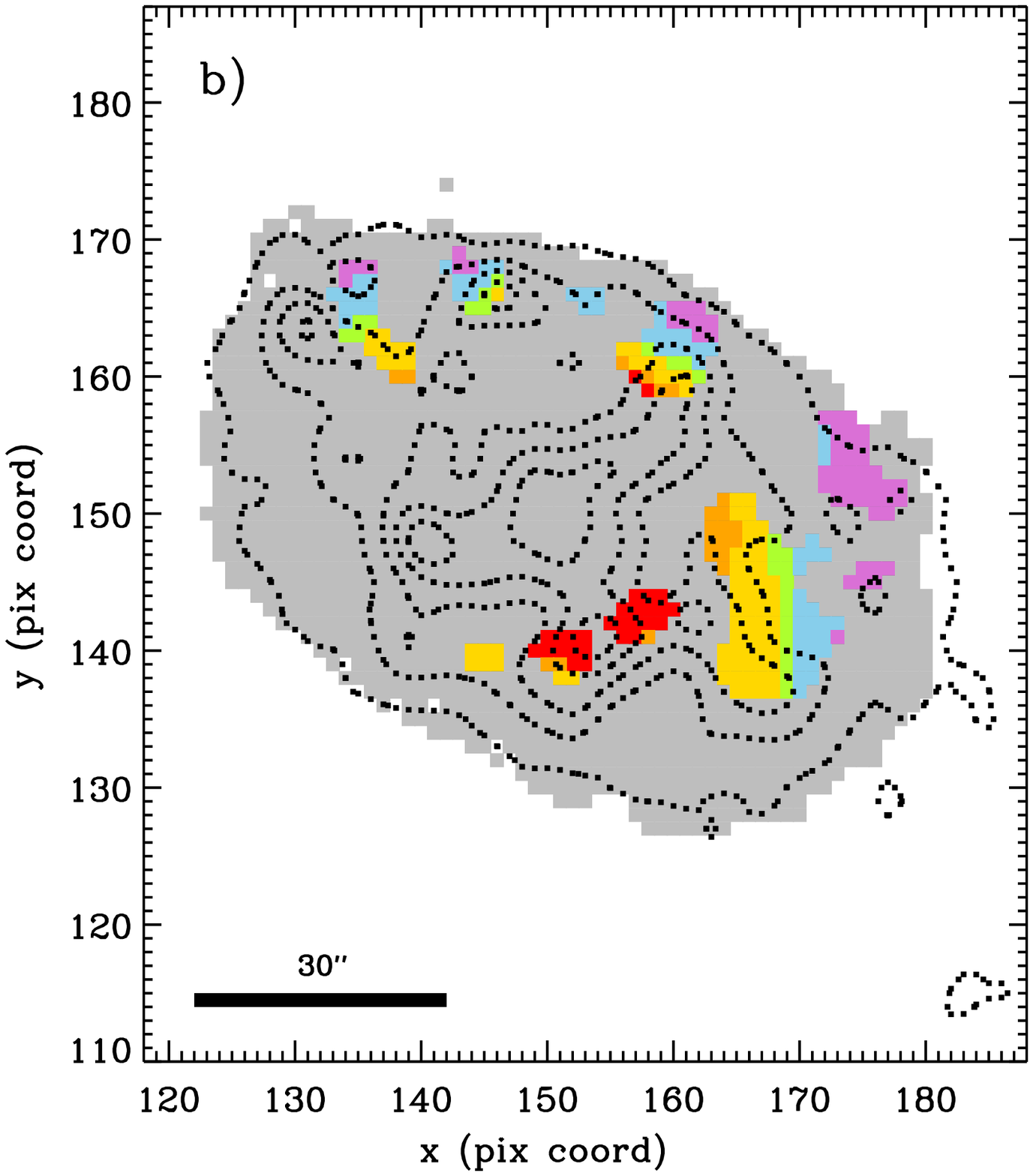,width=0.45\textwidth}
}
\caption{
\emph{a}) A pCMD of $\mu_V$ vs.\ (\uu\,$-$\,3.6\,\mum) for observed images.
The pixels dominated by the light from younger stellar populations
(``younger pixels'') with (\uu\,$-$\,3.6\,\mum)\,$\lesssim$\,0.72\,mag are
plotted in different colors according to the branch-like feature (see text)
that they appear to be part of.  The average uncertainty as a function of
magnitude is indicated by the error bars along the right side of this panel.
A reddening vector with \av\,=\,0.5 mag\,arcsec$^{-2}$ is plotted at the
lower left.
\emph{b}) Spatial distribution of the color-coded and light gray pixels
selected in (\emph{a}). Dotted contours trace the \emph{Spitzer} 8.0\,\mum\
PAH emission.  The younger pixels form \emph{contiguous} regions; they are
\emph{not} randomly distributed.  Different branch-like features in panel
(\emph{a}) correspond to regions at both systematically different
galactocentric radius and with systematically different PAH emission.  The
younger pixels on the red-coded branch correspond to regions with strong PAH
emission, and likely significant \av, that are also relatively close-in;
the purple pixels are faint at 8.0\,\mum, suffer little extinction, and
are found at larger distances from the center.
}\label{obsel}
\end{figure*}

\figref{obsel}a shows a pCMD of $\mu_V$ vs.\ (\uu$-$3.6\,\mum) of the 
observed images at \emph{GALEX} resolution.  The photometric uncertainties
plotted along the right side of the figure are calculated using 
data-points within a horizontal slice of $\mu_V$\,$\pm$\,0.1~mag\,arcsec$
^{-2}$ at each point.  A reddening vector corresponding to 
\av\,=\,0.5~mag\,arcsec$^{-2}$ and assuming the LMC extinction curve is 
drawn at the lower left.  The stellar populations in NGC\,959 are 
unresolved due to a combination of its distance, the coarse pixel-scale, 
and a large PSF.  As a result, \figref{obsel}a does not show clear 
separations among different stellar populations, as are seen for partially 
resolved stellar populations \citep[e.g.,][]{eskridge03}.  Nevertheless, 
some branches and features in this pCMD are still recognizable.  
For simplicity, in the following sections, we will refer to \emph{pixels in 
which the flux is dominated by light from younger stellar populations} as 
``younger pixels'', and \emph{pixels dominated by older stellar 
populations} as ``older pixels''.  Visual inspection of \figref{obsel}a 
(in particular, different slopes for groups of branching pixels on left and 
right sides and groups of pixels forming a ``shoulder-like'' distributions 
at $\mu_V$\,$\simeq$\,21~mag\,arcsec$^{-2}$ in the pCMD) suggests an 
empirical division between the ``younger'' and ``older'' pixels at 
(\uu$-$3.6\,\mum)\,=\,0.72~mag --- close to the color indicated by a 
dash-dotted line at (\uu$-$3.6\,\mum)\,=\,0.76~mag in \figref{pcmd_model}, 
where young stellar populations start dominating the mass-fraction of a 
mixed stellar population within a resolution element and suffering a total 
extinction of \av\,=\,1.0~mag\,arcsec$^{-2}$.

Before we proceed, we first perform several visual checks to ensure this
separation of younger and older pixels is indeed acceptable.  At colors
bluer than (\uu$-$3.6\,\mum)\,=\,0.72~mag, groups of pixels --- or
``branch-like features'' --- seem to have similar slopes as the reddening
vector, suggesting that they represent young stellar populations affected
by dust.  This agrees with the notion \citep[e.g.,][]{vanhouten61,
knapen91, regan04, barmby06} that dust is strongly associated with active
and recent SF-regions.  On the redder side of the pCMD, the brighter
pixels ($\mu_V$\,$\lesssim$\,21.9~mag\,arcsec$^{-2}$) form a distribution
with a positive slope (i.e., perpendicular to the reddening vector),
indicating that a mechanism other than dust
extinction might be affecting the fluxes in these pixels.  Since the
younger stellar populations tend to be much brighter than the older
stellar populations (see \figref{model}), for the lower surface
brightness ($\mu_V$) pixels, the fraction of light from younger stellar
populations as well as its corresponding dust extinction would be smaller.
The theoretical tracks in \figref{pcmd_model} suggest this would result
in a pile-up of points along a fairly vertical line, e.g., the dash-dotted
line at (FUV\,$-$\,3.6\,\mum)\,=\,0.76~mag.  We therefore conclude that our
separation of younger and older pixels is reasonable, at least to first
order.  We will perform several more checks to confirm that this
separation is appropriate (see below).  At this point and at the present
resolution, we do not see any special features indicating possible pixels
dominated by the light from ``extremely'' young stellar populations (i.e.,
$t$\,$\lesssim$\,10--100~Myr, see \figref{betamap}).  Considering the
effect of light-blending, the light from extremely young stellar
populations would most likely be diluted, and hence the separation of
pixels simply into younger and older pixels seems to be sufficient for
this galaxy (and galaxies viewed at similar linear resolution).

\figref{obsel}b shows the spatial distribution of the younger and older 
pixels selected in \figref{obsel}a.  The \emph{Spitzer}/IRAC 8.0\,\mum\ 
PAH emission --- an indicator of nearby SF-activity \citep[e.g.,][]{
helou04, calzetti05, calzetti07} --- is overplotted in dotted contours.  
The distributions of younger pixels and 8.0\,\mum\ PAH emission do not 
exactly overlap, but are shifted slightly with respect to one another.  
Since the directions of these shifts are not constant, they are not the 
result of astrometric error in the World Coordinate System (WCS) of the 
images, but are genuine features, as shown by \citet{calzetti05}.  Analysis 
of the PAH emission and its distribution relative to SF-regions is beyond 
the scope of this paper, and will be deferred to a future paper.  The 
most important result obtained from \figref{obsel}b is that --- while no 
coordinate information is used to select the younger pixels in 
\figref{obsel}a --- the selected pixels are grouped together into cohesive 
regions that coincide, or are close to peaks in the IRAC 8.0\,\mum\ 
emission.  
Moreover, different branch-like features in \figref{obsel}a turn out to
correspond to regions at both systematically different distances from the
galaxy center and systematically different PAH surface brightness.  The
younger pixels that form the high-surface brightness ($\mu_V$) feature
that is color-coded red in \figref{obsel}a originate mostly from two
regions that are located near one of the strongest peaks in the 8.0\,\mum\
emission and likely suffer significant extinction.  Their high surface
brightness appears mostly due to the relatively small distance from the
galaxy center and exponential decline in surface brightness of the disk
of NGC\,959 \citep[e.g.,][]{heraudeau96, taylor05}.  Younger pixels on
progressively lower surface brightness features (color-coded in order:
orange, gold, green, blue and purple) correspond to regions with
progressively larger distance from the center and fainter PAH emission
(smaller \av).

\begin{figure*}
\centerline{\psfig{file=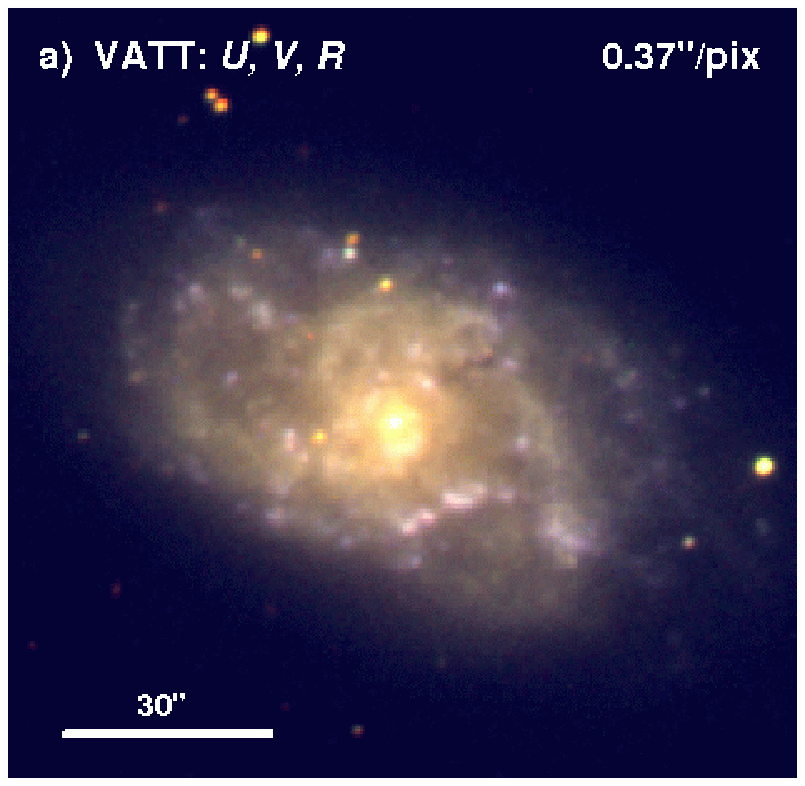,height=0.45\textwidth}
 \hspace{10mm}
 \psfig{file=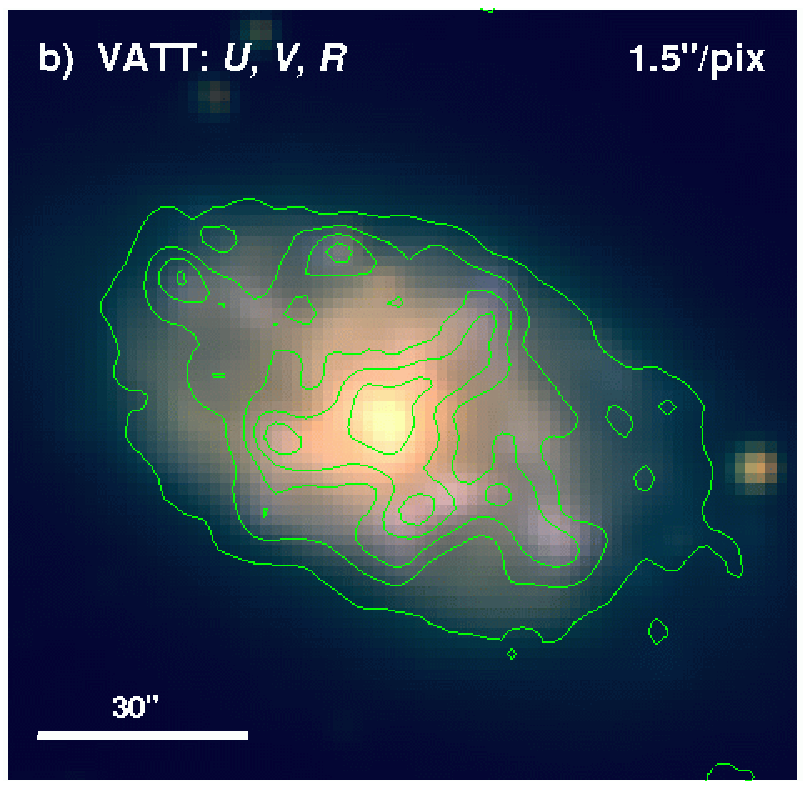,height=0.45\textwidth}}
\caption{
\emph{UVR} color composite images of NGC\,959 at two different
spatial resolutions.  \emph{a}) A color composite image constructed
at the original pixel-scale of the VATT CCD, 0\farcs37 pixel$^{-1}$
with a seeing of 1\farcs3 FWHM.  Dust lanes and stellar populations with
different color are readily discernable in this image.  \emph{b}) A
color composite image constructed from images that are resampled to
the \emph{GALEX} pixel-scale of 1\farcs5~pixel$^{-1}$, and convolved to
the \emph{GALEX} PSF of $\sim$\,5\farcs3~FWHM.  IRAC 8.0\,\mum\
emission is indicated by green contours.  Compared to the higher
resolution image, dust lanes and stellar populations in this lower
resolution image are not as evident.  However, the blue young stellar
populations remain clearly visible in the lower resolution image.
}\label{colimg}
\end{figure*}

To further confirm that our selection of the younger pixels is \emph{not} a 
random result, we also visually compare \figref{obsel}b to color composite 
images of NGC\,959.  \figref{colimg} shows \emph{UVR} color composites at 
two different spatial resolutions.  Both are composed from the same \uu-, 
\vv-, and \rr-band images observed at the VATT \citep{taylor05}.  
\figref{colimg}a is presented at the original pixel-scale of 
0\farcs37~pixel$^{-1}$, with the PSFs in all images matched to 
$\sim$\,1\farcs3~FWHM.  In \figref{colimg}b, the pixel-scale and resolution 
were matched to that of the \emph{GALEX} NUV image, i.e., 1\farcs5~pixel$
^{-1}$ and 5\farcs3~FWHM.  \emph{Spitzer}/IRAC 8.0\,\mum\ contours (green) 
are overlaid in the latter image for a comparison to the pixel-map 
(\figref{obsel}b).  We find that the spatial distribution of the selected 
\emph{younger} pixels clearly follows that of \emph{bluer} regions in 
NGC\,959.

\subsection{Selecting the Theoretical \vv-to-3.6\,\mum\ Flux Ratio}

\begin{figure}
\centerline{\psfig{file=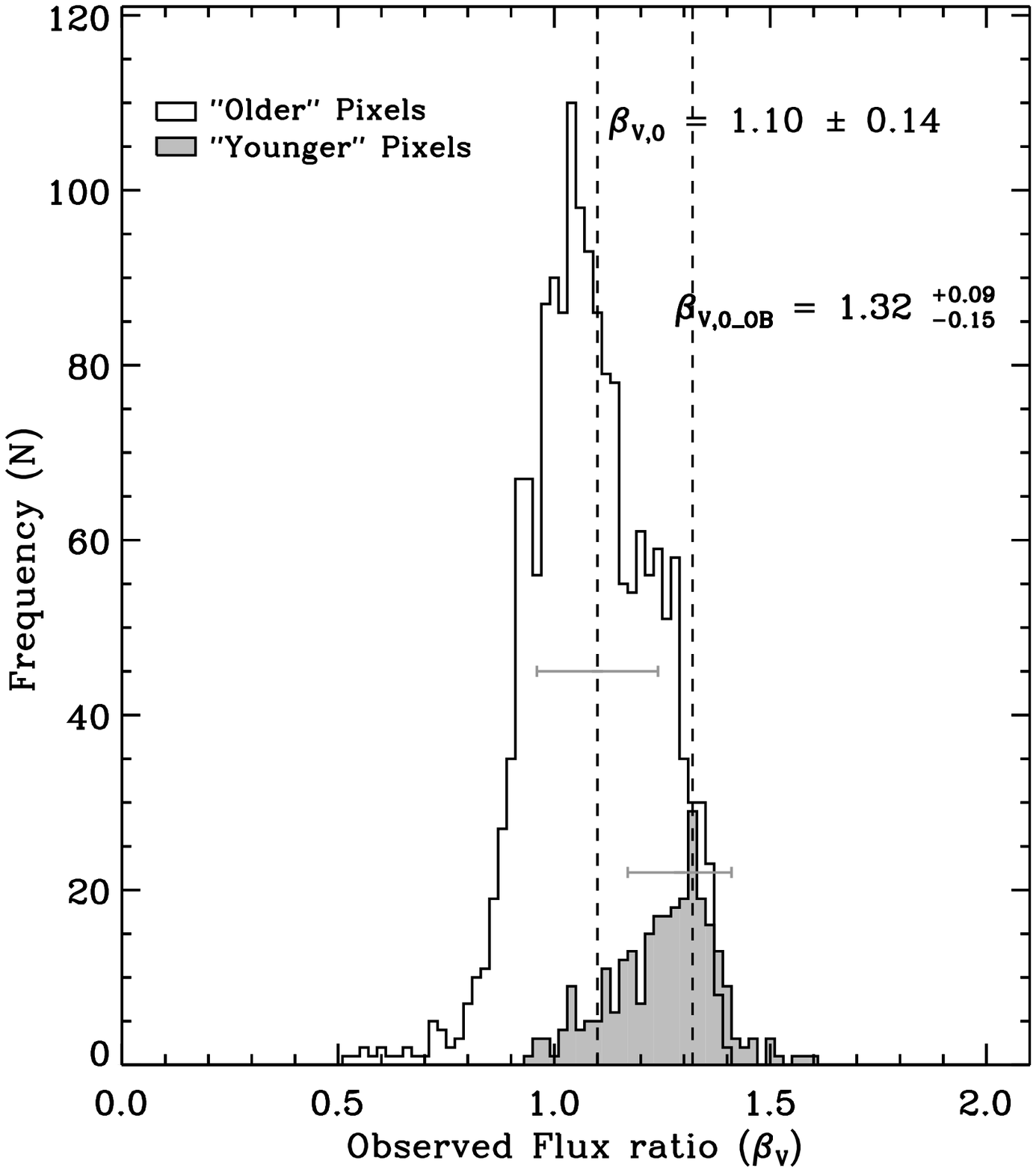,width=0.45\textwidth}}
\caption{
Histograms of observed \vv-to-3.6\,\mum\ flux ratio, \betav, for individual
pixels dominated by the light from older stellar populations (``older
pixels'') and younger stellar populations (``younger pixels''), as selected
in \figref{obsel}.  For both populations, the low-ratio tail of the
distribution --- presumably
due to dust extinction --- is more extended than the tail on the high-ratio
side.  We adopt an extinction-free flux ratio, \bzero, of
1.10\,$\pm$\,0.14 for the older pixels and of 1.32\,$^{+0.09}_{-0.15}$ for
the younger pixels.  Once  2$\sigma$ clipped mean and peak values are
selected for older and younger pixels, respectively, corresponding
uncertainties are determined from the FWHM of \betav\ distributions.
Because of light-blending from different populations contributing to the
flux in a given pixel, the \betav\ values for the young stellar population
are significantly reduced compared to those calculated for SSP models
in \figref{betamap}.
}\label{beta}
\end{figure}

Having separated the ``younger'' and ``older'' pixels, we now want to 
estimate the theoretical extinction-free flux ratio for each pixel.  As 
mentioned above, however, the theoretical \betav\ value cannot be simply 
selected from \figref{betamap}, due to the smoothing over stellar 
populations and subsequent blending of light.  For pixels dominated by the 
light from older stellar populations, this should not be a major problem, 
since their theoretical flux ratios do not change much with age (see 
\figref{betamap}).  The problem is the effect on pixels dominated by 
younger stellar populations.  While a young SSP has a theoretical flux 
ratio of $\sim$\,5 (or up to $\sim$\,11 for an extremely young stellar 
population), the extinction-free \betav\ value for mixed stellar 
populations depends strongly on how large a fraction of the light 
originates from underlying and neighboring older stellar populations.  
Since we cannot measure the exact fractions of light from younger and older 
stellar populations, we have to find another way to estimate theoretical 
\betav\ values.   
 
We first calculate the flux ratio of the observed \vv\ and 3.6\,\mum\ images 
for each pixel to characterize the effects of light-blending.  The result is 
plotted in \figref{beta}, where the white-colored histogram represents 
the distribution of \betav\ values for the pixels dominated by the light 
from older stellar populations (``older'' pixels), and the gray-colored 
histogram for those dominated by the light from younger stellar populations 
(``younger'' pixels).  The older pixels have a peak around \betav\,$\simeq
$\,1.0 with a secondary peak, or a shoulder, at \betav\,$\simeq$\,1.25\,.  
This confirms our assumption that light-blending has only a minor effect on 
older pixels.  The younger pixels display a peak at \betav\,$\simeq$\,1.32, 
which is much lower than the theoretical ratio (4\,$\lesssim$\,\betav\,$
\lesssim$\,7, see \figref{betamap}).  This indicates that younger pixels  
are significantly affected by the light from underlying and neighboring 
older stellar populations.  An important feature of the distribution for 
the younger pixels is that the observed range of \betav\ values is narrow 
and concentrated within 0.9\,$\lesssim$\,\betav\,$\lesssim$\,1.6\,.  This 
implies the effect of light-blending is rather uniform and consistent for 
these pixels.  Stated differently, the contaminating older stellar 
population is distributed much more smoothly than the younger stellar 
population (see \figref{colimg}, \figref{img_v} and 
\figrefm{img_fuv}{img_r}).

Another important feature of \figref{beta} is that the tail of the 
distribution toward lower \betav\ values is larger than the tail toward 
higher values.  Since dust extinction is the primary cause for the 
reduction of the \vv-band flux, the pixels with anomalously low \betav\ 
values are most likely located in the high-extinction regions within the 
galaxy.  While both groups of pixels have low-\betav\ tails, the relative 
size of the tail, compared to the size of the main distribution, is much 
larger for the younger pixels.  This indicates that dust extinction is 
more significant for the younger pixels.

Because interstellar dust is usually concentrated in relatively small 
regions \citep[e.g.,][]{deo06} --- while most other regions suffer minimal 
dust extinction --- we should be able to estimate the intrinsic dust-free 
\vv-to-3.6\,\mum\ flux ratio, \bzero, for both stellar populations from 
\figref{beta}.  While the peak of the distribution for the older pixels 
is at \betav\,=\,1.04, this value occurs toward the lower end of the 
distribution.  A statistical analysis with 2$\sigma$ clipping shows 
that the mean is at \bzero\,=\,1.10 with a standard deviation of 0.14.  
Even though the exact value of the intrinsic flux ratio varies from 
pixel-to-pixel due to the different stellar population components in each 
pixel, we adopt this value of \bzero\,=\,1.10\,$\pm$\,0.14 as the 
dust-free flux ratio for the older pixels.  Unlike the older pixels, the 
\betav\ distribution of the younger pixels is more asymmetric, with a 
much larger tail toward lower \betav\ values.  After 1$\sigma$ clipping, 
the statistical mean is \betav\,=\,1.29, which is smaller than the peak 
value of \betav\,=\,1.32\,.  Because younger stellar populations are more 
affected by dust than older stellar populations, which is also apparent 
from the much larger and wider lower-\betav\ tail, we elected to use 
$\beta_{V,{\rm 0\_OB}}$\,=\,1.32\,$^{+0.09}_{-0.15}$ (the peak \betav\ 
value) as the extinction-free flux ratio for the younger pixels.  The 
added subscript, OB, indicates that the value is for the pixels dominated 
by the light from younger stellar populations, but does not imply that 
such pixels have no light contributed by older stellar populations.

\subsection{Calculating the Flux Difference}

\begin{figure*}
\centerline{\psfig{file=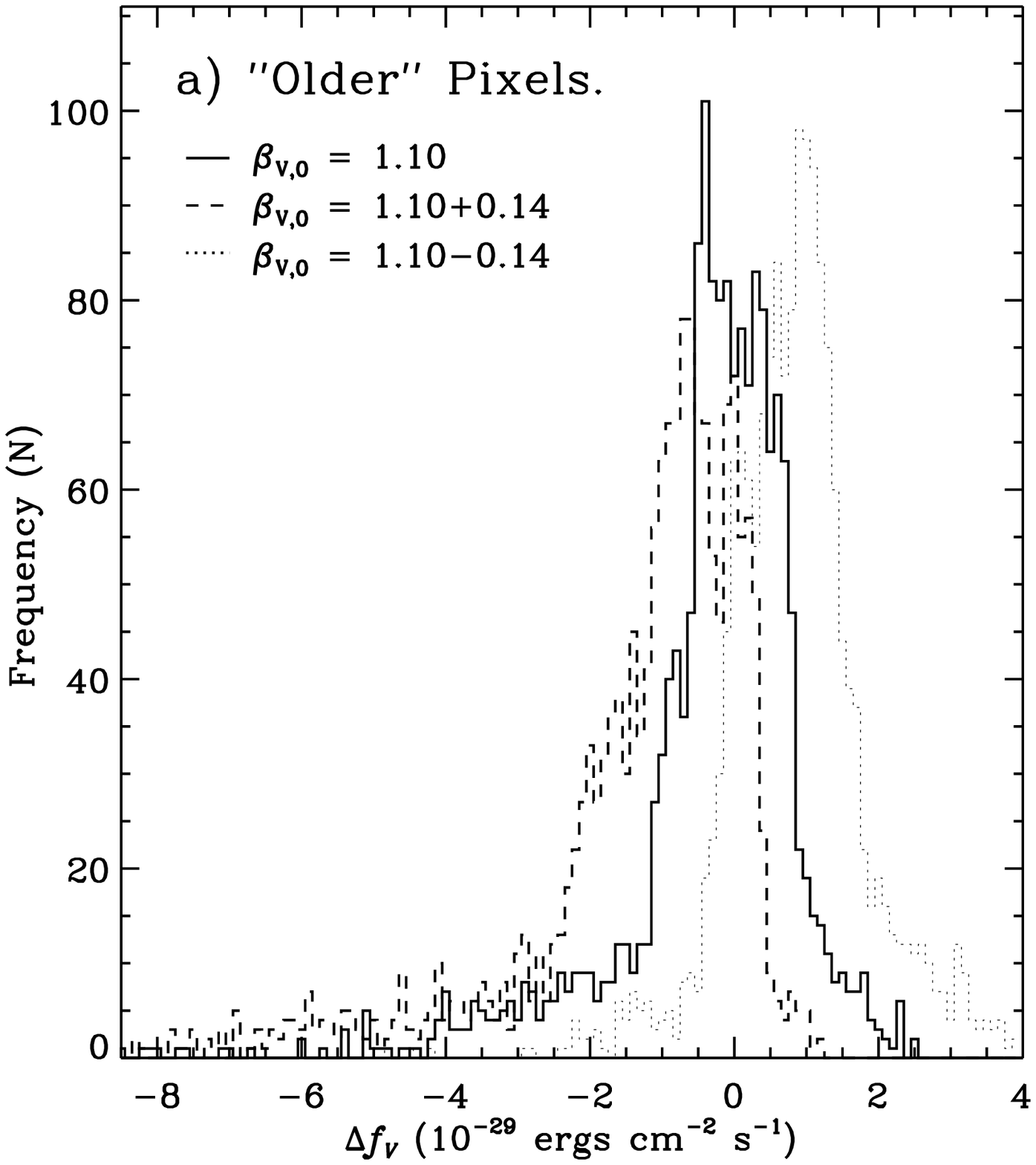,width=0.45\textwidth}
\hspace{10mm}
\psfig{file=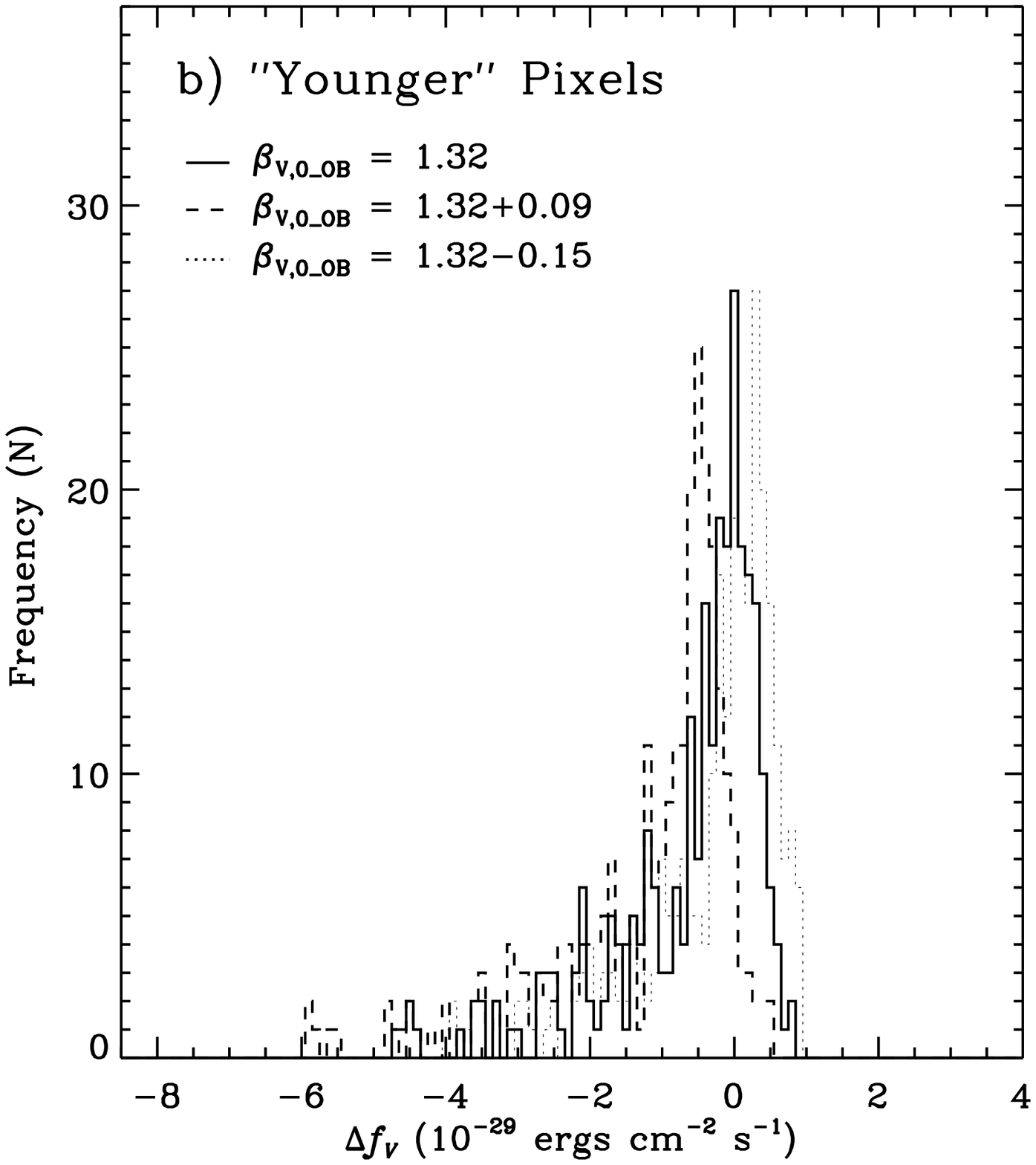,width=0.45\textwidth}}
\caption{
Histograms of $\Delta f_V$, the difference between the observed and
theoretical \vv-band flux for selected pixels with light dominated by
(\emph{a}) older and (\emph{b}) younger stellar populations.  In each panel,
the solid histogram is for the adopted value of \bzero, while the
dashed (dotted) histogram assumes a value for \bzero\ that corresponds
to the upper (lower) bound of the uncertainty range determined in
\figref{beta}.
}\label{fluxdiff}
\end{figure*}

Having determined appropriate extinction-free flux ratios, the next step is
to calculate the \vv-band flux difference, $\Delta f_V$, for each pixel.
Using the theoretical flux defined in \eqref{f_0}, the flux difference is
calculated as:
\begin{equation}\label{eq_dfv}
  \Delta f_V = [f_V - (\beta_{V,{\rm 0}}\times f_{\rm 3.6\,\mu m})]\,,
\end{equation}
where, $f_V$ and $f_{\rm 3.6\,\mu m}$ are the observed pixel-fluxes in the
optical \vv\ and IRAC 3.6\,\mum\ bands, and \bzero\ is the estimated 
extinction-free flux ratio.  The distribution of $\Delta f_V$ values is 
shown in \figref{fluxdiff} for both younger and older pixels.  The solid 
histograms are for the $\Delta f_V$ values calculated from our adopted 
values for \bzero, while the dashed (dotted) histograms are computed for
\bzero\ values at the upper (lower) bound of the quoted 1$\sigma$
uncertainty range in \figref{beta}.
While \figref{beta} shows simply the level of the \vv-band flux relative to 
the flux in 3.6\,\mum\ filter, \figref{fluxdiff} shows the \emph{absolute} 
difference between the observed and estimated theoretical \vv-band 
pixel-fluxes.  Regardless of the selection of \bzero, both histograms --- 
for both younger and older pixels --- show that the number of pixels with 
positive $\Delta f_V$ goes to zero quickly, and that there are distinct 
tails extending to large negative values of $\Delta f_V$.  For the younger 
pixels, the effect of the uncertainty in the selection of \bzero\ is small 
(i.e., the three histograms show similar distribution in 
\figref{fluxdiff}b).  The older pixels, however, are affected more 
(\figref{fluxdiff}a).  Since older stellar populations tend to have less 
dust intermixed, and hence suffer less extinction, the dashed histogram for
the larger value of \bzero\ produce an excess of pixels with large values
of \av.  On the other hand, the dotted histogram for the smaller value of 
\bzero\ will produce many pixels with an unphysical excess of visual flux,
indicating that this value must be a robust lower bound on \bzero.  This
confirms that \bzero\,=\,1.10 is a reasonable and appropriate value for
the older pixels. 

\begin{figure}
  \centerline{\psfig{file=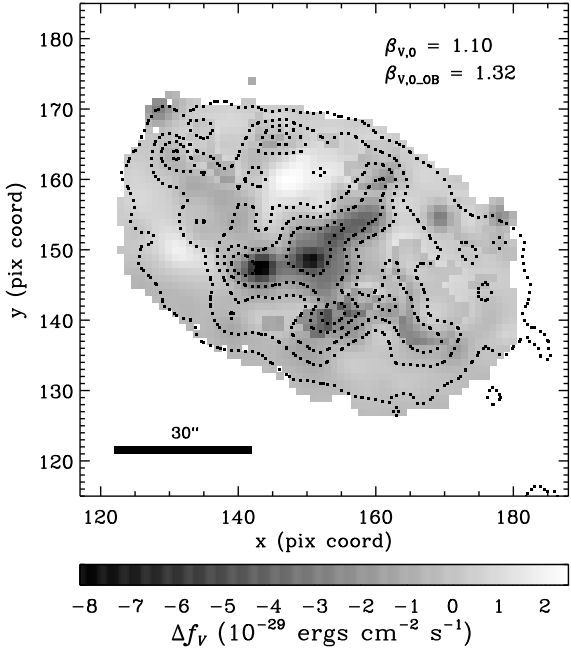,width=0.45\textwidth}}
\caption{
Pixel map of the flux difference, $\Delta f_V$, rendered at the resampled
pixel-scale of 1\farcs5~pixel$^{-1}$.  Darker gray-scale levels indicate
a larger deficiency of observed \vv-band flux compared to the modeled
intrinsic flux.  Black dotted contours trace the IRAC 8.0\,\mum\ PAH
emission.  While it is hard to visually trace the distribution of dust in
\figref{colimg}b, the locations of the dust lanes in \figref{colimg}a
generally agree with the distribution of the flux-deficient pixels in the
present figure.
}\label{dfmap}
\end{figure}

Next, we check the spatial distribution of the calculated $\Delta f_V$ 
values.  If the distribution of negative $\Delta f_V$ follows genuine 
galactic features, such as SF-regions, spiral arms, and PAH emission, then 
this strengthens our argument that our method largely traces the dust 
extinction.  The spatial distribution of $\Delta f_V$ is shown as a 
pixel-coordinate map in \figref{dfmap}, with the IRAC 8.0\,\mum\ emission 
overlaid as dotted contours.  This map demonstrates that the distribution of 
$\Delta f_V$ values is not random at all, but closely associated with 
genuine galactic structures.  Visual comparison of this map to the color 
composite image in \figref{colimg}b confirms that pixels with a large 
deficiency of observed \vv-band flux are not simply corresponding to pixels 
appearing darker (lower surface brightness).  
Instead, pixels with $\Delta f_V$\,$<$\,0 are distributed around the regions 
that appear bluer, as well as near the center of the galaxy.  The higher 
resolution image of \figref{colimg}a shows that some pixels with larger 
negative values of $\Delta f_V$ (darker gray in \figref{dfmap}) trace the 
visible dust lanes, seen in silhouette, and the bluer SF-regions.  This 
demonstrates that even though some of the dust is not visually conspicuous 
(as in the lower resolution image in \figref{colimg}b), our method is 
capable of estimating dust extinction and its spatial distribution from 
observations in only two broadband filters --- with a third filter 
(\uu-band) serving only to robustly separate younger pixels from older 
pixels (\figreftwo{pcmd_model}{obsel}).  Since regions with visible or 
plausible dust content are recovered 
well by our method, this gives us confidence that regions for which this
method indicates a low dust content could also be real.

\begin{figure}
\centerline{\psfig{file=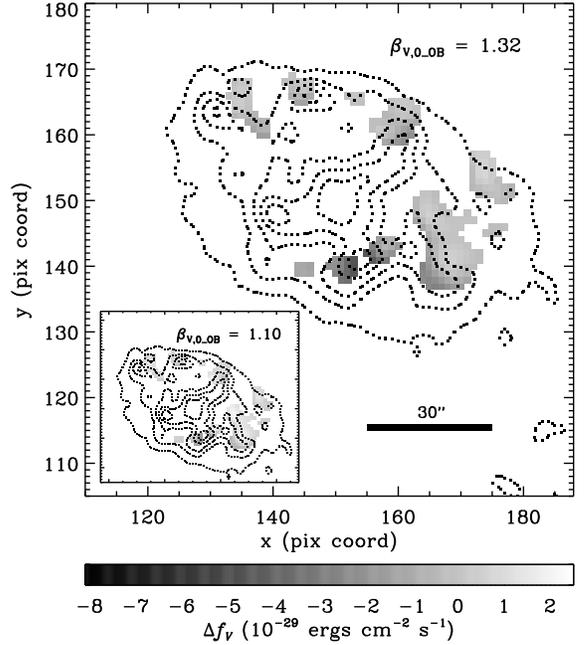,width=0.45\textwidth}}
\caption{
Comparison of the flux difference, $\Delta f_V$, inferred for pixels
dominated by younger stellar populations.  The main panel shows the result
when the younger pixels are treated separately, adopting
$\beta_{V,{\rm 0\_OB}}$\,=\,1.32\,.  Here, most pixels show that they are
missing \vv-band flux, as expected.
The inset panel shows the result for no such special treatment (i.e.,
assuming $\beta_{V,{\rm 0\_OB}}$\,=\,\bzero\,=\,1.10, as for pixels
dominated by old stellar populations).  Here, most of the younger pixels
show an unphysical \emph{excess} of observed \vv-band flux.  Black dotted
contours trace 8.0\,\mum\ PAH emission.
}\label{difference}
\end{figure}

Before we proceed to estimate the amount of dust extinction in each pixel,
there is one more check that we must perform to support our separate 
treatment of the younger pixels.  The main panel of \figref{difference} 
shows a map of $\Delta f_V$ for the younger pixels for our adopted value of
$\beta_{V,{\rm 0\_OB}}$\,=\,1.32, while the inset shows the result when 
these pixels are treated as older pixels with 
$\beta_{V,{\rm 0\_OB}}$\,=\,\bzero\,=\,1.10\,.  The lighter shade of gray 
of the pixels in the inset panel indicates 
that most of them have positive $\Delta f_V$, suggesting that the observed 
\vv-band flux is equal to or larger than expected from the 3.6\,\mum\ data, 
i.e., that the younger pixels miss either no \vv-band flux or show an 
unphysical \emph{excess} of observed \vv-band flux.  Since young stellar 
populations are usually associated with relatively large amounts of dust 
\citep[e.g.,][]{vanhouten61, knapen91, regan04, barmby06}, a value of 
\bzero\,=\,1.10 is clearly too small for the younger pixels.  Our separate 
treatment of the younger pixels, adopting 
$\beta_{V,{\rm 0\_OB}}$\,=\,1.32\,$^{+0.09}_{-0.15}$ as outlined above, is
therefore appropriate.  For that larger value of \bzero, the younger pixels
are found to suffer small-to-moderate amounts of extinction by dust, 
consistent with the known presence of dust at or near these regions of 
recent star formation and producing consistent results with previous 
studies.

\subsection{Measuring the Dust Extinction}

We can now estimate the most likely amount of dust extinction in each pixel.
With \bzero\,=\,1.10 (and $\beta_{V{\rm ,0\_OB}}$\,=\,1.32) and AB-magnitude
zeropoint, $V_{\rm zp,AB}$\,=\,48.59~mag, \eqref{av_calc} becomes:
\begin{equation}\label{av_obs}
  A_V = m_V - [-2.5\log(\beta_{V,{\rm 0}}\times f_{\rm 3.6\,\mu m}) - V_{\rm zp,AB}]\,.
\end{equation}
We cannot simply apply \eqref{av_obs} to all pixels, however, since some 
pixels 
have \betav\,$>$\,\bzero\ (see \figref{beta}) and  $\Delta f_V$\,$>$\,0 (see 
\figref{fluxdiff}), i.e., seemingly implying an unphysical negative dust 
extinction, \av.  Because the \bzero\ value is the \emph{estimated} 
extinction-free flux ratio, rather than the true dust-free ratio, it is 
possible that some pixels have an observed \betav\ value that is larger than 
\bzero.  We therefore have to be careful how we treat these pixels in our 
analysis. 

As mentioned above, the pixels with positive $\Delta f_V$ (\figref{dfmap}) 
appear darker than neighboring regions in the color composites of
\figref{colimg}.  Considering also that the 8.0\,\mum\ PAH emission is 
weak in these regions (i.e., these pixels are surrounded by outer 8.0\,\mum\ 
contours with large spacing between neighboring contours), these positive 
$\Delta f_V$ pixels are not caused by SF-activity, but are most likely the 
result of underestimating the intrinsic \bzero\ value.  The actual amount of 
dust extinction might also be minimal or zero (\av\,$\simeq$\,0) for these 
pixels.  We adopted a single \bzero\ for younger and for older pixels, yet 
the intrinsic flux ratio for each pixel will vary slightly based on 
different factors.  As a result, some pixels with positive $\Delta f_V$ are 
expected, even in the absence of noise.  For our adopted \bzero\ of 1.10 
(1.32) for older (younger) pixels, most pixels with excess flux have $\Delta 
f_V$\,$\lesssim$\,1.0\,$\times$10$^{-29}$~ergs\,cm$^{-2}$\,s$^{-1}$, and 
only a very small fraction has an excess as large as
$\sim$\,2.0\,$\times$\,10$^{-29}$~ergs\,cm$^{-2}$\,s$^{-1}$ (see 
\figref{fluxdiff}).  In the following, we will therefore assume that these
pixels suffer no extinction, i.e., that \av\,=\,0~mag\,arcsec$^{-2}$.

\subsubsection{Impact of the Uncertainty in \bzero}

\begin{figure*}
\centerline{\psfig{file=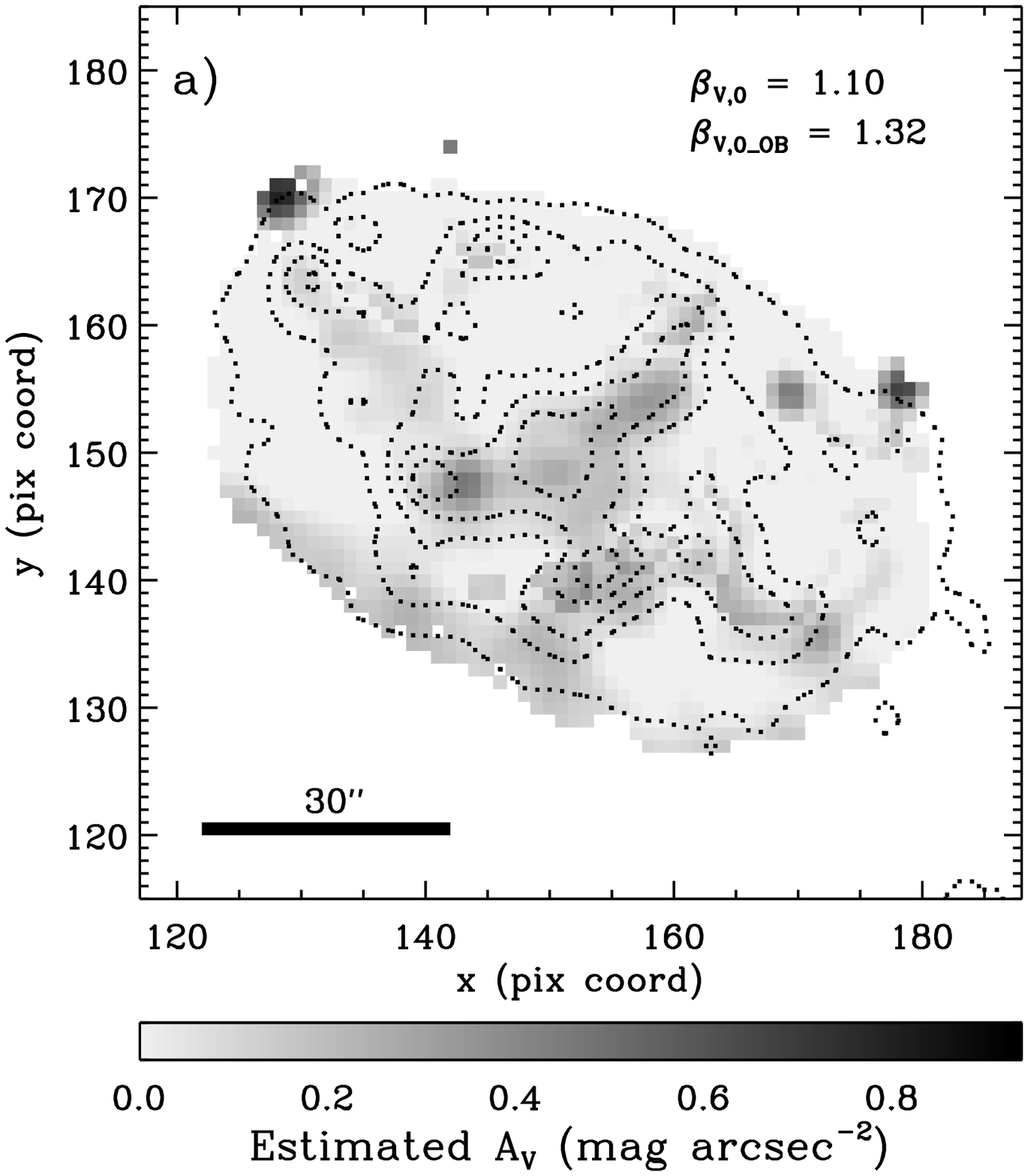,width=0.45\textwidth}
\hspace{1cm}
\psfig{file=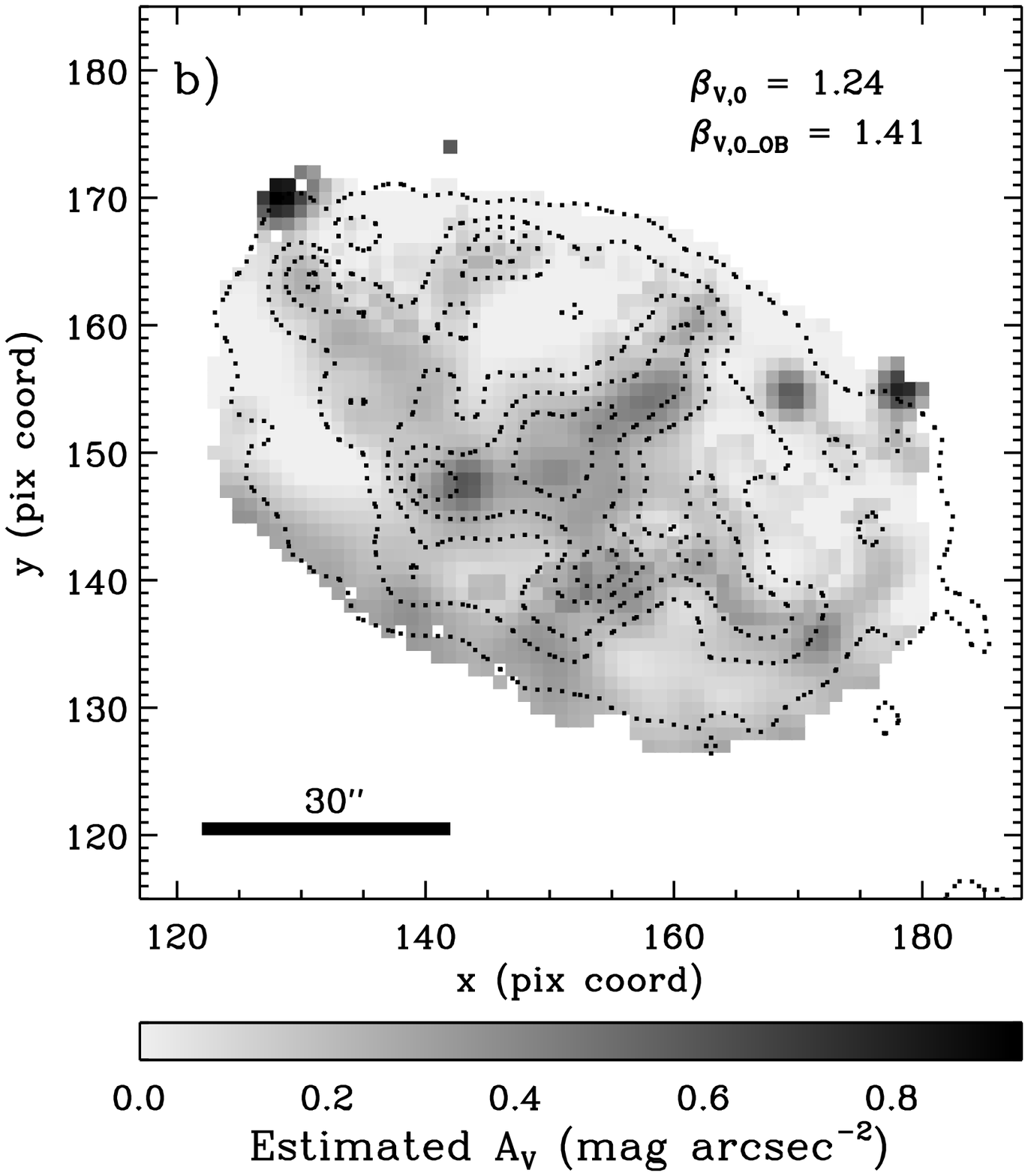,width=0.45\textwidth}}
\caption{
Spatial distribution of \av, inferred for \bzero\ values of (\emph{a}) 1.10
and 1.32 (mean), and (\emph{b}) 1.24 and 1.41 (upper limit), for older and
younger pixels, respectively.  The same gray-scale is used for both
(\emph{a}) and (\emph{b}).  The pixels with the lightest shade of gray have
no detectable extinction (\av\,=\,0~mag\,arcsec$^{-2}$).  The darkest pixels
have a visual extinction \av\,$\simeq$\,0.8~mag\,arcsec$^{-2}$ in (\emph{a})
and $\simeq$\,0.93~mag\,arcsec$^{-2}$ in (\emph{b}).  Dotted contours trace
the 8.0\,\mum\ PAH emission.  These extinction maps indicate that the higher
extinction coincides with SF-regions that appear blue in the color composite
images (see \figref{colimg}), as well as with several dust features
visible in \figref{colimg}a.
}\label{av}
\end{figure*}

\begin{deluxetable*}{lccccc}\label{tbl_1}

\tablecaption{Dust Extinction Based on Different \bzero\ Values\label{table1}}
\tablewidth{0pt}
\tablehead{
Selected & & & $A_{V_{\rm mean}}$\tablenotemark{a} & $A_{V_{\rm max}}$ & 
\% of pixels\\
\bzero\ Level & \ch{\bzero} & \ch{$\beta_{V,\rm 0\_OB}$} & 
(mag arcsec$^{-2}$) & (mag arcsec$^{-2}$) & with \av\,=\,0
}
\startdata
Adopted \bzero\ & 1.10 & 1.32 & 0.064 & 0.80 & 45.0\% \\
Upper ($+$\,$\sigma$) & 1.24 & 1.41 & 0.15 & 0.93 & 16.4\% \\
Lower ($-$\,$\sigma$) & 0.96 & 1.27 & 0.015 & 0.63 & 83.1\% 
\enddata
\tablenotetext{a}{Mean $A_V$ value of all pixels analyzed for the galaxy.}

\end{deluxetable*}

Since neither the relationship between \av\ and $\Delta f_V$ nor the 
relationship between $\Delta f_V$ and \betav\ is linear, the 1$\sigma$ 
errors in the estimated extinction-free flux ratios ($\sigma_{\beta_{V,{\rm 
0}}}$\,=\,$\pm$0.14 and $\sigma_{\beta_{V,{\rm 0\_OB}}}$\,=\,$+$0.09 \& 
$-$0.15) cannot be simply converted to corresponding $\sigma_{\!A_V}$ 
values.  Instead, we will assess \emph{how} the distribution and values of 
\av\ change, as we vary the estimated dust-free flux ratio from the 
adopted value of \bzero\ to \bzero\,$\pm$\,$\sigma_{\beta_{V,{\rm 0}}}$.

\figref{av} shows the spatial distribution of the estimated dust extinction, 
\av, for each pixel in NGC\,959, for two sets of theoretical extinction-free 
flux ratios.  \figref{av}a shows the distribution of \av\ inferred for our 
adopted \bzero\ values of 1.10 and 1.32 for older and younger pixels, while 
\figref{av}b shows the result for \bzero\ values set at the higher end of 
the uncertainty range (i.e., \bzero\,=\,1.24 and 1.41, respectively).  The 
mean and maximum dust extinction for all pixels in \figref{av}a 
(\figref{av}b) are $A_{V,{\rm mean}}$\,=\,0.064 (0.15) and $A_{V,{\rm 
max}}$\,=\,0.80 (0.93) mag\,arcsec$^{-2}$.  As expected, the dust-extinction
map (not shown) for the lower limit, \bzero$-$$\sigma_{\beta_{V,{\rm 0}}}$,
is covered mostly by \av\,=\,0 pixels.  These results are summarized in 
Table~1, where the last column indicates the fraction of pixels with 
\av\,=\,0~mag\,arcsec$^{-2}$ in the image.

Even though the estimated \bzero\ value changes, \figref{av} shows that the 
distribution of dust extinction follows the structures of the galaxy and the 
8.0\,\mum\ contours.  \figref{av} and Table~1 also show that the effect of 
varying the theoretical \bzero\ value is not equal to simply adding or 
subtracting a constant value $\Delta A_V$ to the dust-extinction values
calculated for our adopted value of \bzero.  As \bzero\ change from 1.10 and 
1.32 (\figref{av}a) to 1.24 and 1.41 (\figref{av}b), some pixels near pixels 
with \av\,$\gtrsim$\,0~mag\,arcsec$^{-2}$, which originally were deemed 
extinction-free, now suffer a slight amount of dust extinction.  Other 
pixels that are further away, e.g., pixels in areas that are faint at 
8.0\,\mum, stay at \av\,=\,0~mag\,arcsec$^{-2}$.  This confirms our 
assumption that these pixels have minimal or no dust extinction.

\subsubsection{Analysis at Higher Spatial-Resolution as Confirmation}

\begin{figure}
\centerline{\psfig{file=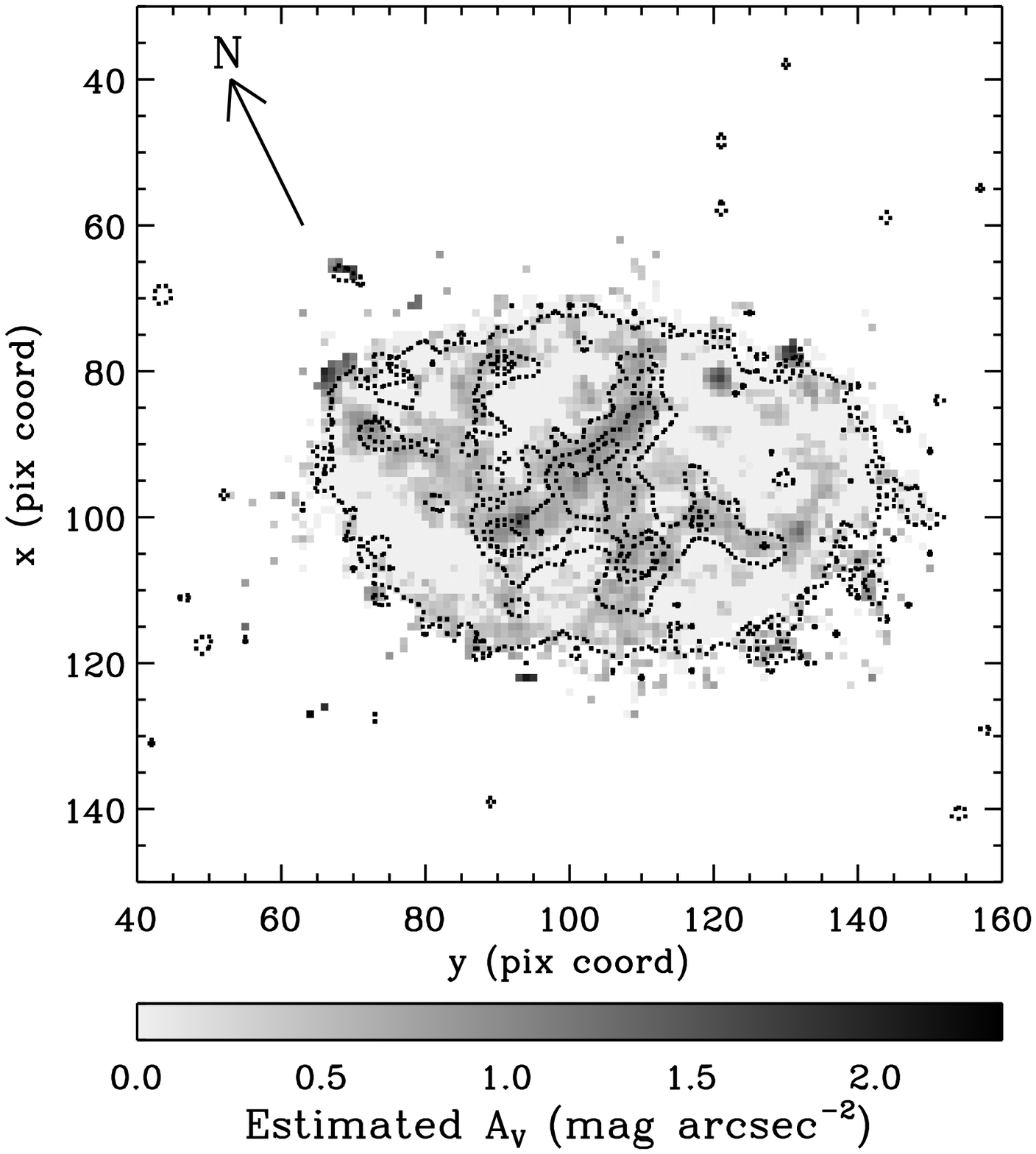,width=0.45\textwidth}}
\caption{
Distribution of dust extinction at the (higher) spatial resolution of the
\emph{Spitzer}/IRAC 3.6\,\mum\ image (1\farcs2~pixel$^{-1}$ and
$\sim$2\farcs2~FWHM).  The dotted contours trace 8.0\,\mum\ PAH emission at
a native resolution of $\sim$\,2\farcs3~FWHM.  This map is
presented in the instrument coordinate system, oriented as indicated at the
upper left.
}\label{av_hr}
\end{figure}

While individual dust features usually are small-scale structures, our
initial analysis was performed at the low spatial resolution of the 
\emph{GALEX} NUV image.  To determine if our result is a true measure of 
dust extinction or not, we repeat the same analysis with \emph{higher} 
spatial resolution images, using only \uu, \vv, and IRAC 3.6\,\mum\ images.  
Since the 3.6\,\mum\ image has the coarsest pixel scale of 1\farcs2~pixel$
^{-1}$ and a PSF with $\sim$\,2\farcs2~FWHM, the ground-based \uu- and 
\vv-band images are registered, resampled, and convolved to match the 
orientation, pixel scale, and resolution of the 3.6\,\mum\ image.  
\figref{av_hr} shows the distribution of estimated \av\ values at this 
higher spatial resolution.  The 8.0\,\mum\ contours are overplotted at 
the native IRAC resolution of $\sim$\,2\farcs3~FWHM.

An important difference between the two spatial resolutions is that, while 
the maximum dust extinction is $A_{V,{\rm max}}$\,$\simeq$\,0.8~mag\,arcsec$
^{-2}$ in \figref{av}a, $A_{V,{\rm max}}$ in \figref{av_hr} reaches $\sim
$\,2.3~mag\,arcsec$^{-2}$.  This jump in \av\ value is expected, since the 
coarser \emph{GALEX} PSF smoothes out the effect of dust extinction and 
reduces the averaged \av\ for each pixel.  Since the ratio of the effective 
areas of the \emph{GALEX} NUV and \emph{Spitzer} 3.6\,\mum\ PSFs is 
(5\farcs3/2\farcs2)$^{2}$\,$\simeq$\,6, the typical extinction per pixel 
should change, to first order, by $\sim$\,1.9~mag\,arcsec$^{-2}$.  Other 
factors, such as uncertainties in estimating \bzero\ and blending of light 
from structures that are unresolved with the coarser while resolved with 
the smaller PSF, also play a role.  The observed shift in \av\ of 
$\Delta A_V$\,$\simeq$\,1.5~mag\,arcsec$^{-2}$ is therefore broadly 
consistent, while the 0.4~mag\,arcsec$^{-2}$ difference illustrates the 
effects of the clumpiness of stars and dust on scales of 2\farcs2--5\farcs3 
($\sim$\,110--250\,pc at the distance of NGC\,959).  The higher spatial 
resolution images are better at tracing smaller dust features and their 
higher extinction values within a single pixel.  Nonetheless, the overall 
distribution of dust extinction is similar in \figreftwo{av}{av_hr}.  While 
the fine details of the measurable dust extinction and its spatial 
distribution depend on the resolution, both extinction maps trace the 
8.0\,\mum\ PAH emission and trace genuine galactic structures.
 
In conclusion, from the series of tests described above, we find that the 
two-dimensional distribution of dust extinction can, indeed, be reliably
estimated by our pixel-based method and the observed \vv-to-3.6\,\mum\
flux ratio.

\subsubsection{Interesting Regions}

\begin{figure}
\centerline{\psfig{file=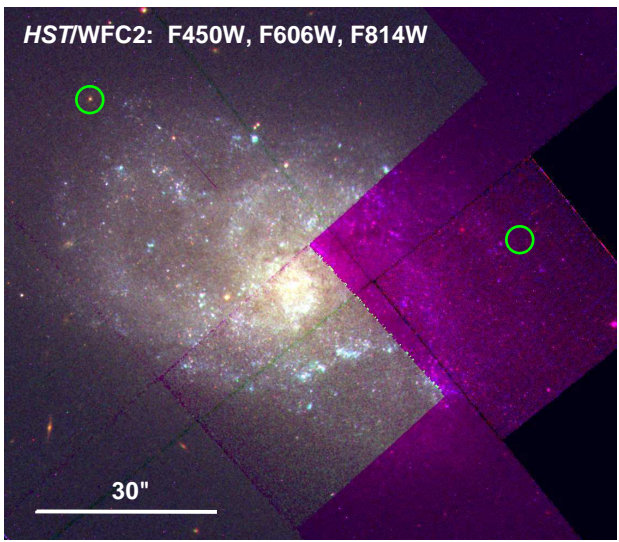,width=0.45\textwidth}}
\caption{
Color composite of Archival \emph{HST}/WFPC2 F450W, F606W, and F814W images.
The green circles indicate the two regions with high apparent \av\ in
\figref{av}.  While there is no object evident within the right green
circle, the circle in the upper left corner contains a red object.  Along
the southern rim of NGC\,959, a faint dust lane is discernable that appears
to trace an outer arm.  The typical extinction along this dust feature is
\av\,$\simeq$\,0.3--0.4~mag\,arcsec$^{-2}$ in \figref{av}.
}\label{hst}
\end{figure}

Having produced maps of the spatial distribution of dust extinction in 
NGC\,959, we now discuss some of the most interesting dust features in
\figref{av} (and \figref{av_hr}).  While the distribution of regions with 
larger \av\ closely traces the SF-regions, a not previously identified bar,
and arm-like structures, there are several regions that draw our attention.
These are: (1) a compact region at the Northeast edge (NE; upper-left) of 
NGC\,959 that appears to suffer very high extinction; (2) another such
high-extinction region at the Northwest edge (NW; upper-right); and (3) an 
extended area of moderate dust extinction along the Southeast (SE; bottom), 
rim of the galaxy.  Since all of these features are also visible in 
\figref{av_hr}, these must be real features.  To visually confirm 
small-scale dust features in NGC\,959, and check their interpretation as 
genuine galactic features as opposed to chance superpositions of unrelated
objects, we created another color composite from Archival \emph{HST}/WFPC2 
F450W, F606W, and F814W images, shown in \figref{hst}.  The two circles in 
this image mark regions (1) and (2).
  
The NE high-\av\ region, region (1), is centered around pixel coordinates 
[$x$,\,$y$]\,$\simeq$\,[128,\,169] in \figref{av}.  Even though nothing 
conspicuous is visible in the low-resolution color composite 
(\figref{colimg}a), the \emph{HST} image (\figref{hst}) reveals a compact, 
bright red source at the center of the marked region.  Without morphological 
indicators, multi-filter photometry, or spectroscopic 
information for this
particular object, it is hard to decide whether this is a reddened stellar 
population within NGC\,959's disk, or a background (foreground) object that 
is visible through (against) the disk.  We do note, however, that the color 
of this object is very similar to that of the edge-on background galaxy that 
is visible at the bottom left of \figref{hst}.  Region (2) is located in the
NW corner of the galaxy around [$x$,\,$y$]\,$\simeq$\,[178,\,155].  There is 
no F606W coverage for this region in \figref{hst}, but there is no object 
discernable in the other two \emph{HST} filters.  At present, we lack 
sufficient information to establish if these regions are truly associated 
with NGC\,959.  Until further evidence is obtained, we will treat these 
regions as a part of the galaxy.

The pixels comprising feature (3), the moderate-extinction region running 
along the SE rim of the galaxy disk, initially did not show up as having 
particularly large-negative $\Delta f_V$ values (see \figref{dfmap}).  Once 
the dust extinction is calculated, these pixels do stand out with 
low--moderate \av\ values, indicating the presence of an extended dust 
structure.  The higher-resolution dust extinction map (\figref{av_hr}) also 
shows that the pixels in this region suffer higher dust extinction than 
inferred for the opposite (NW) rim of the galaxy.  Visual comparison to 
\figref{hst} confirms that a faint dust lane runs along the SE rim,
apparently tracing an outer spiral arm or armlet.  This shows that our dust
extinction measurement using the \vv-to-3.6\,\mum\ flux ratio is sensitive 
to even low amounts of dust extinction per resolution element, whether
inherently smoothly distributed or resulting from a small filling factor.

\subsubsection{Correcting for Dust Extinction}

\begin{figure*}
\centerline{\psfig{file=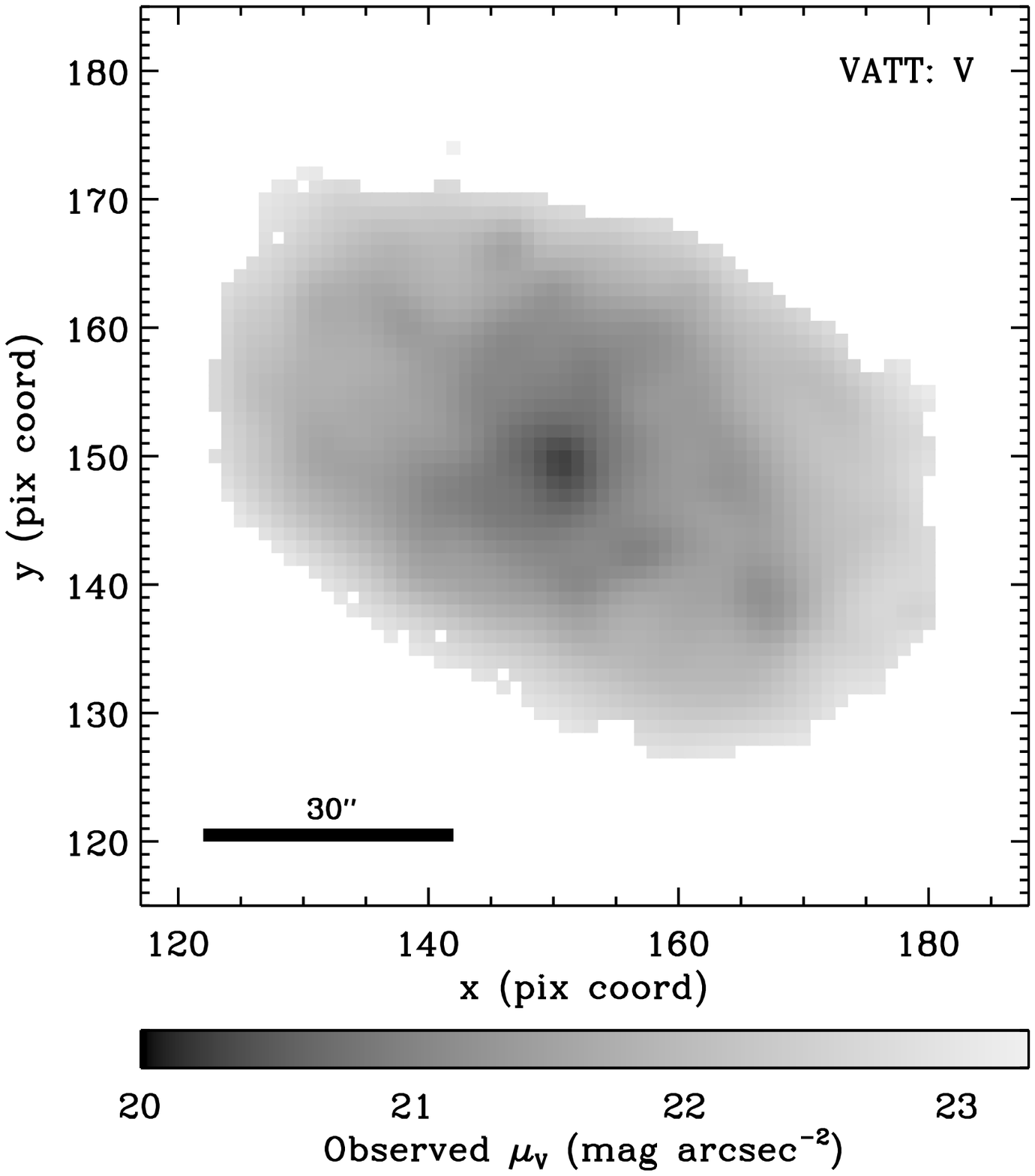,width=0.45\textwidth}
  \psfig{file=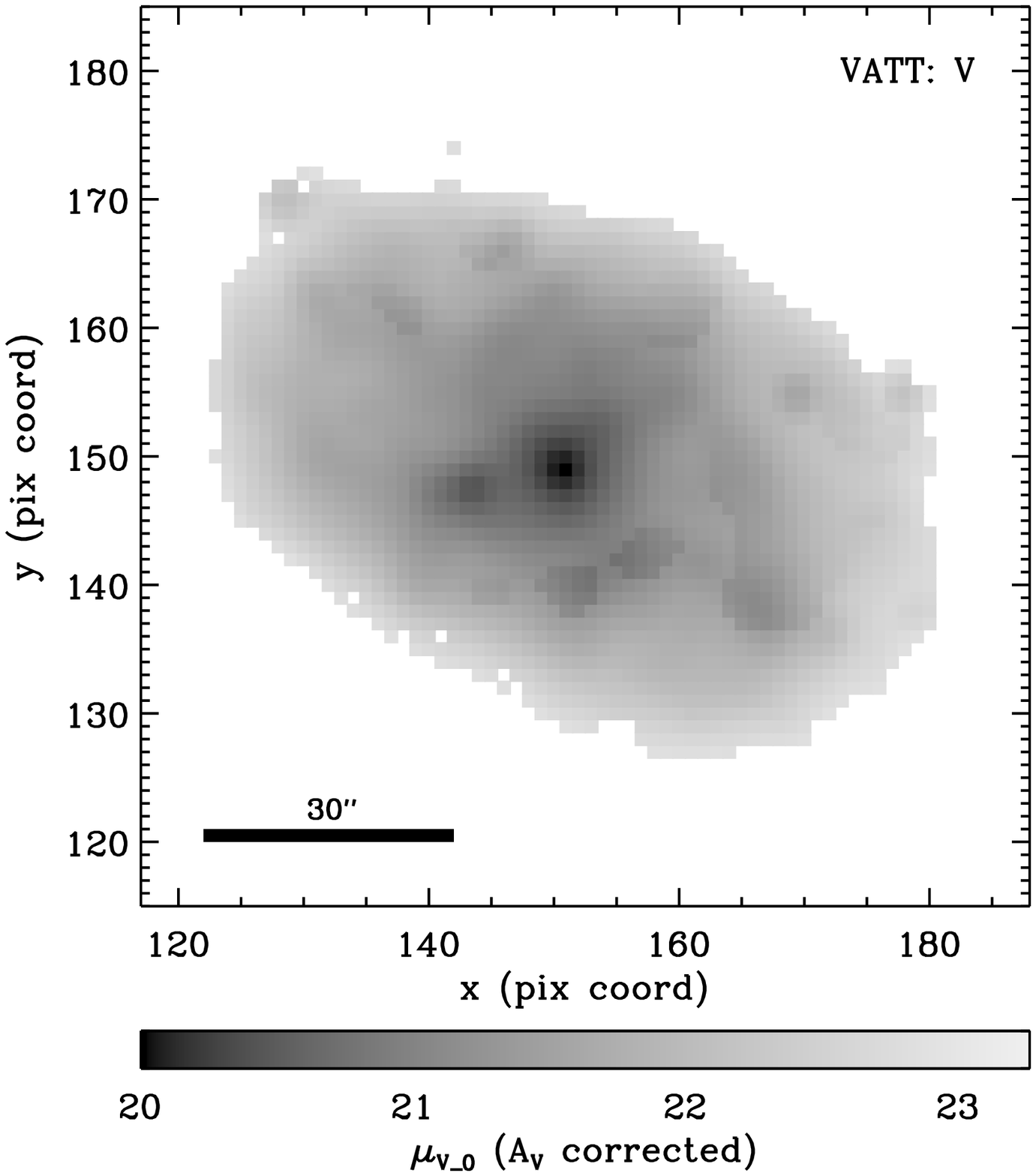,width=0.45\textwidth}}
\centerline{\psfig{file=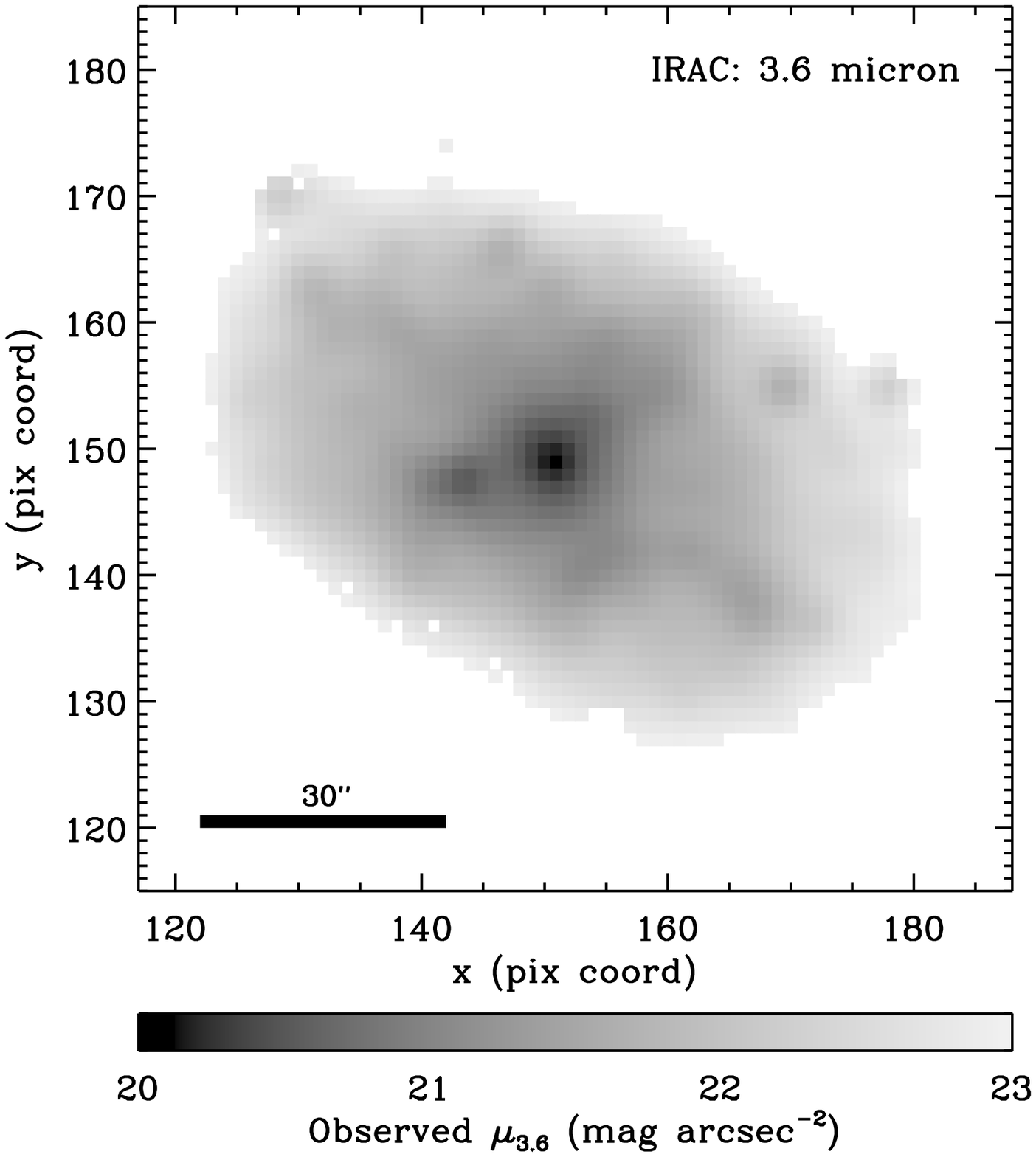,width=0.45\textwidth}
  \psfig{file=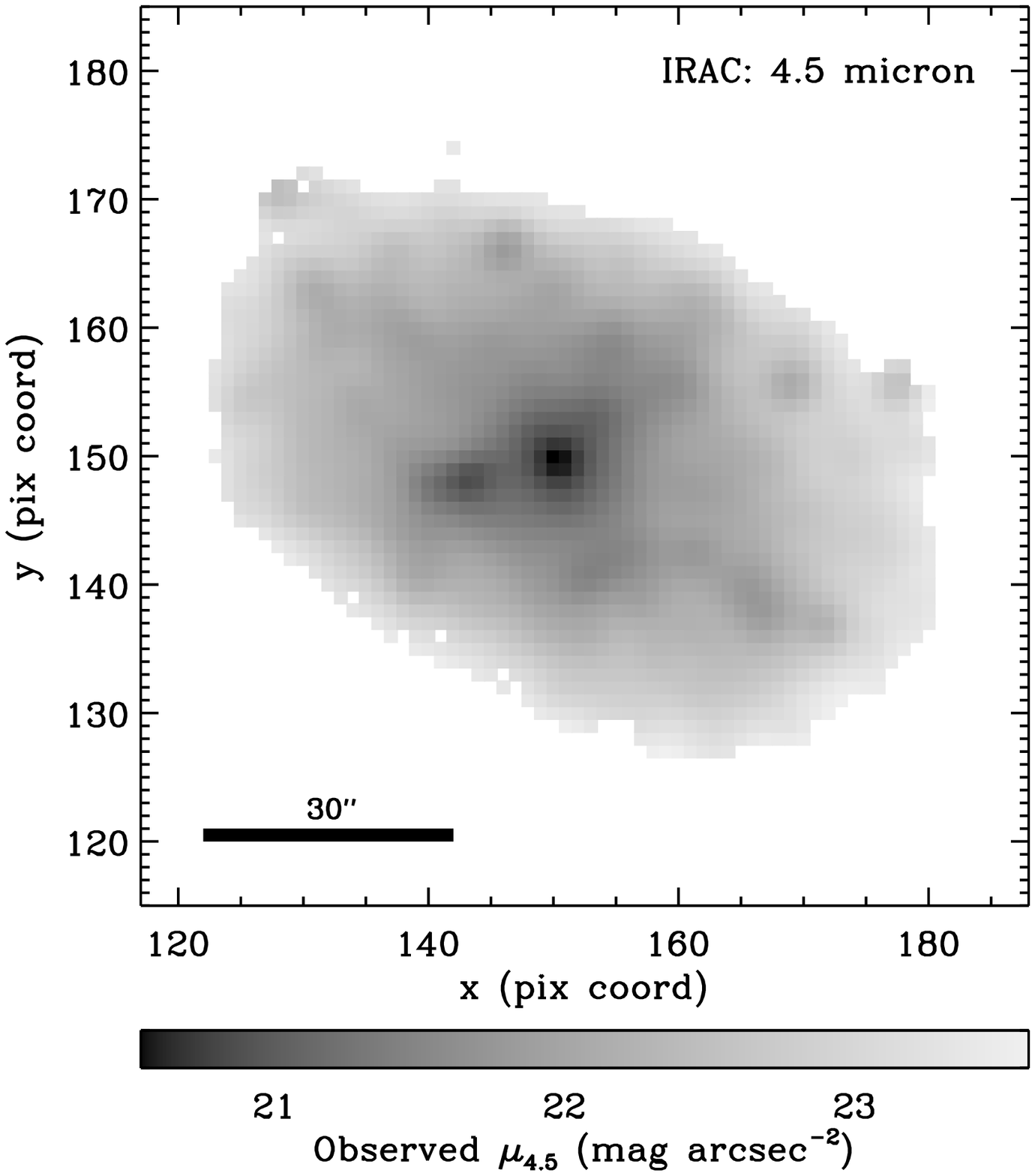,width=0.45\textwidth}}
\caption{
Comparison of observed (\emph{top left}) and extinction-corrected (\emph{top 
right}) \vv-band images and the observed \emph{Spitzer}/IRAC 3.6 and
4.5\,\mum\ images (\emph{bottom}) of NGC\,959 at \emph{GALEX} resolution.
The MIR images are good
tracers of the underlying stellar distribution.  The extinction-corrected
image in \vv\ displays more brighter (darker shade) pixels around blue
SF-regions (see \figref{colimg}) and better resembles the MIR images than
does the observed \vv-band image.
}\label{img_v}
\end{figure*}

Finally, using the calculated \av\ values for each pixel, we correct the 
observed \vv-band image of NGC\,959 to reveal the true underlying stellar 
populations.  The observed (uncorrected) and extinction-corrected images are 
shown at \emph{GALEX} resolution in the top panels of \figref{img_v}, using 
the \emph{same} gray-scale lookup table.  The surface brightness 
distribution before extinction correction (\figref{img_v}a) has lower 
contrast overall and fewer high-contrast features.  Most conspicuous after
applying our dust correction (\figref{img_v}b) is that bluer regions in the 
color composite of \figref{colimg}b become much more prominent, while other 
galactic structures (particularly the central bulge) also become better 
defined.  Especially for areas SE (centered on [$x$,\,$y$]\,$\simeq
$\,[142,\,148]) and NW ([$x$,\,$y$]\,$\simeq$\,[159,\,155]) of the galaxy 
center, the \vv-band surface 
brightness becomes much brighter in areas where dust lanes are evident in 
the color composites (\figref{colimg}a and \figref{hst}).  The bottom 
two panels show the observed \emph{Spitzer}/IRAC 3.6\,\mum\ and 4.5\,\mum\
images (Figs.~13c and 13d), which trace the distribution of the 
underlying older stellar populations \citep[e.g.,][]{regan04, willner04}.  
These MIR images and the \emph{extinction-corrected} \vv-band surface 
brightness distribution show excellent qualitative agreement.  Therefore, as 
\citet{regan00} did with optical--NIR images and radiative transfer 
modeling, we successfully corrected the dust extinction with images in only 
two filters (\vv\ and 3.6\,\mum) --- with a \uu-band image serving only to 
robustly separate pixels dominated by the flux from younger stellar 
populations from those dominated by older stellar populations.

\section{Application to Other Filters}

Given the amount of visual dust extinction, \av, in each pixel, we can 
calculate the extinction in any other filter.   The extinction in a given
filter depends on its throughput as a function of wavelength, as well as on
the metallicity of the stellar populations of the galaxy of interest through
the adopted extinction curve \citep[e.g.,][]{seaton79, koornneef81, 
howarth83, cardelli89, calzetti94, gordon03}.  In a forthcoming paper
(Tamura \etal\ 2009b, in preparation), we will use UV--MIR multi-filter
dust-corrected surface photometry for a detailed analysis of the stellar 
populations within NGC\,959.

\subsection{Extinction Curves}

\begin{figure}
\centerline{\psfig{file=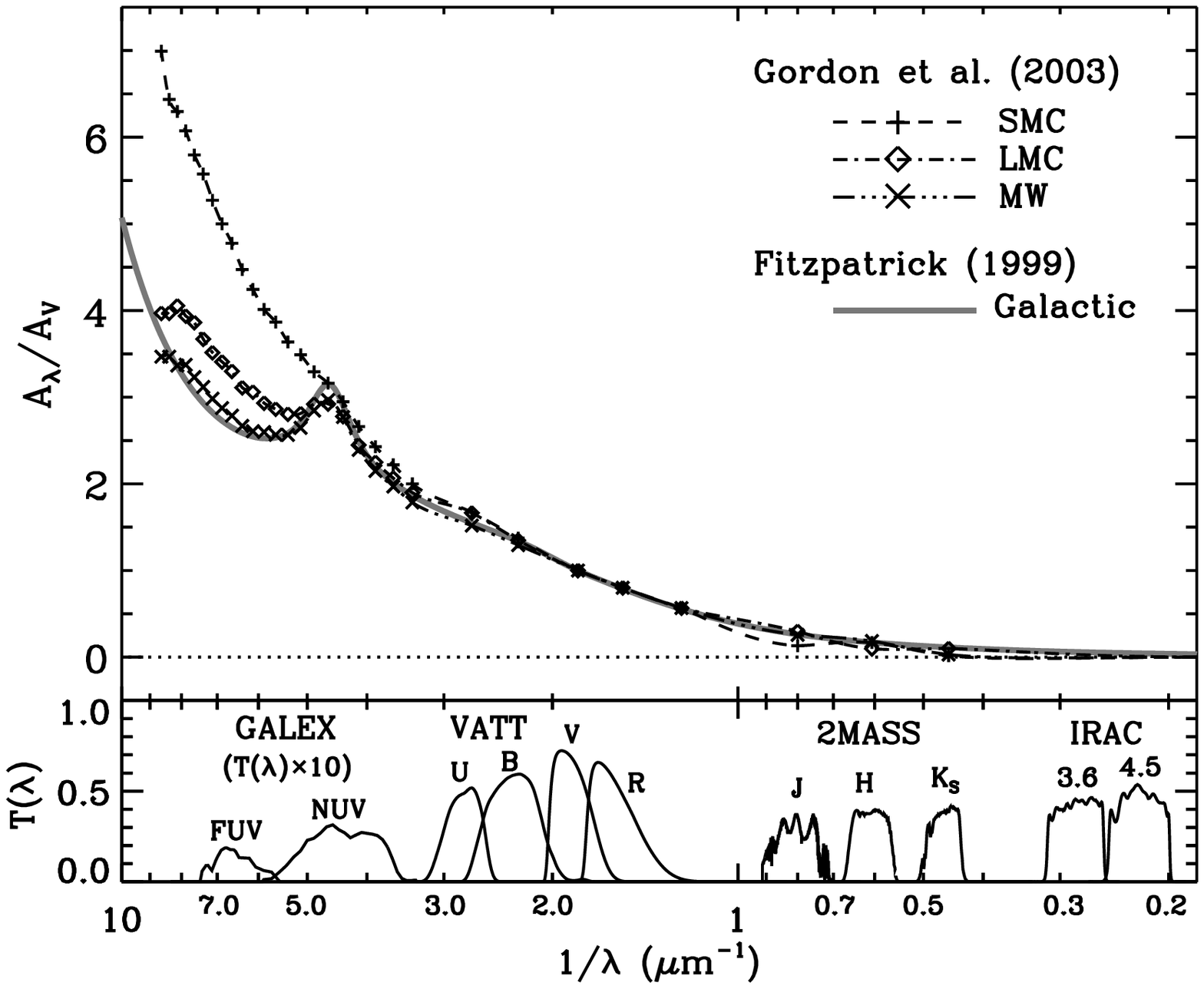,width=0.45\textwidth}}
\caption{
\emph{Top panel}: Extinction curves for SMC (pluses), LMC (open diamonds),
and MW (crosses) metallicity from FUV to MIR wavelengths.  The smooth
extinction curves were interpolated from the data points of \citet{gordon03}.
Also shown is the Galactic extinction curve from \citet{fitzpatrick99},
which we adopt for the NIR and MIR filters.  \emph{Bottom panel}: Total
throughput curves, T($\lambda$), for different telescope-filter combinations
(see also \figref{model}).  The \emph{GALEX} filter curves were scaled up by
a factor of 10 for better visibility.
}\label{extinction}
\end{figure}

In general, dust extinction increases toward shorter wavelengths.  For 
metal-poor stellar populations --- such as those in the Small Magellanic 
Cloud (SMC) --- the dust-extinction curve is largely monotonic as a function 
of wavelength.  The shape of the extinction curve becomes more complicated 
for metal-rich stellar populations.  Especially the extinction at shorter 
wavelengths ($\lambda$\,$\lesssim$\,2200\,\AA) may be significantly affected 
by the 2175\,\AA\ interstellar dust feature and the increased steepness of 
the UV extinction \citep[e.g.,][see also \figref{extinction} 
here]{calzetti94, gordon03}.   

\citet{gordon03} compared the observations of many stars to calibrate the 
extinction curves --- the wavelength dependent extinction relative to that 
at 0.55\,\mum\ --- for SMC, LMC, and MW type dust.  Based on Table~4 (and 
Fig.~10) of \citet{gordon03}, we interpolated the published extinction 
curves on a finer wavelength grid.  In the NIR, \citet{gordon03} have only
one data-point for each of the 2MASS $JHK_s$ filters, and the SMC extinction
curve in particular appears quite uncertain.  To cover longer wavelengths up
to the \emph{Spitzer}/IRAC filters, in the NIR and MIR, we adopted the 
Galactic extinction curve of \citet{fitzpatrick99}, rescaled for 
$R$\,=\,$A_V/E(B\!-\!V)$\,=\,3.1\,.  The adopted extinction curves for SMC, 
LMC, and MW metallicities are plotted in \figref{extinction}.  For 
wavelengths shorter than the 2175\,\AA\ interstellar dust feature, the SMC 
and MW type extinctions differ by $\sim$\,2~mag.  The bottom panel of 
\figref{extinction} shows the relevant filter throughput curves for 
comparison.  The amount of dust extinction in each filter, $A_{\rm filter}$, 
is calculated as a ratio to the extinction in \vv: 
\begin{equation}\label{eq_af_av}
  A_{\rm filter}/A_V = 
    \frac{\int_{\lambda}^{} 
      T_{\rm filter}(\lambda)\,[A(\lambda)/A_V]\,d\lambda}
  {\int_{\lambda}^{} T_{\rm filter}(\lambda)\,d\lambda}\,,
\end{equation}
where, $T_{\rm filter}(\lambda)$ is the throughput curve for each filter, 
and $[A(\lambda)/A_V]$ denotes an extinction curve (\figref{extinction}). 
Table~2 summarizes the computed dust extinction for the different extinction
curves.  For the 2MASS $JHK_s$ filters, the extinction calculated from the 
SMC, LMC, and MW extinction curves of \citet{gordon03} differs significantly 
(indicated with asterisks in Table~2).  Since the extinction at longer
wavelengths is progressively insensitive to metallicity, we conclude that 
these values must be highly uncertain.  Instead, we use the extinction curve 
of \citet{fitzpatrick99} to provide an upper limit to the dust extinction in 
the filters longward of 1\,\mum\ (parenthesized values in Table~2).  Since 
the extinction is small ($A_{\rm filter}/A_V$\,$<$\,0.1), following previous 
studies \citep[e.g.,][]{fazio04, regan04}, we assume that there is no 
measurable dust extinction in the IRAC filters.

An important assumption for the extinction curves described above is that 
the dust is distributed in the form of a ``diffuse-ISM'' or in a 
diffuse-screen geometry, which is applicable only for nearby stars and star 
clusters \emph{within} our own Galaxy.  In extragalactic objects, the dust 
appears to be distributed in smaller clumps of much higher density 
\citep[e.g.,][]{deo06} intermixed with the stars.  Even though the effect of 
dust extinction is averaged (or smoothed out) within an aperture or a 
resolution element --- a single pixel in our pixel-based analysis --- the 
properties of dust extinction (e.g., the 2175\,\AA\ feature and the 
steepness of the UV-extinction curve) are governed by the actual geometry of 
the dust distribution within a galaxy.  This means that two regions with the 
same \emph{average} visual extinction \av\ can have different amounts of 
extinction $A_{\lambda}$ at another wavelength, either because of 
differences in \emph{metallicity} or because of different \emph{dust 
geometries}.  

Two extreme cases of dust geometry are: (1) a uniform thin slab (i.e., 
commonly referred to as a ``diffuse ISM''); and (2) dense clumps covering a 
small fraction of a resolution element.  While we would like to perform a 
detailed study of the dust geometry and its effect on the extinction curve 
for extragalactic objects, this is beyond the scope of the present paper.  
Here, we briefly discuss the effects of the different dust 
geometries studied by \citet{whittet01, whittet04}.  \citet{whittet01} 
studied a total of 27 sight-lines (stars) toward the dark clouds in the 
Taurus region.  One of the results from their study is that 
$R_V$\,=\,$A_V/E(B\!-\!V)$ changes from a ``normal'' to a ``dense cloud''
regime once the extinction exceeds a threshold value of 
$A_{\rm th}$\,$\simeq$\,3.2~mag.  \citet{whittet04} subsequently studied the
effect of dust geometries in detail, which included a thin ``diffuse 
screen'' and a diffuse screen with an embedded ``dense cloud'' (see their 
Fig.~1).  For a detailed analysis and discussion, we refer the reader to 
their papers.  The main effect of the ``dense cloud'' geometry on the 
extinction curve is to weaken or remove 
the 2175\,\AA\ extinction bump \citep[see Fig.~5 of][]{whittet04}
while having little effect on the extinction 
curve at other wavelengths, which remains at the same level as for the mean 
``diffuse ISM'' \citep[Fig.~2 of][]{whittet04}.  The extinction curves 
recreated from \citet{gordon03} and \citet{fitzpatrick99}, and the 
calculated extinction values $A_{\rm filter}/A_V$ (\figref{extinction} and 
Table~2) are therefore treated as the upper limit to the dust extinction 
from different dust geometries.  Since the 2175\,\AA\ feature is covered 
only by the \emph{GALEX} NUV filter, the uncertainty associated with 
different dust geometries is assumed to be minimal in all filters except the 
\emph{GALEX} NUV.  Since the \emph{GALEX} NUV filter is not used in our 
method, we will defer further analysis of the NUV filter to subsequent 
papers (Tamura \etal\ 2009b,c, in preparation).

\begin{deluxetable}{lccccc}\label{table2}

\tablecaption{Dust Extinction ($A_{\rm filter}/A_V$) \label{table2}}
\tablewidth{0pt}
\tablehead{
\ch{Instrument} &  $\lambda_{\rm center}$ & $x$ & & & \\
\ch{\& Filter} & (\mum) & (\mum$^{-1}$) & \ch{SMC} & \ch{LMC} & 
\ch{MW}
}
\startdata
GALEX FUV    & 0.153 & 6.536 & 4.56 & 3.16 & 2.70 \\
GALEX NUV    & 0.227 & 4.405 & 2.91 & 2.59 & 2.53 \\
VATT $U$     & 0.360 & 2.778 & 1.65 & 1.64 & 1.50 \\
VATT $B$     & 0.437 & 2.288 & 1.37 & 1.35 & 1.29 \\
VATT $V$     & 0.542 & 1.845 & 1.00 & 1.00 & 1.00 \\
VATT $R$     & 0.642 & 1.558 & 0.81 & 0.81 & 0.81 \\
2MASS $J$    & 1.235 & 0.810 & 0.14$^*$ & 0.29$^*$ & 0.36$^*$ \\
             &       &       &          &          & ($\leq$0.26) \\
2MASS $H$    & 1.662 & 0.602 & 0.15$^*$ & 0.10$^*$ & 0.18$^*$ \\
             &       &       &          &          & ($\leq$0.17) \\
2MASS $K_s$  & 2.159 & 0.463 & 0.03$^*$ & 0.09$^*$ & 0.04$^*$ \\
             &       &       &          &          & ($\leq$0.11) \\
IRAC 3.6\,\mum & 3.550 & 0.282 & \nodata & \nodata & ($\leq$0.06) \\
IRAC 4.5\,\mum & 4.493 & 0.223 & \nodata & \nodata & ($\leq$0.04) \\
IRAC 5.8\,\mum & 5.791 & 0.173 & \nodata & \nodata & ($\leq$0.03) \\
IRAC 8.0\,\mum & 7.872 & 0.127 & \nodata & \nodata & ($\leq$0.02) 
\enddata
\tablecomments{An asterisk indicates that the uncertainty is significant 
compared to the actual extinction values.  The values in parenthesis are 
calculated using the Galactric extinction curve from \citet{fitzpatrick99}. 
}

\end{deluxetable}

\subsection{A Dust-Free View of NGC\,959}

\begin{figure*}
\centerline{\psfig{file=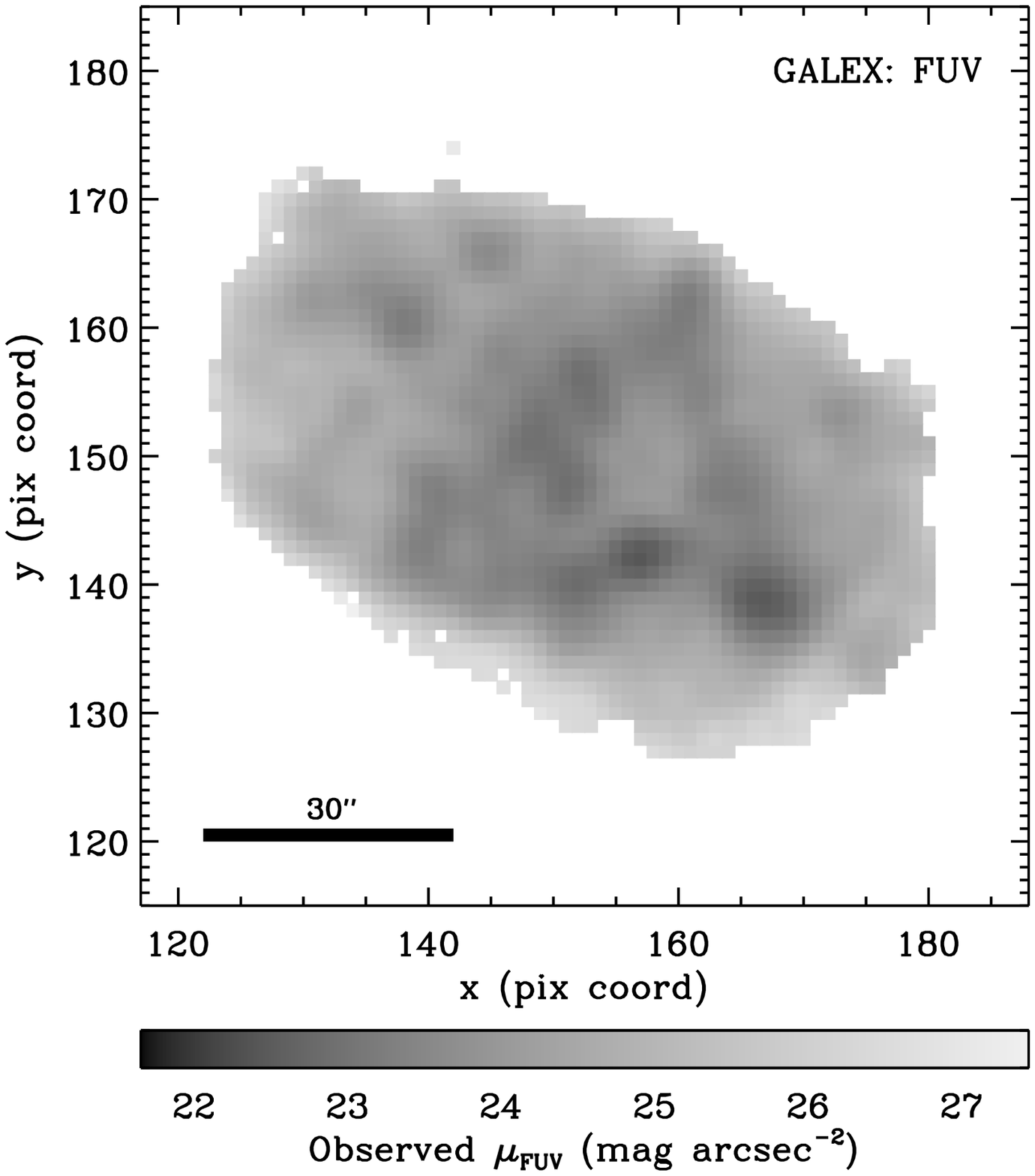,width=0.45\textwidth}
  \psfig{file=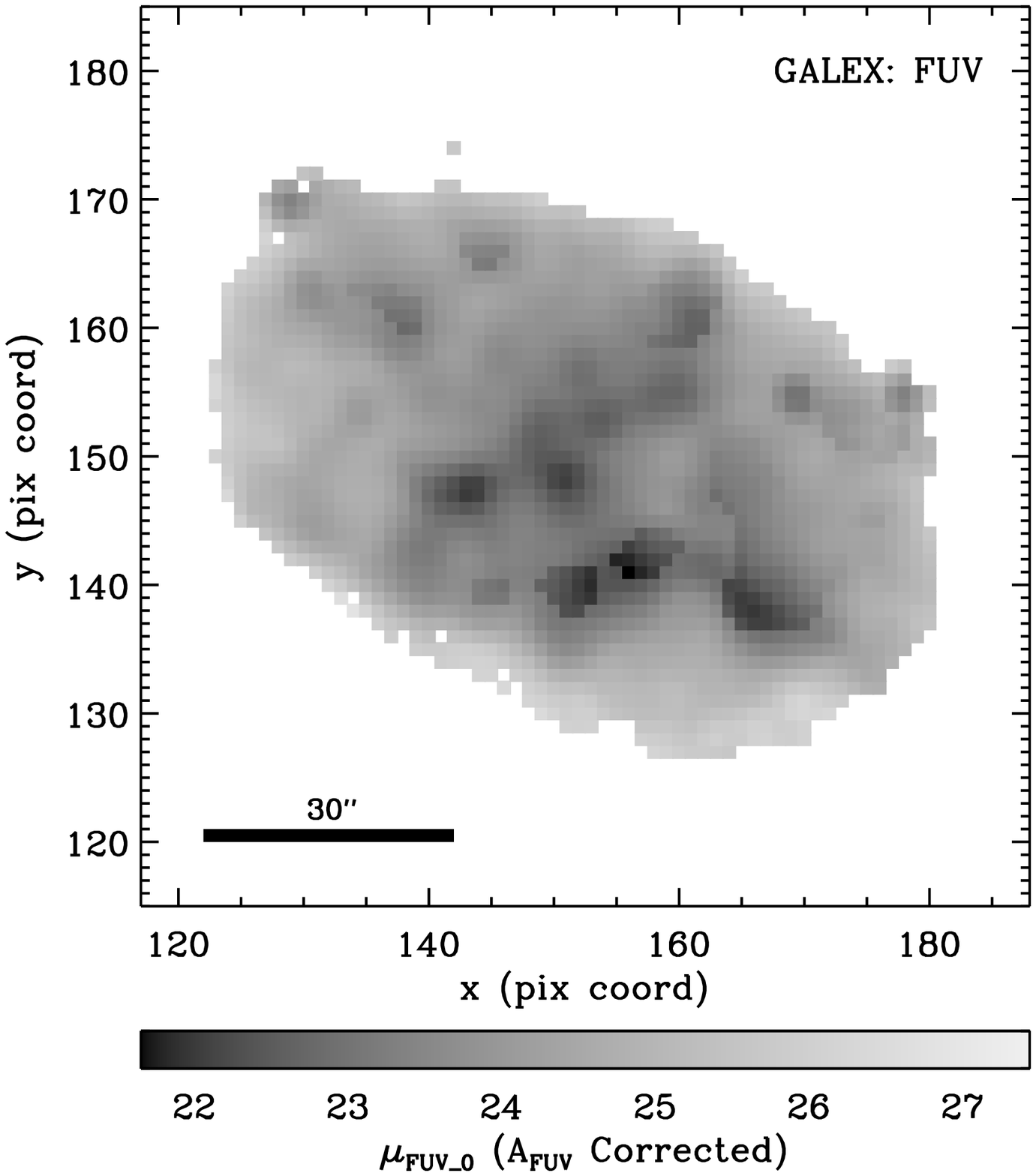,width=0.45\textwidth}}
\centerline{\psfig{file=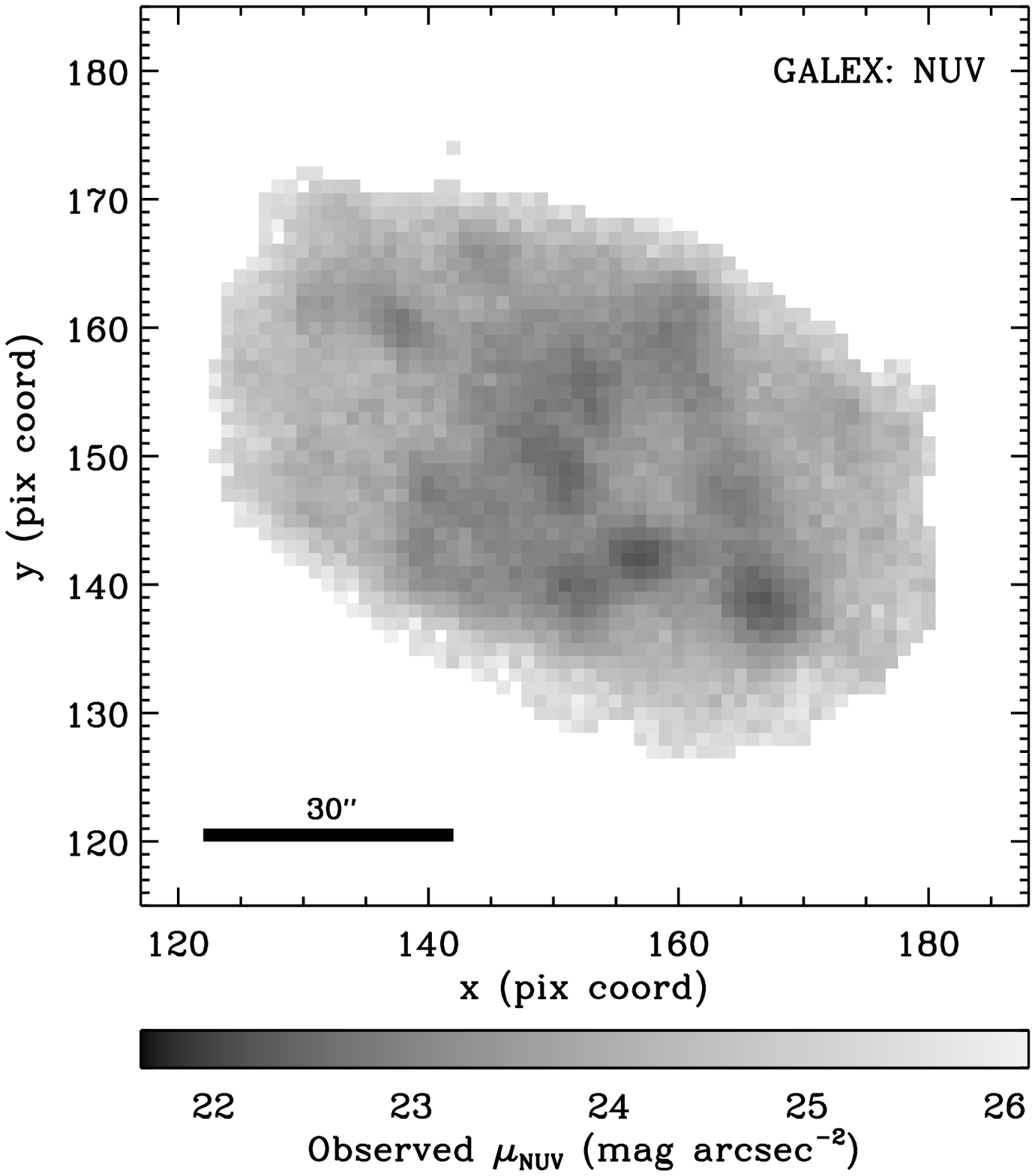,width=0.45\textwidth}
  \psfig{file=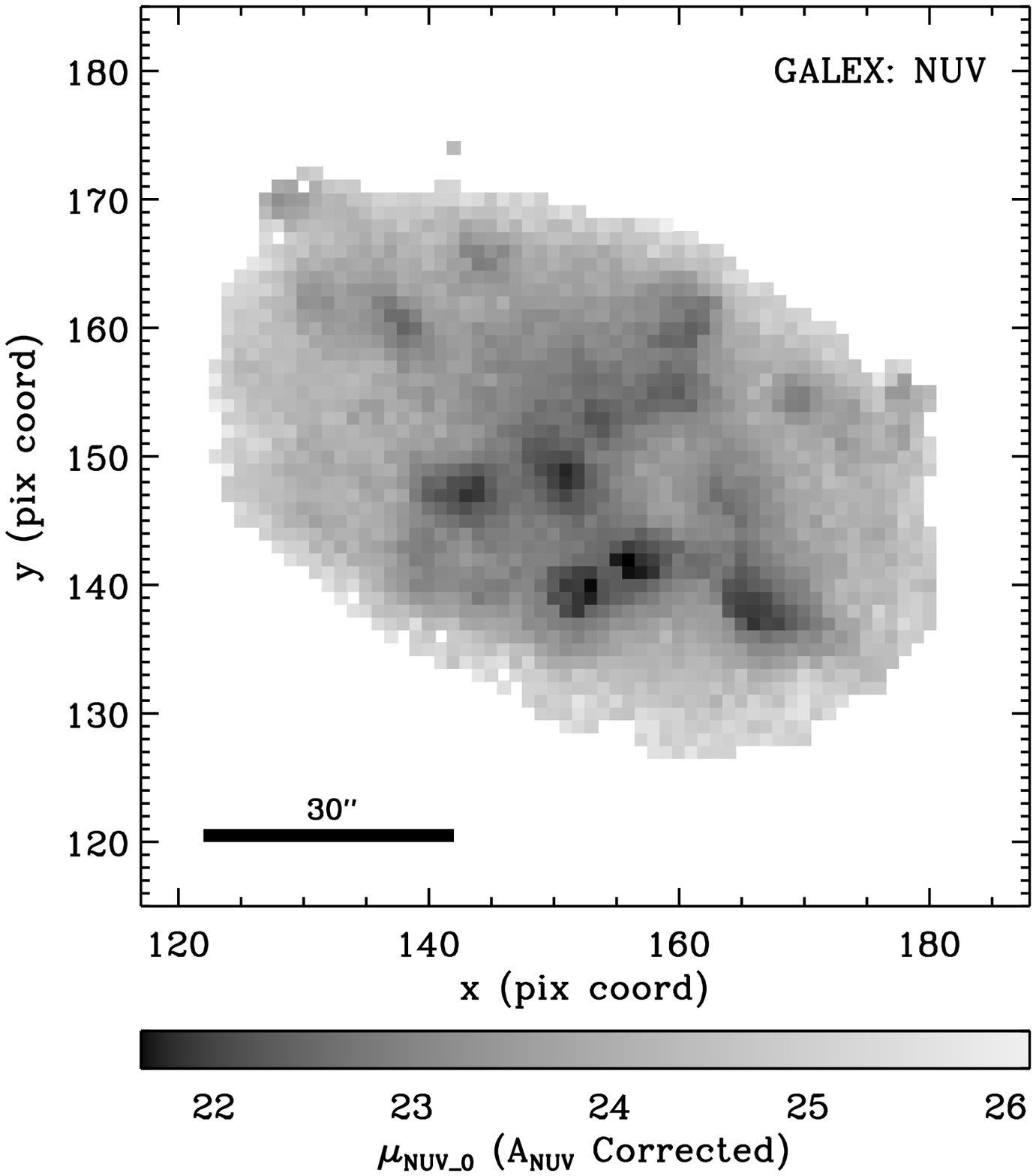,width=0.45\textwidth}}
\caption{
Comparison of observed (\emph{left}) and extinction-corrected (\emph{right})
images in the \emph{GALEX} FUV (\emph{top}) and NUV (\emph{bottom}) filters.
}\label{img_fuv}
\end{figure*}

\begin{figure*}
\centerline{\psfig{file=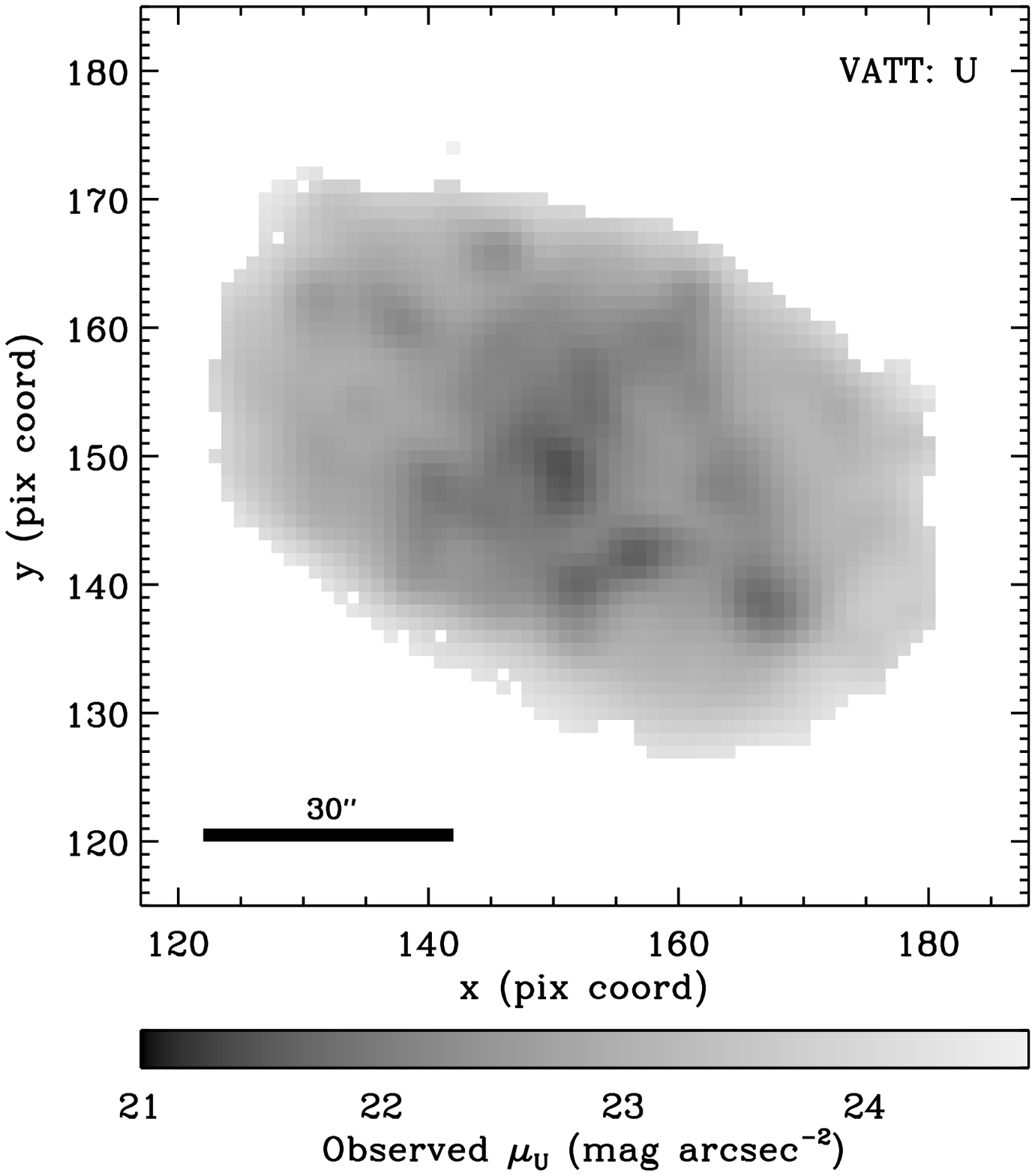,width=0.45\textwidth}
  \psfig{file=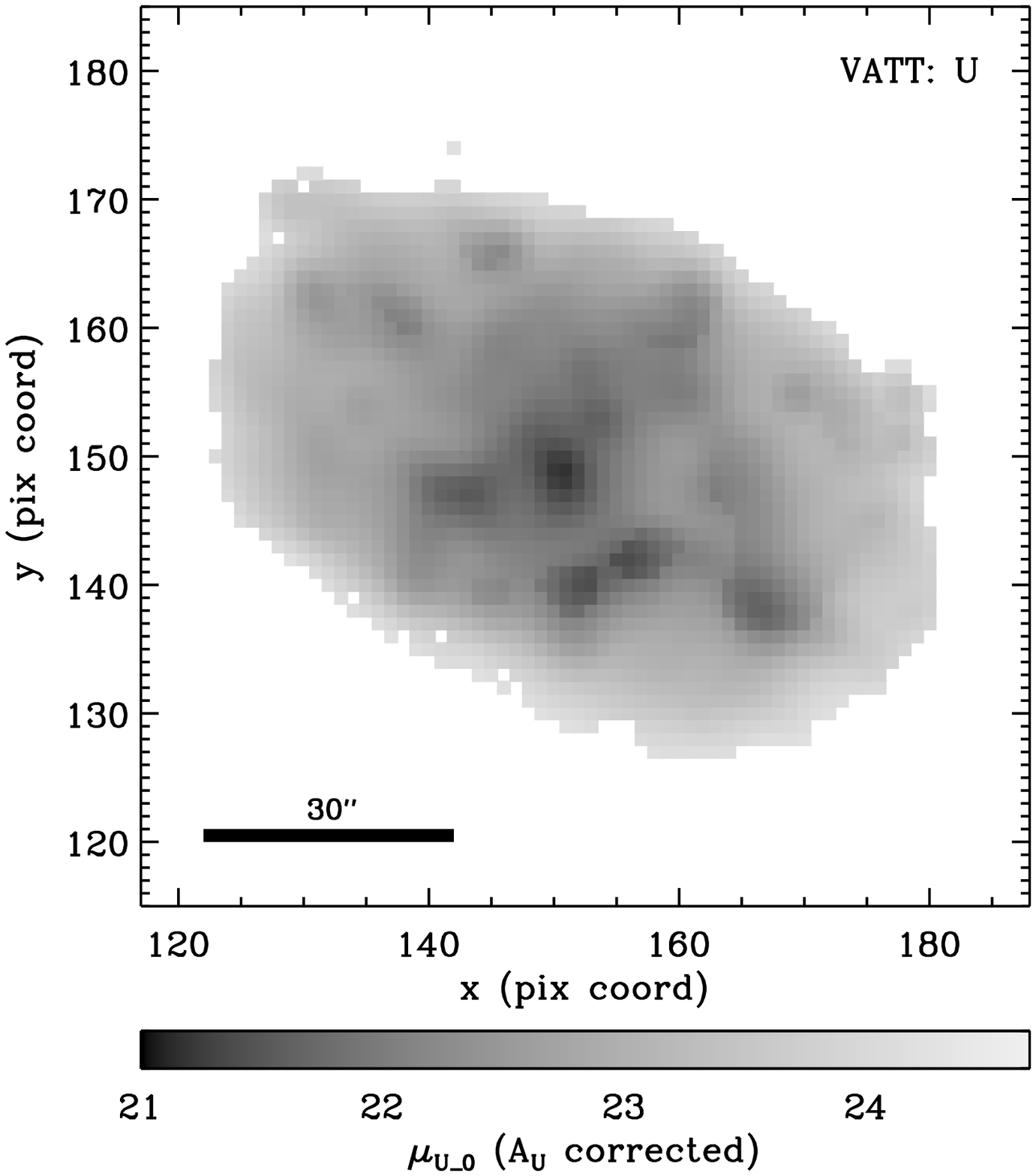,width=0.45\textwidth}}
\centerline{\psfig{file=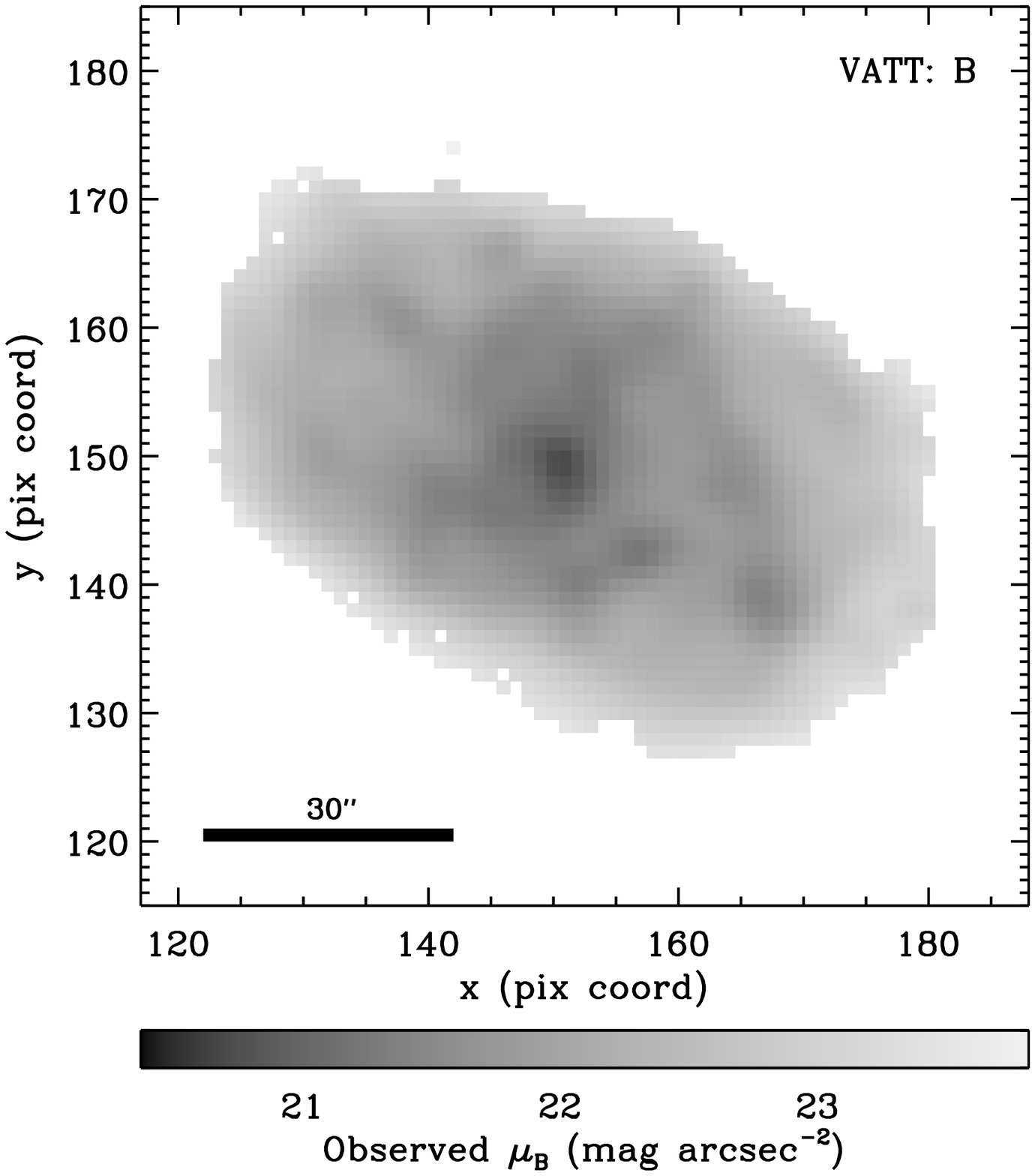,width=0.45\textwidth}
  \psfig{file=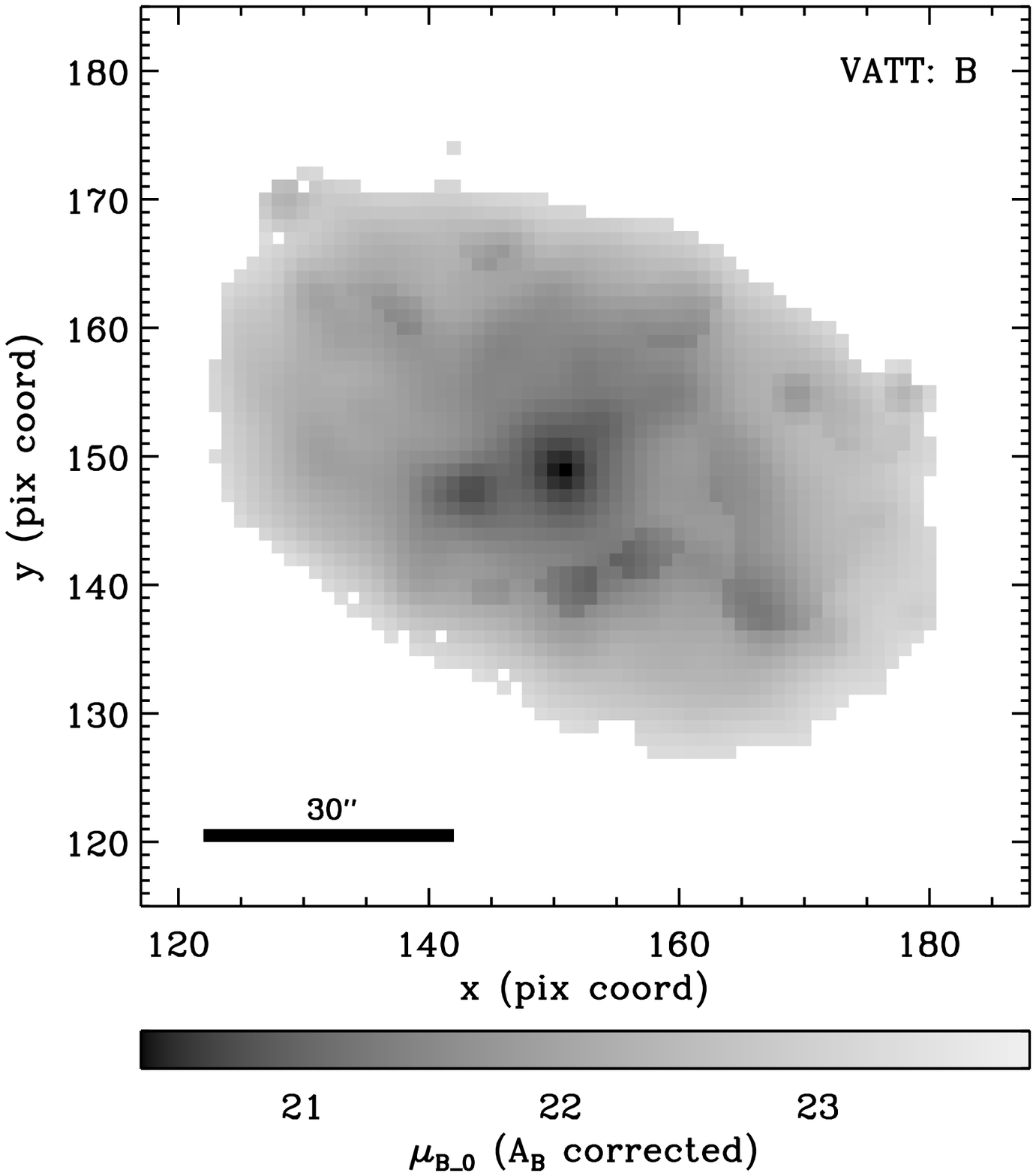,width=0.45\textwidth}}
\caption{
Comparison of observed (\emph{left}) and extinction-corrected (\emph{right})
images in the VATT \uu\ (\emph{top}) and \bb\ (\emph{bottom}) filters
at \emph{GALEX} resolution.
}\label{img_u}
\end{figure*}

Since NGC\,959 is classified as an Sdm galaxy in the RC3, we assume that it 
has an average metallicity between the SMC and the MW value.  We therefore 
adopt the LMC extinction curve to estimate the extinction in filters other 
than \vv.  Using the LMC extinction ratios $A_{\rm filter}/A_V$ from 
Table~2, we scale the \vv-band extinction for each pixel as:
\begin{equation}\label{eq_av}
  A_{\rm filter}=A_V \times (A_{\rm filter}/A_V)\,,
\end{equation}
where, \av\ was computed using \eqref{av_obs}.  \figrefm{img_fuv}{img_r} 
show the \emph{uncorrected} (left panels) and the 
\emph{extinction-corrected} images (right panels) for NGC\,959 from the FUV 
to \rr\ at \emph{GALEX} NUV resolution.  In all filters, the SF-regions 
become more clearly visible in the extinction corrected images.  
Compared to the optical \vv-band and the MIR 3.6 and 4.5\,\mum\ images 
(\figref{img_v}), in which the highest surface brightness is reached in the 
galaxy center, the brightest regions in the extinction-corrected 
\emph{GALEX} FUV and NUV images are distributed all over the galactic disk.  
Since these FUV and NUV filters are especially sensitive to young stellar 
populations (see \figref{model}), this indicates that most of the recent 
star formation occurred in the galaxy disk, and \emph{not} in its nuclear 
region.  The color composites of \figref{colimg} and \figref{hst} show that 
the extinction-corrected FUV and NUV images are clearly tracing stellar 
populations that appear bluer than other regions.  The images in \uu, \bb,
and \rr\ (\figreftwo{img_u}{img_r}) show the transition of the dominant 
emission from young stellar populations in the SF-regions to older stellar 
populations in the bulge and center of the galaxy (\figref{colimg}a).  
While the extinction corrected image in \uu\ is still sensitive mostly to 
young stellar populations, the \rr-band image shows that the older stellar 
populations in the small bulge of NGC\,959 become the dominant light source 
at redder wavelengths, as expected.

Among the FUV--\rr\ filters, the largest morphological change as a result
of the dust-extinction correction occurs in the optical images, especially
in the \bb-band image.  In the 
FUV, NUV, and \uu, the light from the younger 
stellar populations dominates even without applying our dust-extinction 
correction.  Correcting for dust extinction therefore strengthens already 
discernable galactic structures, but does not dramatically change the 
apparent morphology \citep[cf.,][]{windhorst02}.  The \rr-band image, on the 
other hand, is dominated by light from older stellar populations and suffers
to a lesser degree from the effects of dust.  The \bb-band samples both 
younger and older stellar populations (see \figref{model}) and suffers a 
larger dust extinction than \vv\ (see \figref{extinction} and Table~2).  
After correction for extinction, light from younger stellar populations that 
is initially largely hidden behind the dust becomes visible, causing the 
galaxy morphology to change  relatively more in \bb\ than in other filters.

\begin{figure*}
\centerline{
  \psfig{file=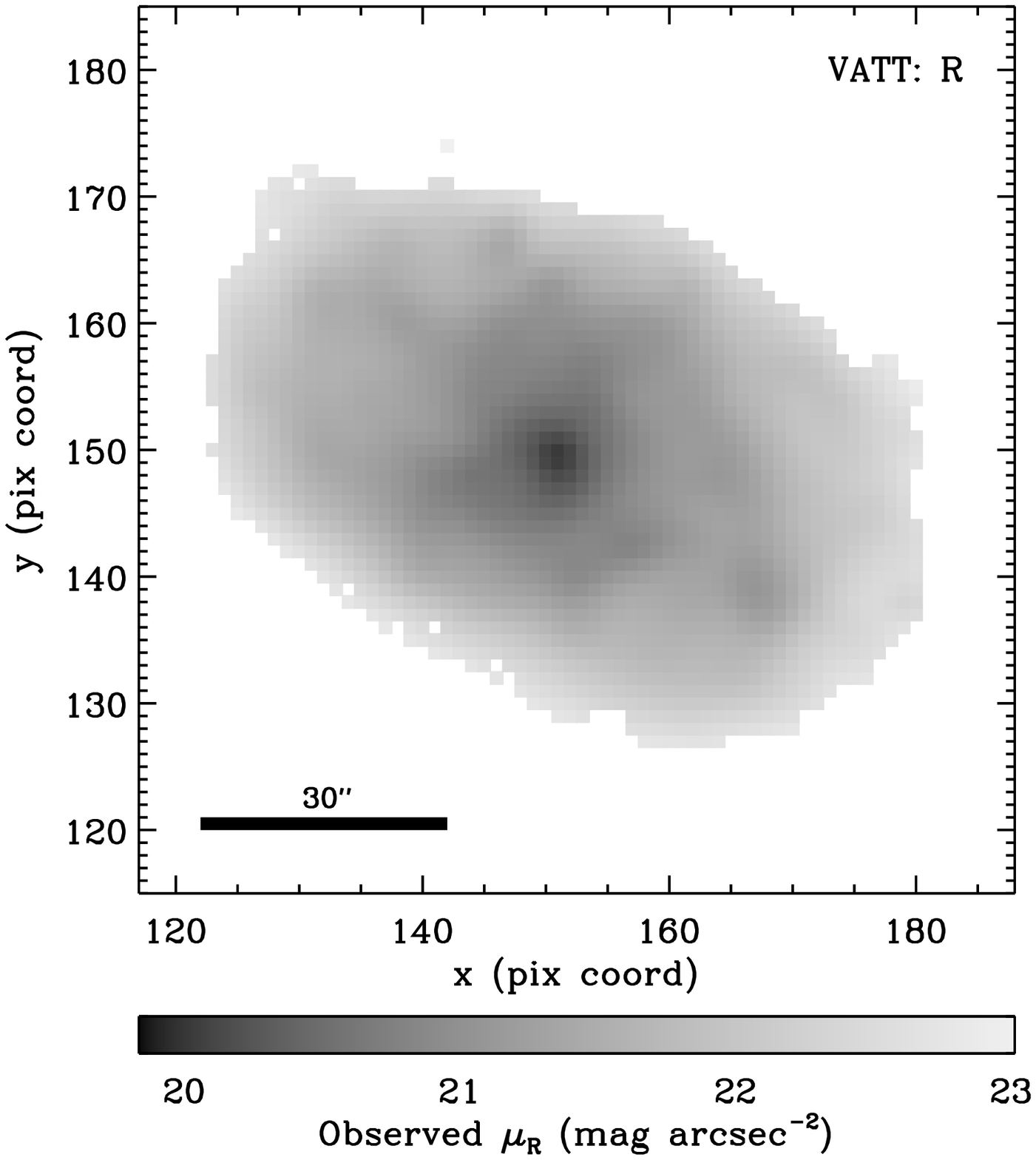,width=0.45\textwidth}
  \psfig{file=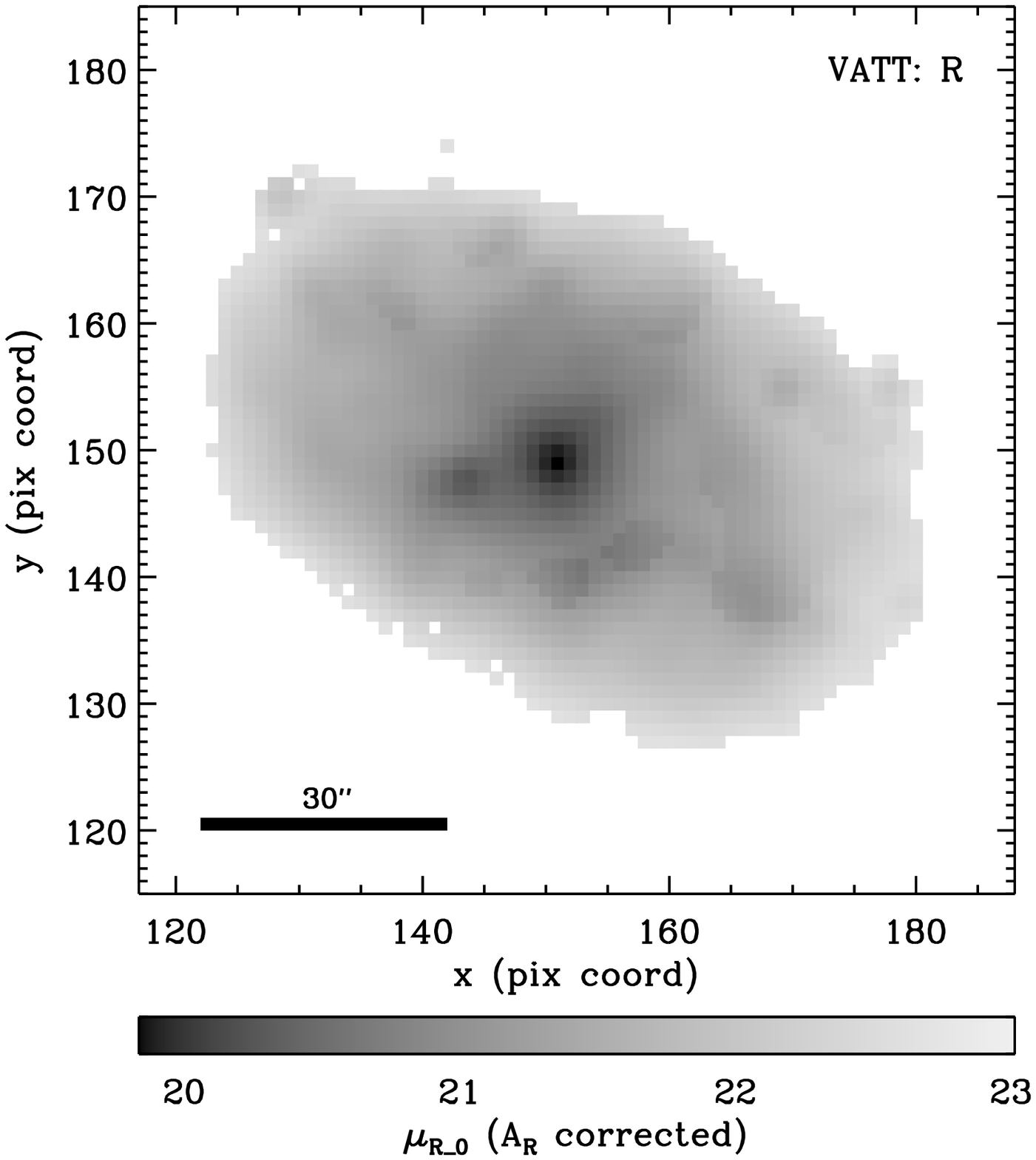,width=0.45\textwidth}}
\caption{
Comparison of observed (\emph{left}) and extinction-corrected (\emph{right})
images in the VATT \rr\ filter at \emph{GALEX} resolution.
}\label{img_r}
\end{figure*}

Previous studies have compared the morphological appearance in \bb- and NIR 
$H$- or $K_s$-bands \citep[e.g.,][]{block99, elmegreen99, block01, buta01, 
eskridge02, block04, seigar05}.  \citet{eskridge02} compared the 
morphological classifications in \bb\ and $H$ for $\sim$\,200 spiral
galaxies from the Ohio State University (OSU) Bright Spiral Galaxy Survey.   
They found a relatively good correlation between the classifications in the 
optical and NIR (see their Fig.~2 and Table~2).  On average, the $H$-band 
classification was found to be $\sim$\,1 T-type earlier than the optical one.  
Other studies, however, found no correlation between optical Hubble type 
\citep{hubble26} and dust-penetrated morphological classes for smaller 
samples of 14--36 galaxies observed in $K_s$ \citep[e.g.,][]{block99, 
elmegreen99, block01, buta01, block04, seigar05}.  Since the Hubble types 
are based on blue photographic plates, the Hubble classification can be 
significantly affected by dust extinction \citep[e.g.,][]{windhorst02}.  
$K_s$-band images suffer only 10\% of the extinction in \vv\ 
\citep[\figref{extinction} and, e.g.,][]{martin90} and therefore show the 
stellar distribution with 
much smaller effects from dust.  Therefore, if the amount of dust in (and in 
front of) a galaxy is significant, the optical \bb-band is affected 
accordingly, possibly resulting in drastically different morphology than 
suggested by the NIR classification.  As this was seen in only a small 
subset of the OSU Bright Spiral Galaxy Survey samples \citep{eskridge02}, 
the galaxy samples used in the $K_s$-band studies by, e.g., \citet{block99} 
may have selected dustier galaxies.

The apparent \bb-band morphology of NGC\,959 before and after extinction 
correction (bottom panels of \figref{img_u}) does not differ as drastically 
as the optical versus $K_s$ morphologies reported in the above studies.  But
some regions, such as the SE side of the bulge at 
[$x$,\,$y$]\,$\simeq$\,[142,\,148], become much more prominent after 
extinction correction and the distribution of light becomes similar to that 
seen at 3.6 and 4.5\,\mum\ (see bottom panels of \figref{img_v}).  This 
means that if a galaxy contains a large amount of dust along major 
structures, such as a bar or spiral arms, it is possible that the \bb-band 
morphology can change significantly after correction for dust.

\section{DISCUSSION}

Correction for dust extinction is an important, yet challenging issue when 
studying stellar populations in galaxies, because extinction has a similar 
effect as stellar population age \citep[e.g.,][]{gordon97} and metallicity 
\citep[e.g.,][]{worthey94, kaviraj07b}.  Failure to correct for extinction 
will render analyses of stellar populations highly uncertain.  Commonly, the 
FIR/UV flux ratio is used to measure ``the'' extinction within a galaxy, 
given as a single averaged extinction value \citep[e.g.,][]{buat96, 
calzetti00, boselli03, kong04}, or as a one-dimensional (i.e., azimuthally 
averaged) radial extinction profile \citep[e.g.,][]{boissier04, boissier05}.  
This method assumes that the distribution of dust is relatively uniform 
across the entire galaxy, or representable by a simple radial extinction 
gradient.  In reality, the distribution of dust is complex, following 
galactic structures such as SF-regions, bars, and spiral arms.  It may also 
be affected by nearby companions or satellite galaxies.  Therefore, a more 
detailed analysis of the spatial distribution of dust is needed.  To obtain 
a two-dimensional distribution of dust extinction, some studies 
\citep[e.g.,][]{scoville01, calzetti05} use ratios of Hydrogen recombination 
lines such as H$\beta$/H$\gamma$, \ha/\hb, or \ha/Pa$\alpha$, that have 
known intrinsic values \citep{osterbrock89}.  This method is applied to 
measure dust extinction in \HII\ regions within some of the nearest 
galaxies, but is applicable over only a small fraction of an entire galaxy.  
Our method, based on the observed \vv-to-3.6\,\mum\ ratio and model SEDs, is
able to map the \emph{full} two-dimensional distribution of dust.  
\figref{av} and \figref{av_hr} demonstrate that the actual distribution of 
extinction within NGC\,959 does not follow a simple radial trend and that  
dust is not concentrated only in the most actively star-forming regions; it 
is present throughout, tracing the complicated structures from the galaxy 
center all the way to the outer regions of the disk, where the S/N in the
images becomes the limiting factor.  

\begin{figure*}
\centerline{\psfig{file=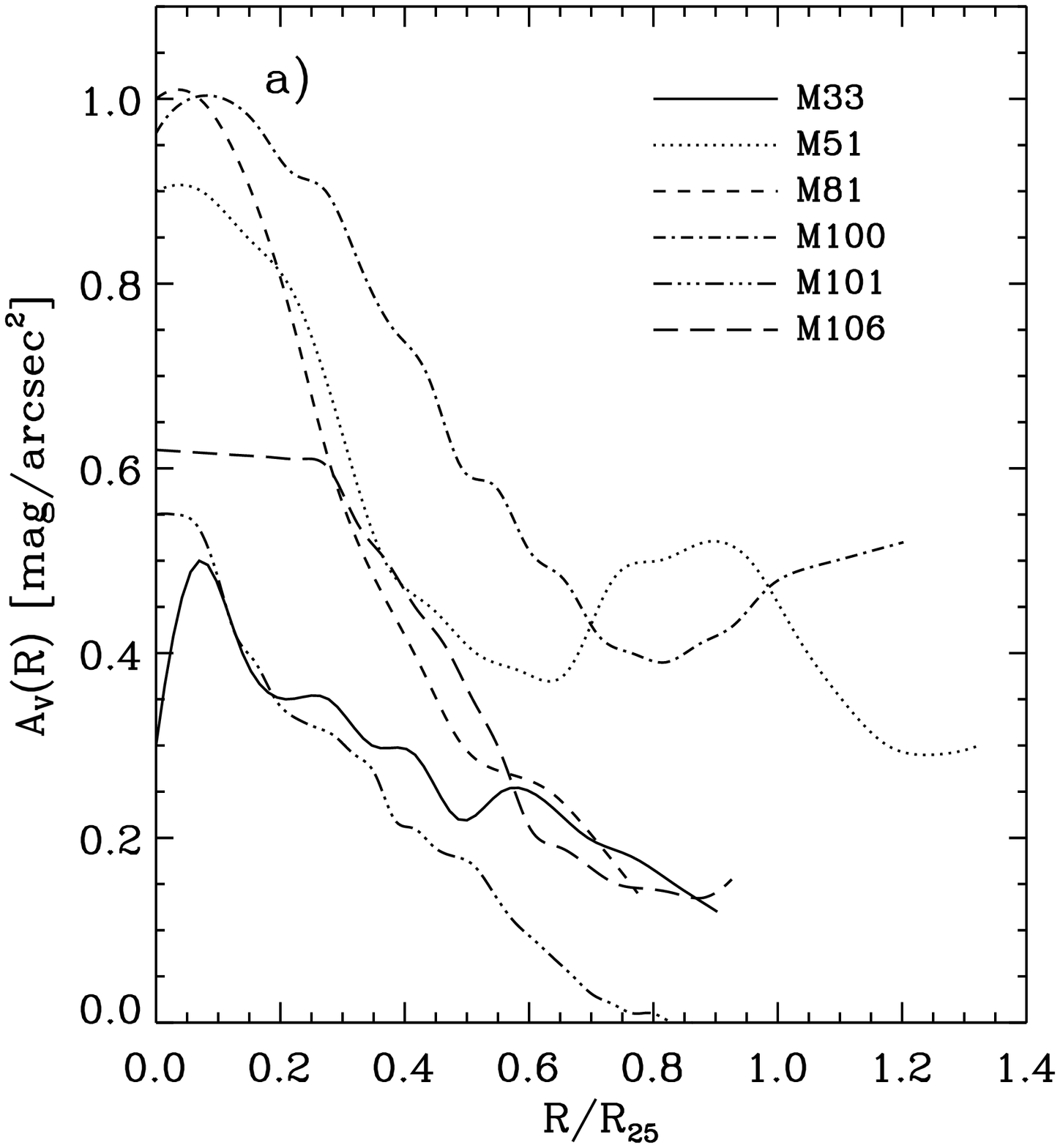,width=0.45\textwidth}
 \psfig{file=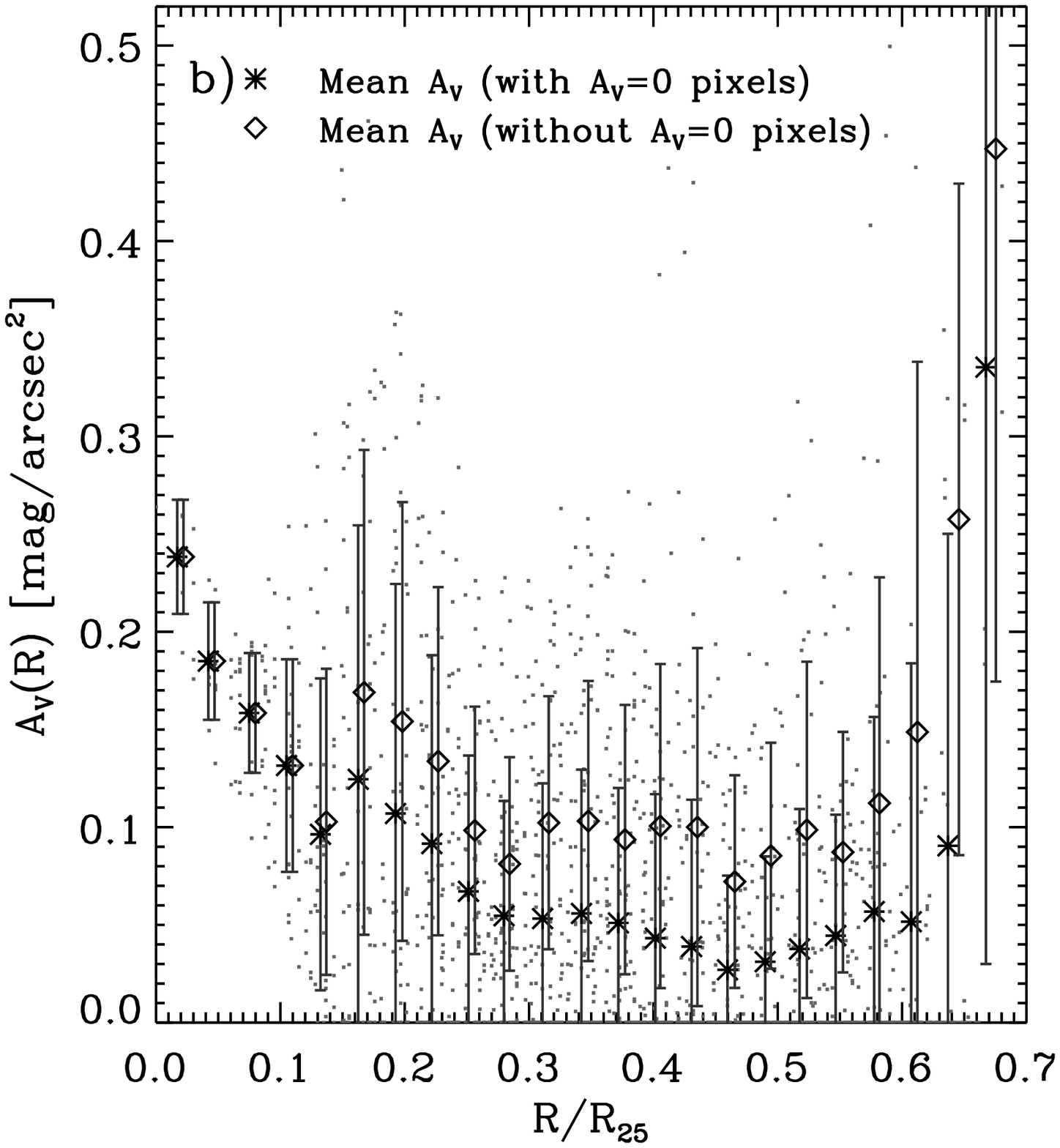,width=0.45\textwidth}}
\caption{
\emph{a}) Radial extinction profiles for the six nearby spiral galaxies from
\citet{boissier04}, with distances expressed in units of $R_{\rm 25}$.
While some galaxies show a bump or an upturn in the outermost regions of
their disk, the extinction generally decreases from the center of a galaxy
outward.  \emph{b}) Visual extinction, \av, as a function of distance from
the center of NGC\,959.  Each dot represents the extinction measured in a
single pixel.  Averages within 0.1\,kpc annuli are plotted as asterisks when
including pixels with \av\,=\,0~mag\,arcsec$^{-2}$, and as open diamonds when
excluding such pixels.  The error bars represent the standard deviation
within each annulus.  Twelve pixels with
\av\,$\gtrsim$\,0.5~mag\,arcsec$^{-2}$ and $R$\,$\gtrsim$\,1.3\,kpc
($R$\,$\gtrsim$\,0.6\,$R_{\rm 25}$),
are not shown here.  Note that the very outskirts of NGC\,959 may suffer
from the light-blending from pixels with S/N\,$\lesssim$\,3 in one or more
of the filters, hence rendering the \av\ measurements more uncertain.
}
\label{rad_av}
\end{figure*}

At this point, we would have liked to compare our results for NGC\,959 with 
extinction measurements from other methods.  While \citet{esipov91} and 
\citet{taylor05} have studied this galaxy, they did not analyze the internal 
dust extinction, so we cannot directly compare our results to prior work.  
However, radial extinction profiles have been analyzed in other galaxies 
\citep[e.g.,][]{jansen94, boissier04, boissier05, calzetti05, holwerda09}.  
\citet{jansen94} used the special geometry offered by a nearly edge-on disk
and a large bulge to demonstrate that the maximum extinction in the dust
lanes of two galaxies (UGC\,3214 and UGC\,3065) decreased outward with 
distance from the minor axis.  \citet{boissier04} used the azimuthally 
averaged FIR/UV flux ratio to measure radial extinction profiles for six 
nearby late-type spiral galaxies.  Their results are reproduced in 
\figref{rad_av}a, where radius is expressed in units of $R_{25}$, the major 
axis radius at the $m_B$\,=\,25.0~mag\,arcsec$^{-2}$ isophote as listed in
the RC3.  Each galaxy shows a general trend of decreasing extinction from 
the center to the outer regions of the galaxy.  \citet{boissier05} studied 
the radial profiles of extinction in M\,83 using different methods --- the 
Total-IR/FUV luminosity ratio, the UV spectral slope, and the Balmer 
decrement.  Their Fig.~2 of shows that all three methods give 
similar results:
a general decrease of extinction with  radius, with a small upturn at the 
outer edge of M\,83's disk.  \citet{calzetti05} measured the 
H$\alpha$/Pa$\alpha$ ratio in \HII\ regions to study the distribution of 
dust extinction in M\,51.  Their Fig.~14 shows the distribution of 
individual extinction measurements as a function of radial distance from the 
center of M\,51.  While there are some relatively highly extincted \HII\ 
regions at outer radii --- corresponding to the bump at $R/R_{25}$\,$\sim 
$\,0.8 in the profile for M\,51 in our \figref{rad_av}a --- the distribution 
does follow the general decreasing trend with increasing radius.  On the 
other hand, \citet{holwerda09} used an occulting galaxy pair to measure the 
distribution and amount of dust via the optical depth against the background 
galaxy, providing measurements that extend beyond the optically visible disk
of the foreground galaxy (see their Fig.~2).  They showed that large amounts 
of dust can exist even in the outermost parts of spiral galaxies, where 
these dust features are usually undetectable by other means (see their 
Figs.~11 and 12).

If we see a similar radial trend for our estimated dust extinction in 
NGC\,959 as in the studies above, it would lend additional credence to the 
reliability of our method.  \figref{rad_av}b shows the radial distribution 
of \av\ for each pixel in NGC\,959.  The galaxy center is located at 
[$x$,\,$y$]\,=\,[151,\,148], and the radius is expressed in units of 
$R_{25}$.   Also indicated are mean extinction values computed within 
0.1\,kpc bins in radius, with (asterisks) and without (open diamonds) 
including pixels with \av\,=\,0~mag\,arcsec$^{-2}$.  The error bars 
represent standard deviations for the distribution of \av\ values in each 
such bin.  While \av\ for individual pixels spans a wide range at each 
radius, the azimuthally averaged extinction, $A_{V,\rm mean}(R)$, clearly 
decreases from the center outward.  At the outermost radial bin, however, 
the average extinction shoots up to 
$A_{V,\rm mean}(R)$\,$>$\,0.3.  This is caused 
mostly by the pixels with high \av\ within the ``interesting regions''
discussed in \S3.6.3 (see also \figref{av}), which may or may not be 
associated with NGC\,959.  The larger uncertainties for the outermost bins 
also reflect the fact that these pixels only marginally exceed our minimum 
S/N criterion in one or more of the filters, hence possibly affecting the 
reliability of \betav\ and \av.  Yet, some pixels with relatively high \av\ 
(0.2\,$\lesssim$\,\av\,$\lesssim$\,0.5) indicate the existence of dust in 
the outermost regions of NGC\,959, perhaps analogous to the findings of 
\citet{holwerda09}.  The extensive tests described in previous sections, as 
well as the general agreement with results from other studies --- although
they involved different galaxies --- give us confidence that our method 
produces reliable measurements of the spatial distribution of extinction by 
dust within a galaxy.

An important lesson from the application of our method to NGC\,959 is that 
we are able to map the two-dimensional distribution of dust extinction even 
from the low-resolution images, which show \emph{no} conspicuous dust 
features silhouetted against the stars in \figref{colimg}b.  Our 
initial expectation was that the extinction map created with our method 
might be relatively featureless, with specific dust features smoothed out.  
Instead, a map with a complicated pattern that follows the galactic 
structure emerged (see \figref{av}).  A similar, but more detailed dust 
distribution is recovered when the analysis is repeated on images at the 
higher IRAC 3.6\,\mum\ resolution (\figref{av_hr}).  Pixels with large \av\ 
in \figreftwo{av}{av_hr} clearly trace the location of silhouetted dust
features in the higher resolution VATT (\figref{colimg}a) or \emph{HST} 
(\figref{hst}) color composites of NGC\,959.  These results stress that the 
contrast between regions with high and relatively low extinction can be 
large within a galaxy disk.  At resolutions of 2\farcs2 
(\emph{Spitzer}/IRAC) and 5\farcs3 (\emph{GALEX}) or $\sim$110--250\,pc at 
the distance of NGC\,959, our method is able to reliably and meaningfully 
generate a two-dimensional distribution of the dust extinction.  Such modest
resolutions are accessible and our method will be applicable in galaxies 
well beyond the Local Group, where the individual dust features may no 
longer necessarily be spatially resolved.  

Our method will also be useful to study dust extinction in galaxies at much 
larger distances.  Once \emph{HST}/WFC3 and \emph{JWST} are in operation, 
galaxies at $z$\,$\simeq$\,0.3--0.4 will be easily accessible for such 
studies, and galaxies at even larger redshifts might be reachable.
WFC3\footnote[13]{
  Space Telescope Science Institute (STScI), Wide Field Camera~3: \\
  \texttt{http://www.stsci.edu/hst/wfc3}
} 
has two imaging channels: UVIS covers 0.2--1.0\,\mum\ with a pixel-scale of 
0\farcs04 pixel$^{-1}$, while NIR covers 0.9--1.7\,\mum\ at 0\farcs13 
pixel$^{-1}$.  \emph{JWST}\footnote[14]{
  NASA, The \emph{James Web Space Telescope}: \\
  \texttt{http://www.jwst.nasa.gov}
}
will also have two imaging cameras: NIRCam, covering 0.6--5\,\mum\ at 
0\farcs032 or 0\farcs065 pixel$^{-1}$ (Short versus Long Wavelength 
Channel), and MIRI, which covers 5--27\,\mum\ at $\sim$\,0\farcs11 
pixel$^{-1}$.  At $z$\,$\gtrsim$\,0.3, the rest-frame \vv\ and 3.6\,\mum\
bands 
shift progressively further into the near- to mid-IR regime covered by 
\emph{HST}/WFC3 and \emph{JWST}, allowing one to apply our method to large
numbers of intermediate redshift galaxies.  The only limiting factor is the 
apparent size of the galaxy at these redshifts.  Up to $z$\,$\simeq$\,1.6,
the 
angular scale becomes smaller \citep{wright06} and each pixel samples a 
larger surface area within a galaxy.  At $z$\,$\simeq$\,0.4, the central 
wavelengths of the \vv\ and 3.6\,\mum\ bands shift to 0.77\,\mum\ and 
5.04\,\mum, where UVIS and MIRI are expected to deliver resolutions of 
0\farcs08 and 0\farcs195 FWHM, respectively.  This corresponds to 
$\sim$\,420\,pc and $\sim$\,1\,kpc, where the latter sets the relevant 
resolution for our method.  While dust lanes will certainly not be visible,
we expect our method to still produce meaningful maps of the variations in 
extinction on scales of $\sim$\,1\,kpc, as long as a galaxy is at least 
several kpc in diameter.

Another application of our method is to investigate the result of 
\citet{taylor05}, that the color in the outer regions becomes redder in the 
majority of late-type spiral and irregular galaxies.  This reddening may be 
caused either by a change in stellar populations or be due to the presence 
of dust, which is usually only apparent in higher resolution images when 
silhouetted against a relatively bright stellar background \citep[as in,
e.g.,][]{holwerda09}.  In NGC\,959, a nearly face-on late-type spiral 
galaxy, we found evidence for the existence of a moderate amount of 
extinction in the outermost regions of its disk (\figref{av}).  
Our method will also be useful to study the dust content of elliptical and 
lenticular galaxies.  In recent years, large amounts of dust were 
discovered to exist in elliptical and lenticular galaxies, as well as in 
the halo of spiral galaxies \citep[e.g.,][]{kaviraj07a, oosterloo07a, 
oosterloo07b, emonts08}.  Since very little SF-activity is ongoing in 
these galaxies, methods based on the FIR/UV flux ratio or on Hydrogen 
recombination lines are not as useful as they are for actively star-forming 
galaxies.  Our method, which only depends on images in \vv\ and 3.6\,\mum\ 
filters, is well-suited for a study of the distribution of dust within 
early-type galaxies.  We will present a more detailed study of dust 
distribution in spiral galaxies and in a small number of elliptical and
lenticular galaxies --- a total of 45 galaxies of all types --- in a 
subsequent paper (Tamura \etal\ 2009c, in preparation).

A potential future use of our method is to estimate the distribution of dust 
in simulated galaxy models \citep[e.g.,][]{croton06}.  A detailed 
two-dimensional analysis of the dust extinction in a large number of 
galaxies would help understand the properties of dust for different types of 
galaxies or galactic structures, such as SF-regions, bars, spiral arms, and 
inter-arm regions.  Current simulations are able to model the 
extinction-free SEDs for many galaxies, while the treatment of internal dust 
extinction is still broad-stroke.  For such models, it will be useful to 
construct a database of galaxies of different morphological type and mass,
to constrain age, metallicity, and the amount and spatial distribution of 
the dust.

\section{CONCLUSIONS}

In this paper, we presented a new method for estimating the extinction by 
dust within galaxies by comparing the observed \vv-to-3.6\,\mum\ flux ratio, 
\betav, to theoretical SED models.  Using a pixel-based analysis, our method 
is able to estimate the two-dimensional distribution of dust extinction 
within a galaxy.  As a proof of concept, we applied this method to NGC\,959,
a nearby late-type spiral galaxy.  From a pCMD, constructed using an 
additional \uu-band image, we robustly selected pixels dominated either by 
the light from younger stellar populations, or from older ones.  Since their 
intrinsic \vv-to-3.6\,\mum\ flux ratios differ, they were treated separately 
in our analysis.  We presented a two-dimensional map of the visual 
extinction, \av, that closely resembles the observed distribution of 
SF-regions and underlying galactic structures (including a newly identified
bar), and which traces the distribution of 8.0\,\mum\ PAH emission.
Although dust features are inevitably smoothed out to some extent due to
light-blending and the low spatial resolution of the images, we were able to
construct a two-dimensional extinction map with sufficient detail to 
delineate the structure
of dust features within the disk of NGC\,959.  We then presented original
and extinction-corrected views of NGC\,959 from the FUV through MIR.
Through a series of tests, we demonstrated the validity of our results and 
method.

Our method has several advantages over other methods based on, e.g., the
FIR/UV flux ratio, UV spectral slope, or Hydrogen recombination line 
ratios.  At its core, our method only depends on images in two relatively 
common broadband filters, \vv\ and 3.6\,\mum\ ($L$-band), and is therefore 
applicable continuously across the face of a galaxy.  We exploit the fact 
that the wavelength-dependent extinction by interstellar dust reaches a 
minimum near 3.6\,\mum\ while it increases toward shorter wavelengths.  
In \vv-band, we are sensitive to dust extinction, but fairly insensitive 
to age and metallicity effects compared to UV--blue filters.  
We demonstrated that the intrinsic \vv-to-3.6\,\mum\ flux ratio, \bzero, 
is well-behaved over a wide range in stellar age and metallicity.  While 
\bzero\ depends on age more strongly than on metallicity, this ratio stays 
relatively constant for older ($t$\,$\gtrsim$\,500~Myr) stellar 
populations, and occupies a relatively narrow range for younger stellar 
populations.  As a result, we can simply 
compare the observed and intrinsic \vv-to-3.6\,\mum\ flux ratios (after 
taking mixing or superposition of stellar populations into account) to 
estimate the amount of dust extinction, \av, in each pixel.  To translate 
\av\ to bluer filters, knowledge of the metallicity become more important 
(or, alternatively, the uncertainty increases) due to the 
metallicity-dependence of the extinction curve.  
This simplicity allows our method --- which is mostly automated with only
a few manual parameter-settings for each galaxy --- to be applied to a
large number of galaxies in a very short time.

Since our method does not 
require visual confirmation or identification of individual dust features, 
it is applicable to any galaxy beyond the Local Groups, if rest-frame \vv\ 
and 3.6\,\mum\ images are available with at least several resolution 
elements across that galaxy.  This offers the possibility of applying our 
method to \emph{HST}/ACS, \emph{HST}/WFC3, and \emph{JWST} NIRCam and MIRI 
images to study the two-dimensional distribution of dust not only in the 
local universe, but also for higher redshift galaxies.

\vspace{0.5cm}

{\small This work was funded by NASA/ADP grant NNX07AH50G.  R.A.W.\ was 
supported in part by NASA/JWST grant NAG\,5-12460.  We thank 
V.A.Taylor-Mager for providing the data observed at the Vatican Advanced 
Technology Telescope (VATT): the Alice P.~Lennon Telescope and the Thomas 
J.~Bannan Astrophysics Facility.  We thank the referee for a careful reading 
and constructive comments that helped improve the paper significantly.  We
also thank Daniela Calzetti, Seth Cohen, Paul Eskridge, Nimish Hathi, and 
Russell Ryan for their help, comments, and discussion.  This study has 
made use of the NASA/IPAC Extragalactic Database (NED), which is operated 
by the Jet Propulsion Laboratory, California Institute of Technology, 
under contract with NASA, and has used NASA's Astrophysics Data System 
(ADS) bibliographic services.  Additional observations made with the 
NASA/ESA \emph{Hubble Space Telescope} were obtained from the data 
archive at STScI, which is operated by AURA, Inc., under NASA contract 
NAS\,5-26555.}



\end{document}